\documentclass[reprint,superscriptaddress,amsmath,amssymb,aps,pre]{revtex4-1}

\usepackage{amsmath}
\usepackage{graphicx}
\usepackage{dcolumn}
\usepackage{bm}
\usepackage{color}
\usepackage{bbold}
\usepackage{soul}
\usepackage{subfigure}
\newcommand{\newc}{\newcommand}
\newc{\beq}{\begin{equation}}
\newc{\eeq}{\end{equation}}
\newc{\kt}{\rangle}
\newc{\br}{\langle}
\newc{\beqa}{\begin{eqnarray}}
\newc{\eeqa}{\end{eqnarray}}
\newc{\longra}{\longrightarrow}

\makeatletter
\let\Hy@backout\@gobble
\makeatother

\begin{document}

\title{Geometric determination of heteroclinic and unstable periodic orbit classical actions}

\author{Jizhou Li}
\affiliation{Department of Physics and Astronomy, Washington State University, Pullman, WA, USA 99164-2814}
\author{Steven Tomsovic}
\affiliation{Department of Physics and Astronomy, Washington State University, Pullman, WA, USA 99164-2814}

\date{\today}

\begin{abstract}
Semiclassical sum rules, such as the Gutzwiller trace formula, depend on the properties of periodic, closed, or homoclinic (heteroclinic) orbits.  The interferences embedded in such orbit sums are governed by classical action functions and Maslov indices.  For chaotic systems, the relative actions of such orbits can be expressed in terms of phase space areas bounded by segments of stable and unstable manifolds, and Moser invariant curves.  This also generates direct relations between periodic orbits and homoclinic (heteroclinic) orbit actions.  Simpler, explicit approximate expressions following from the exact relations are given with error estimates.  They arise from asymptotic scaling of certain bounded phase space areas. The actions of infinite subsets of periodic orbits are determined by their periods and the locations of the limiting homoclinic points on which they accumulate.
\end{abstract}


\maketitle

\section{Introduction}
\label{Introduction}

The properties of sets of rare classical orbits can be extremely important in the study of chaotic dynamical systems~\cite{Poincare99}.  For example, classical sum rules over unstable periodic orbits describe various entropies, Lyapunov exponents, escape rates, and the uniformity principle~\cite{So07}.  The information which enters these classical summations are the stability properties and densities.  Such sets of orbits are also linked to the properties of the analogous quantized systems through the derivation of semiclassical sum rules.  A few cases are given by periodic~\cite{Gutzwiller71,Balian71,Berry76} and closed orbit sum rules~\cite{Du88a,Du88b,Friedrich89} that determine quantal spectral properties, and homoclinic (heteroclinic) orbit summations~\cite{Tomsovic91b,Tomsovic93} generating wave packet propagation approximations.  The interferences in such sum rules are governed by the orbits' classical action functions and Maslov indices, and thus this information takes on greater importance in the context of the asymptotic properties of quantum mechanics.  Various resummation techniques have been given to work with series which are often divergent in nature~\cite{Tanner91,Cvitanovic89,Berry90}.  Other studies exploring a fuller understanding of the interferences have also been carried out~\cite{Argaman93,Ozorio89,Bogomolny92,Sieber01,Muller04,Turek05,Muller05}.   Our interest in this paper is establishing a framework for understanding the relationships between periodic and homoclinic (heteroclinic) orbit actions and their action correlations.

A periodic orbit in a two degree-of-freedom system becomes either a single fixed point or an invariant set of points visited periodically in a two dimensional Poincare surface of section.  It is sufficient to concentrate on symplectic mappings on a plane and study the unstable fixed or periodic points under their application.  The fixed points play an advantageous role in this work due to convergence theorems in normal form coordinates.  The normal form transformation was first proved by Moser to converge inside a disk-shaped neighborhood of the fixed point denoted by $D_{0}$ hereafter~\cite{Moser56}.  Later, da Silva Ritter $\mathit{et.}$ $\mathit{al.}$ extended the convergence zone along the stable and unstable manifolds out to infinity~\cite{Silva87}.

Within the convergence zone are Moser invariant curves, which are images of invariant hyperbolas.  As  already noted by Birkhoff~\cite{Birkhoff27,Moser56,Harsoula15}, the self or mutual intersections between such invariant curves can support periodic orbits with arbitrarily large periods.  These periodic orbits accumulate alternatively on one or multiple homoclinic (heteroclinic) points in a homoclinic (heteroclinic) tangle.  In the limit of the orbital period going to infinity, the invariant curves become infinitely close to the stable and unstable manifolds of the fixed points.  The periodic orbits of such kind are said to be satellite to their respective homoclinic (heteroclinic) points~\cite{Silva87,Ozorio89}.  Da Silva Ritter $\mathit{et.}$ $\mathit{al.}$ developed a method for the numerical computation of satellite orbits supported by such curves in the quadratic map~\cite{Silva87}.  Therefore, every periodic orbit inside the convergence zone must be satellite to some homoclinic points, with its classical action closely related to that of the homoclinic orbit.  Recent work shows that the size of the convergence zone can be quantified in terms of the outermost Moser curves~\cite{Harsoula15}, i.e.~ones with the largest $QP$ normal form coordinates product, and the convergence zone can be numerically estimated using the outermost Moser curves as boundaries~\cite{Contopoulos15}.

Assuming a system is fully chaotic, the convergence zone should cover most, if not all, of the accessible phase space.  In that case, nearly all of the periodic orbits lie on Moser invariant curves, and each one can be treated as a satellite orbit of some particular set of hyperbolic fixed points.  Even if the system is not fully chaotic, the convergence zone can cover nearly all of the available phase space.  Figure 4 of~\cite{Harsoula15} gives an excellent example of the convergence zone covering almost all of the complex region of the homoclinic tangle of the H\'{e}non map~\cite{Henon69}, avoiding only a small region inside the last KAM curve.  Thus, a study of satellite orbits may often encompass nearly all periodic orbits of the system; i.e.~satellite orbits are not typically a small subset of the periodic orbits.  

In the quantum Baker's map wave packet autocorrelation functions can equivalently be expressed as a sum over periodic fixed points or homoclinic orbit segments with an exact one-to-one correspondence between terms~\cite{Oconnor92}.  Similarly, there is the same, though not exact, correspondence for the stadium billiard as there may be problems with orbits which approach bifurcations points too closely, i.e.~some of the orbits that come too close to the joint between the straight edge and curved hard walls~\cite{Tomsovic93}.   Thus, it is of significant interest to understand how the homoclinic (heteroclinic) and periodic orbits are related.

This work develops a framework for expressing the actions of satellite orbits in terms of the relative actions of homoclinic (heteroclinic) orbits, phase space areas bounded by stable and unstable manifolds, and Moser invariant curves.  These areas scale down with increasing periods, and the determination of the action of a leading satellite periodic orbit with small period is sufficient to approximate satellite orbits with larger periods; the numerical calculation of individual orbits becomes unnecessary to an excellent approximation.  As a final remark, note that Maslov indices can be incorporated into this framework, but are not considered in this paper in order to focus on the classical actions.  Previous studies of Maslov indices can be found in~\cite{Creagh90,Esterlis14,Mao92}.   

This paper is organized as follows.  Section~\ref{Homoclinic and heteroclinic orbits} sets the notation and basic definitions of homoclinic (heteroclinic) orbits and their actions.  Section~\ref{Heteroclinic orbit actions} is a generalization of the  MacKay-Meiss-Percival action principle~\cite{MacKay84a} for heteroclinic orbits, and expresses their actions as phase space integrals.  Section~\ref{Action difference between hyperbolic fixed points} concerns relative actions between two hyperbolic fixed points, and expresses them as phase space areas bounded by segments of the stable and unstable manifolds.  Section~\ref{Satellite periodic orbit actions} studies the satellite periodic orbits, and expresses their actions using phase space areas bounded by segments of the Moser invariant curves together with stable and unstable manifolds.  An approximation for orbits with large periods is also given, together with numerical verification.  Section~\ref{Implications Sieber-Richter} use the results from Sec.~\ref{Satellite periodic orbit actions} to derive an exact expression for the relative actions between Sieber-Richter orbit pairs~\cite{Sieber01}.  Some basic information on homoclinic (heteroclinic) tangles~\cite{Wiggins92,Easton86,Rom-Kedar90}, the MacKay-Meiss-Percival action principle~\cite{MacKay84a,Meiss92}, and normal form theory with satellite period orbits~\cite{Silva87} can be found in Appendices A and B.                           

\section{Homoclinic (heteroclinic) orbits and relative actions}
\label{Homoclinic and heteroclinic orbits}

This section lays out the paper's notation and a few basic concepts of homoclinic (heteroclinic) orbits in classical dynamical systems.  

\subsection{Homoclinic (heteroclinic) orbits}
\label{hho}

\begin{figure}[ht]
\includegraphics[width=8cm]{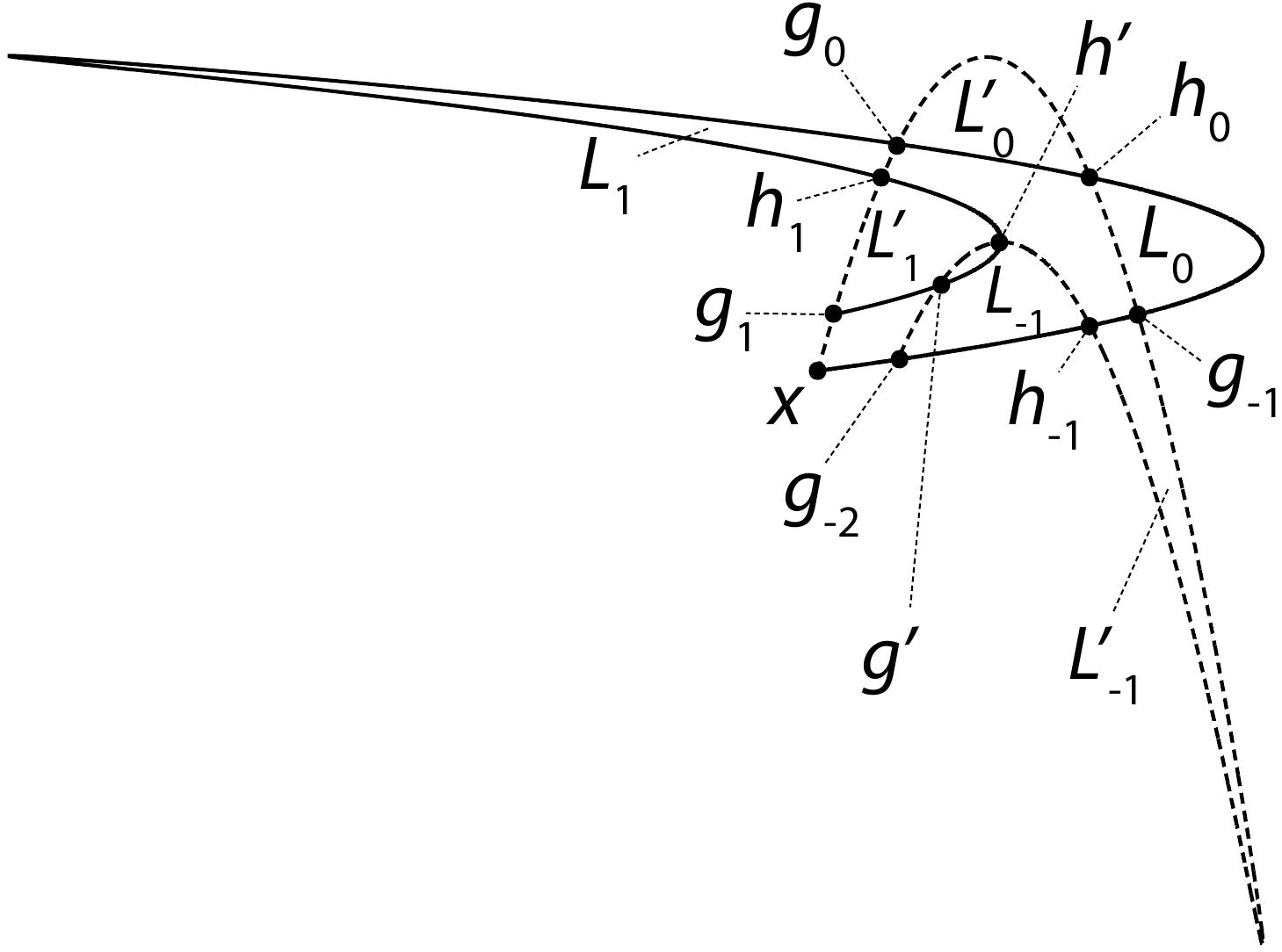}
 \caption{Example partial homoclinic tangle from the H\'{e}non map~\cite{Henon69}.  The unstable (stable) manifold is the solid (dashed) curve.  There are two primary homoclinic orbits $\lbrace h_{0}\rbrace$ and $\lbrace g_{0}\rbrace$.  The lobe regions $L_{0}$ and $L'_{0}$ form a turnstile and govern the transport.  In an open system, the lobes $L_{i}$ ($L'_{-i}$) may extend out to infinity never to re-enter the complex region for $i\geq 1$. 
\label{fig:Homoclinic_tangle}}
\end{figure}

Let $M$ be an analytic and area-preserving map on the $2$-D phase space $(q,p)$, and $x=(q,p)$ be a hyperbolic fixed point under $M$ with stability exponent $\mu$.  Denote the unstable and stable manifolds of $x$ by $U(x)$ and $S(x)$ respectively.  Typically, its unstable and stable manifolds intersect infinitely many times and form a complicated pattern called a homoclinic tangle~\cite{Poincare99,Easton86,Rom-Kedar90} as partially shown in Fig.~\ref{fig:Homoclinic_tangle}.  
The intersection points belong to both $U(x)$ and $S(x)$ for all times.  The homoclinic orbit, denoted by $\lbrace h_{0}\rbrace$, is the bi-infinite collection of images:
\begin{equation}\label{eq:orbit}
\begin{split}
\lbrace h_{0} \rbrace&=\lbrace M^{-\infty}h_{0},\cdots,M^{-1}h_{0},h_{0},Mh_{0},\cdots,M^{\infty}h_{0}\rbrace \\ 
&=\lbrace h_{-\infty},\cdots,h_{-1},h_{0},h_{1},\cdots,h_{\infty}\rbrace
\end{split}
\end{equation}
where both $h_{-\infty}$ and $h_{\infty}$ converge to $x$.  If the unstable and stable segments connecting $x$ with $h_{0}$ intersect only at $h_{0}$, then $\lbrace h_{0}\rbrace$ is a primary homoclinic orbit.  There must be at least two such orbits~\cite{Wiggins92}, such as $\lbrace  h_0\rbrace$ and $\lbrace g_0\rbrace$ in Fig.~\ref{fig:Homoclinic_tangle}.  Of particular interest are the unstable segments $U[g_{i-1},g_{i}]$ and stable segments $S[g_i,g_{i-1}]$, which enclose the so-called ``lobe regions" $L_{i}$ and $L^{\prime}_{i}$, which are extensively studied in transport problems~\cite{Bensimon84,MacKay84a,Rom-Kedar90,Wiggins92}.  The region bounded by $U[x,g_0]$ and $S[x,g_0]$ is called the complex region, which is the main region of interest in transport theory.  More recent works on the topological behavior of lobes resulted in what is termed homotopic lobe dynamics \cite{Mitchell03a,Mitchell03b,Mitchell06}, which gives rise to fractals in escape time graphs, and has been applied to problems such as ionization of hydrogen atoms \cite{Mitchell04} and escape from a vase-shaped cavity \cite{Novick12a,Novick12b}.  Notice that for open systems such as the H\'{e}non map, any point outside the complex region will escape to inifinity, and thus the lobes $L_{i}$ and $L^{\prime}_{-i}$ with $i\geq 1$ will extend to infinity and never come back into the complex region. This ensures that there are no homoclinic points on segments $U(g_i,h_{i+1})$ and $S(g_i,h_i)$.  The homoclinic points in such systems are distributed only on segments $U[h_i,g_i]$ and $S[h_i,g_{i-1}]$.

\begin{figure}[ht]
\centering
{\includegraphics[width=8cm]{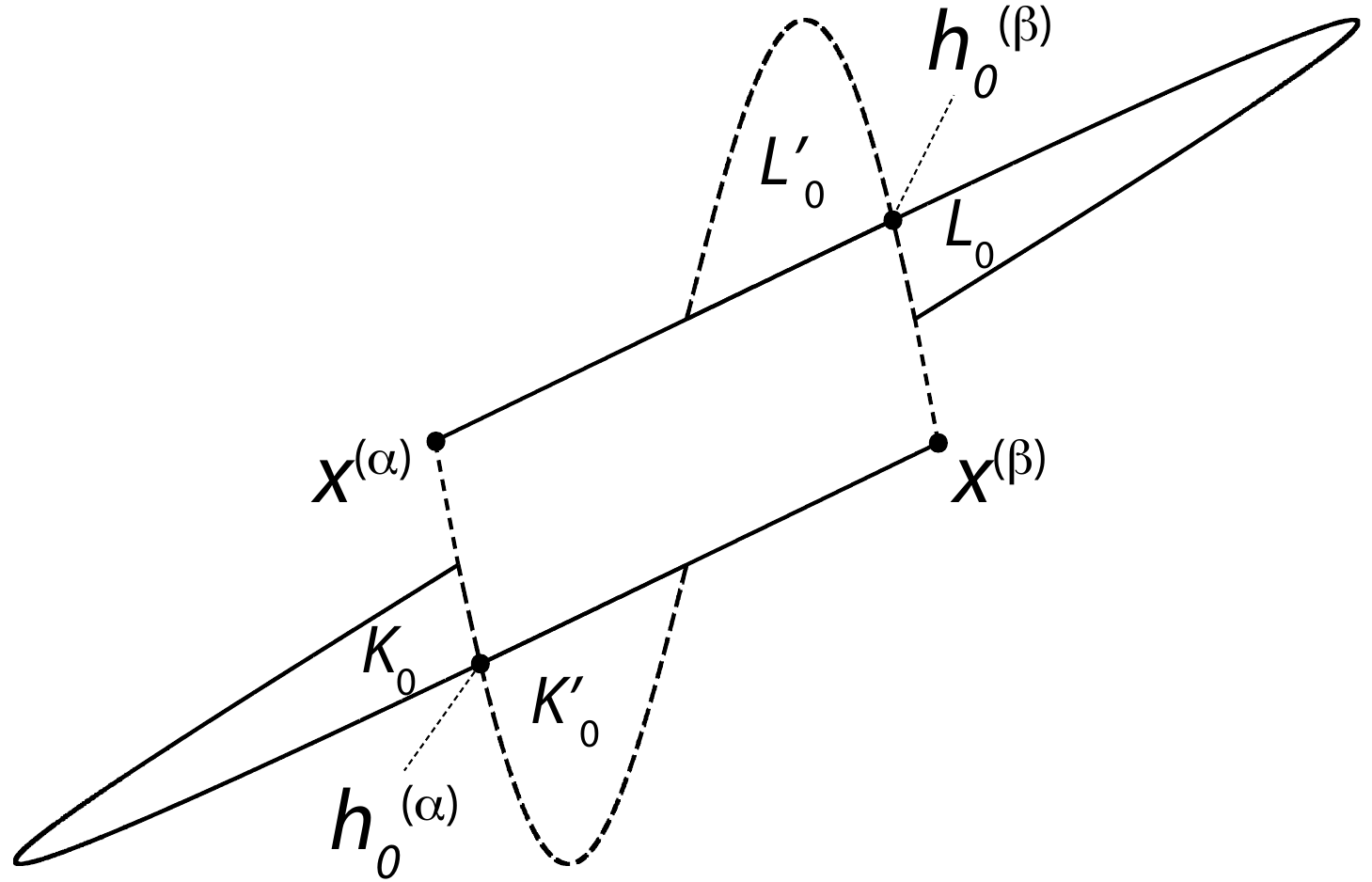}}
 \caption{Schematic partial heteroclinic tangle between $x^{(\alpha)}$ and $x^{(\beta)}$.  The unstable (stable) manifolds are plotted in solid (dashed) curves.  The ``parallelogram" region bounded by segments $U[x^{(\alpha)},h^{(\beta)}_{0}]$, $S[h^{(\beta)}_{0},x^{(\beta)}]$, $U[x^{(\beta)},h^{(\alpha)}_{0}]$ and $S[h^{(\alpha)}_{0},x^{(\alpha)}]$ is the complex region.  The lobes $\{L_{0}$,$L'_{0}\}$ and $\{K_{0}$,$K'_{0}\}$ form the irreducible structures that can be mapped successively to form the entire tangle.}
\label{fig:Heteroclinic_tangle}
\end{figure}

A more general scenario is to have two hyperbolic fixed points with their own stable and unstable manifolds intersecting one another, forming a heteroclinic tangle~\cite{Wiggins92}.  Consider $x^{(\alpha)}$ and $x^{(\beta)}$, with their unstable [stable] manifolds $U(x^{(\alpha)})$ [$S(x^{(\alpha)})$] and $U(x^{(\beta)})$ [$S(x^{(\beta)})$]; see Fig.~\ref{fig:Heteroclinic_tangle}.
The intersecting stable and unstable manifolds of different fixed points generate heteroclinic orbits.  In Fig.~\ref{fig:Heteroclinic_tangle}, $\lbrace h^{(\alpha)}_{0}\rbrace$ and $\lbrace h^{(\beta)}_{0}\rbrace$ have the limiting points:
\begin{equation}
\label{eq:limiting points heteroclinic orbit}
\begin{split}
&h^{(\alpha)}_{\infty},\ h^{(\beta)}_{-\infty}\to x^{(\alpha)}\\
&h^{(\alpha)}_{-\infty},\ h^{(\beta)}_{\infty}\to x^{(\beta)}.
\end{split}
\end{equation}
Unlike homoclinic tangles, there may be only one primary heteroclinic orbit; an example is shown ahead.  Homoclinic and heteroclinic orbits play an important role in chaotic dynamics as they provide clues for the entire structure of the chaotic region.  As shown in~\cite{Birkhoff27,Silva87}, infinite families of satellite periodic orbits accumulate on the homoclinic (heteroclinic) orbits, and the determination of the periodic orbit actions rely on those of the homoclinic (heteroclinic) orbits.  

\subsection{Relative actions}
\label{Relative actions}

The mapping $M$ can be viewed as a canonical transformation that maps a point $(q_{n},p_{n})$ to $(q_{n+1},p_{n+1})$ while preserving the symplectic area, therefore a generating (action) function $F(q_{n},q_{n+1})$ can be associated with this process such that~\cite{MacKay84a,Meiss92}:
\begin{equation}
\begin{split}
&p_{n}=-\partial F/\partial q_{n}\\ 
&p_{n+1}=\partial F/\partial q_{n+1}.
\end{split}
\end{equation}
The total action of an orbit ${\cal F}$ is the sum of the generating functions:
\begin{equation}
\label{eq:full orbit action in general}
{\cal F}=\sum_{n=-\infty}^{\infty}F(q_{n},q_{n+1})
\end{equation}
and is divergent in general.  However, the MacKay-Meiss-Percival action principle~\cite{MacKay84a,Meiss92} can be applied to obtain well defined action differences for particular pairs of orbits.  An important and simple case is the relative action between a fixed point $x$ and any of its homoclinic orbits $\lbrace h_{0}\rbrace$,  which turns out to be equal to an area bounded by unstable and stable manifold segments as
\begin{eqnarray}
\label{eq:relative action homoclinic}
\Delta {\cal F}_{\lbrace h_{0}\rbrace  x} &=& \sum_{n=-\infty}^{+\infty}[F_{\lbrace h_{0}\rbrace}(q_n,q_{n+1})-F_x(q,q)] \nonumber \\
&=& \int\limits_{U[x,h_{0}]}p\mathrm{d}q+\int\limits_{S[h_{0},x]}p\mathrm{d}q = \oint_{US[xh_{0}]} p\mathrm{d}q \nonumber \\
&=& {\cal A}^\circ_{US[xh_{0}]}
\end{eqnarray}
where $U[x,h_{0}]$ is the segment of the unstable manifold from $x$ to $h_{0}$, and $S[h_{0},x]$ the segment of the stable manifold from $h_0$ to $x$.  The $\circ$ superscript from the last line indicates that the area is interior to a path that forms a closed loop, and the subscript indicates the path: $US[xh_{0}]=U[x,h_{0}]+S[h_{0},x]$.  As usual, clockwise enclosure of an area is positive, counterclockwise negative.  $F_{\lbrace h_{0}\rbrace}(q_n,q_{n+1})$ denotes the generating function along $\lbrace h_0\rbrace$ that maps $h_n$ to $h_{n+1}$, and $F_x(q,q)$ denotes the generating function of $x$ in one iteration.  Likewise, a second important case is for homoclinic orbit pairs, which results in 
\begin{eqnarray}
\label{eq:homoclinic action difference}
\Delta{\cal F}_{{\lbrace h^\prime_0\rbrace}{\lbrace h_0\rbrace}} &=& \sum_{n=-\infty}^{\infty}[ F_{\lbrace h^\prime_0\rbrace}(q_{n},q_{n+1}) - F_{\lbrace h_0\rbrace} (q_{n},q_{n+1})] \nonumber \\
& = & \int\limits_{U[h_{0},h^\prime_{0}]}p\mathrm{d}q+\int\limits_{S[h^\prime_{0},h_{0}]}p\mathrm{d}q =  {\cal A}^\circ_{US[h_0h^\prime_{0}]} \nonumber \\
\end{eqnarray}
where $U[h_{0},h^\prime_{0}]$ is the segment of the unstable manifold from $h_{0}$ to $h^\prime_{0}$, and $S[h^\prime_{0},h_{0}]$ the segment of the stable manifold from $h^\prime_{0}$ to $h_{0}$.  See Appendix~\ref{MacKay-Meiss-Percival} for further details.  

It is also desirable to have geometric relations for the differences of any pair of periodic orbits.  Since they may not have the same period, comparing each over its primitive period relative to a fixed point suffices.  For an $l$-period orbit, i.e.~$M^l(x_0) =x_l=x_0$ and $\lbrace x_0 \rbrace = \lbrace x_0,x_1,... x_l\rbrace$
\begin{equation}
\label{eq:twoperiodicorbits}
\Delta {\cal F}_{\lbrace x_0 \rbrace x} = \sum_{n=0}^{l-1}[ F_{\lbrace x_0 \rbrace}(q_n,q_{n+1}) - F_{x}(q,q)]\ .
\end{equation}
However, ahead it is shown that the geometric form also requires homoclinic orbits and Moser invariant curves.

\section{Relative heteroclinic orbit actions}
\label{Heteroclinic orbit actions}

\begin{figure}[ht]
\centering
{\includegraphics[width=8cm]{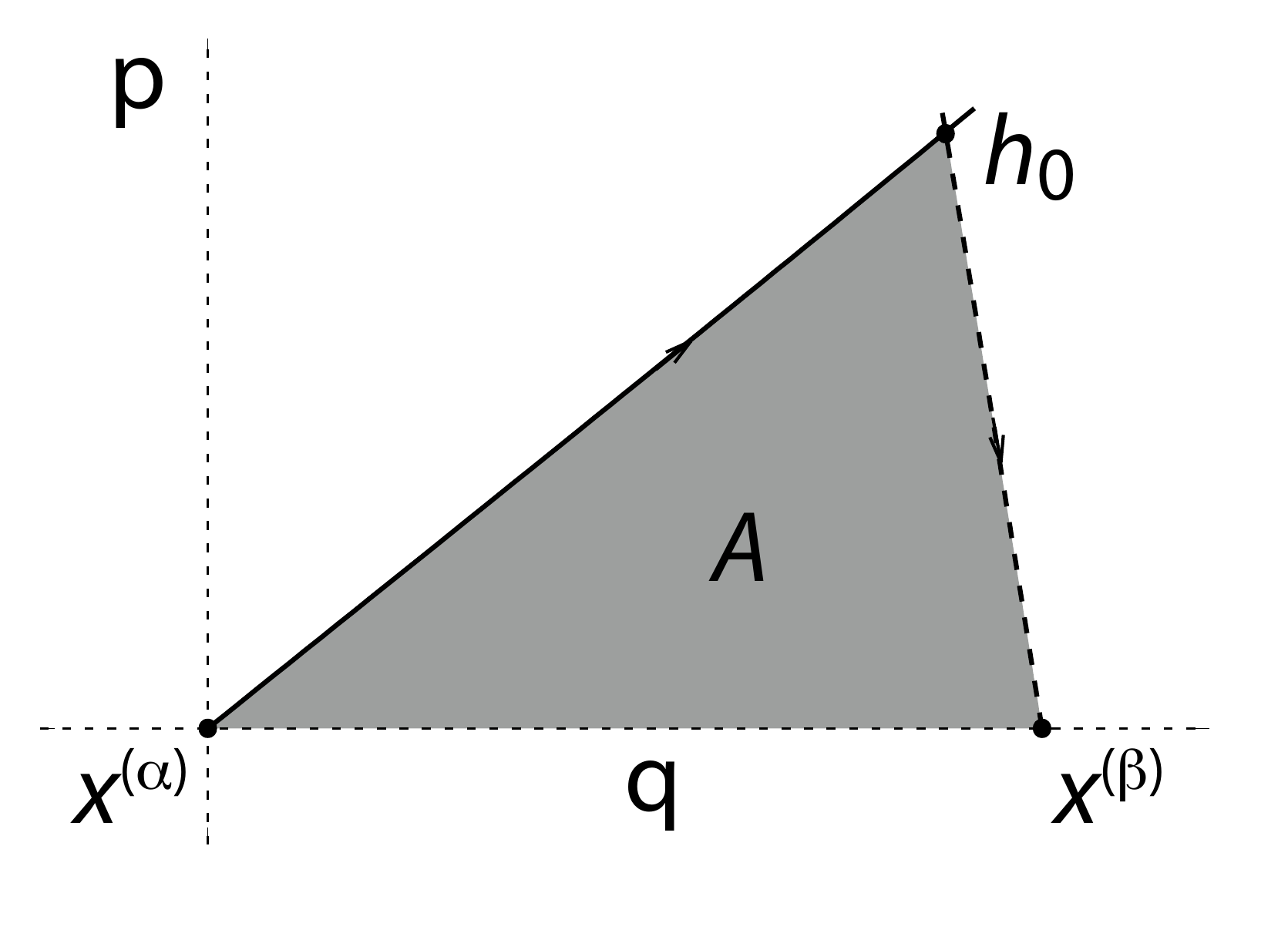}}
 \caption{A primary heteroclinic orbit of the standard map.  $x^{(\alpha)}=(0,0)$ and $x^{(\beta)}=(0.5,0)$ are fixed points of the map, and $U[x^{(\alpha)},h_0]$ and $S[h_{0},x^{(\beta)}]$ make a primary intersection at $h_{0}$.  The relative action of $\lbrace h_{0}\rbrace$ should be equal to the area $A$.}
\label{fig:Heteroclinic_action_sample}
\end{figure}

Consider two hyperbolic fixed fixed points $x^{(\alpha)}$ and $x^{(\beta)}$, and a heteroclinic intersection $h_{0}^{(\beta)}$; see Fig.~\ref{fig:Heteroclinic_tangle}.  Since the infinite past $h_{-\infty}^{(\beta)}$ and the infinite future $h_{\infty}^{(\beta)}$ are asymptotic to different fixed points, it is convenient to consider the heteroclinic orbit in two semi-infinite halves, where $h_{0}^{(\beta)}$ is the dividing point.  The past orbit relative to $h_{0}^{(\beta)}$ is
\begin{equation}
\label{eq:past orbit}
\lbrace h_{0}^{(\beta)}\rbrace^{-}=\lbrace h_{-\infty}^{(\beta)},\cdots,h_{0}^{(\beta)} \rbrace.
\end{equation}  
The future orbit is similarly
\begin{equation}
\label{eq:future orbit}
\lbrace h_{0}^{(\beta)}\rbrace^{+}=\lbrace h_{0}^{(\beta)},\cdots,h_{\infty}^{(\beta)} \rbrace.
\end{equation}  
The action of the past [future] orbit is given relative to $x^{(\alpha)}$ [$x^{(\beta)}$].  In particular, the relative action between $\lbrace h_{0}^{(\beta)}\rbrace^{-}$ and $\lbrace x^{(\alpha)}\rbrace$ is defined as
\begin{equation}
\label{eq:relative action h0- and x alpha}
\Delta {\cal F}_{\lbrace h_{0}^{(\beta)}\rbrace^{-} x^{(\alpha)}}=\sum_{n=-\infty}^{0}\left[F_{\lbrace h_{0}^{(\beta)}\rbrace}(q_{n-1},q_n)-F_{x^{(\alpha)}}(q,q)\right]
\end{equation}
and similarly for the future orbit history
\begin{equation}
\label{eq:relative action h0+ and x beta}
\Delta {\cal F}_{\lbrace h_{0}^{(\beta)}\rbrace^{+} x^{(\beta)}}=\sum_{n=0}^\infty\left[F_{\lbrace h_{0}^{(\beta)}\rbrace}(q_n,q_{n+1})-F_{x^{(\beta)}}(q,q)\right]
\end{equation}
The total relative action of $\lbrace h_{0}^{(\beta)}\rbrace$ is just the sum of the two relative parts.  Following~\cite{MacKay84a,Meiss92} again yields finally:
\begin{equation}\label{eq:heteroclinic action past+future}
\begin{split}
\Delta {\cal F}_{\lbrace h_{0}^{(\beta)}\rbrace^{-} x^{(\alpha)}}+\Delta &{\cal F}_{\lbrace h_{0}^{(\beta)}\rbrace^{+} x^{(\beta)}}\\
 &= \int\limits_{U[x^{(\alpha)},h_{0}^{(\beta)}]}p\mathrm{d}q+\int\limits_{S[h_{0}^{(\beta)},x^{(\beta)}]}p\mathrm{d}q  \\
&= {\cal A}_{US[x^{(\alpha)}h_{0}^{(\beta)}x^{(\beta)}]}  
\end{split}
\end{equation} 
where the subscript $US[x^{(\alpha)}h_{0}^{(\beta)}x^{(\beta)}]=U[x^{(\alpha)},h_{0}^{(\beta)}]+S[h_{0}^{(\beta)},x^{(\beta)}]$.  Since this path is not closed, the final point is added to the area notation.  The integral gives the algebraic area $A$ in Fig.~\ref{fig:Heteroclinic_action_sample}. 

 This result can be generalized by considering a change in the dividing point $h_0^{(\beta)}$ to some other point $h_k^{(\beta)}$.  The form of Eq.~\eqref{eq:heteroclinic action past+future} must be unchanged.
Thus:
\begin{equation}
\label{eq:heteroclinic action past+future-k}
\Delta {\cal F}_{\lbrace h_{k}^{(\beta)}\rbrace^{-} x^{(\alpha)}}+\Delta {\cal F}_{\lbrace h_{k}^{(\beta)}\rbrace^{+} x^{(\beta)}} =  {\cal A}_{US[x^{(\alpha)}h_{k}^{(\beta)}x^{(\beta)}]}
\end{equation} 
where now the past and future relative actions are defined with respect to $h_{k}^{(\beta)}$ and the unstable and stable manifold integral paths change accordingly.  This simple extension is quite useful ahead.

At this point we would like to make a remark on the difference between the areas defined in Eq.~\eqref{eq:relative action homoclinic} and Eq.~\eqref{eq:heteroclinic action past+future}: upon canonical transformations, the former is a closed area, thus invariant; while the latter is an open algebraic area, therefore not invariant.  This is also consistent with the action functions on the left sides of the equations.  Despite that the action functions are modified by the canonical transformations, the modifications cancel out between successive steps for the relative homoclinic actions, but not for the heteroclinics, the net change from which should match the change in the algebraic area.      

\subsection*{Standard map example}
\label{Standard map example}

Consider the action of a primary heteroclinic orbit of the standard map as an example.  The mapping equations are~\cite{Chirikov79}
\begin{equation}
\label{eq:kicked rotor}
\begin{split}
& p_{n+1} =p_{n}-\frac{K}{2\pi }\sin 2\pi q_{n}  \\
& q_{n+1} =q_{n}+p_{n+1} \quad\quad\ \ \ 
\end{split}
\end{equation}  
where our example is for the parameter $K=8.25$, a value for which the system dynamics are overwhelmingly dominated by chaotic motion.  Perhaps the simplest case is that of the two hyperbolic fixed points $x^{(\alpha)}=(0,0)$ and $x^{(\beta)}=(0.5,0)$.  The first point is hyperbolic with inversion, which adds a new element to relations coming further ahead.  The primary intersection of $x^{(\alpha)}$'s unstable manifold with $x^{(\beta)}$'s stable manifold, and the area $A$ defined in Eq.~(\ref{eq:heteroclinic action past+future}) are drawn in Fig.~\ref{fig:Heteroclinic_action_sample}.
Calculating numerically the left hand side of Eq.~(\ref{eq:heteroclinic action past+future}) using the action function for $\lbrace h_{0}\rbrace$, and the right hand side using a construction of the manifolds gives $A-\Delta {\cal F}_{\lbrace h_{0}\rbrace^{-} x^{(\alpha)}}-\Delta {\cal F}_{\lbrace h_{0}\rbrace^{+} x^{(\beta)}} = 9.4 \times 10^{-15}$, which is as accurate as one could expect using double precision computation.  In this example, the two fixed points both lie on the $p=0$ axis, and the algebraic area defined by Eq.~\eqref{eq:heteroclinic action past+future} is relatively simple.  Examples of more complicated heteroclinic orbits connecting fixed points with nonzero $p$ values, can be found in Fig. 11 of~\cite{Li17}.  

 The area-relative-action relation has the advantage of giving results without the necessity of calculating the heteroclinic orbit.  Only the intersection point $h_0$ and manifold segments are needed.  Otherwise, a long orbit segment of $\lbrace h_{0}\rbrace$ centered at $h_{0}$ must be determined to get high accuracy.  As numerical iterations forward and backward of $h_0$ fail to follow $\lbrace h_{0}\rbrace$ after a logarithmically short time in the precision divided by the Lyapunov exponent, numerical orbits diverge in this example after just a few iterations.  Although techniques can be constructed to evade the divergence problem~\cite{Li17}, Eq.~(\ref{eq:heteroclinic action past+future}) makes it unnecessary.  

\section{Relative actions between hyperbolic fixed points}
\label{Action difference between hyperbolic fixed points}

A very interesting relation derives from comparing Eqs.~(\ref{eq:heteroclinic action past+future},\ref{eq:heteroclinic action past+future-k}).  Subtraction generates a relation between the relative action between two fixed points with an area bounded by unstable and stable manifolds.  Defining
\begin{equation}
\label{eq:defab}
\Delta {\cal F}_{x^{(\beta)}x^{(\alpha)}}(k) =k\cdot\left[F_{x^{(\beta)}}(q,q)-F_{x^{(\alpha)}}(q,q)\right]
\end{equation}
gives
\begin{equation}
\label{eq:difference}
\Delta {\cal F}_{x^{(\beta)}x^{(\alpha)}}(k)= {\cal A}_{US[x^{(\alpha)}h_{k}^{(\beta)}x^{(\beta)}]} - {\cal A}_{US[x^{(\alpha)}h_{0}^{(\beta)}x^{(\beta)}]}\ .
\end{equation}
Since
\begin{equation}
\int\limits_{U[x^{(\alpha)},h_{k}^{(\beta)}]}p\mathrm{d}q-\int\limits_{U[x^{(\alpha)},h_{0}^{(\beta)}]}p\mathrm{d}q=\int\limits_{U[h_{0}^{(\beta)},h_{k}^{(\beta)}]}p\mathrm{d}q
\end{equation}
and similarly for the stable manifold segments, Eq.~\eqref{eq:difference} simplifies to ($\Delta k = k_2 - k_1$)
\begin{eqnarray}
\label{eq:action difference k}
\Delta {\cal F}_{x^{(\beta)}x^{(\alpha)}}(k) &=& {\cal A}^\circ_{US[h_{0}^{(\beta)}h_{k}^{(\beta)}]} \nonumber \\
\Delta {\cal F}_{x^{(\beta)}x^{(\alpha)}}(\Delta k) &=& {\cal A}^\circ_{US[h_{k_1}^{(\beta)}h_{k_2}^{(\beta)}]} \ .
\end{eqnarray}

\begin{figure}[ht]
{\includegraphics[width=8cm]{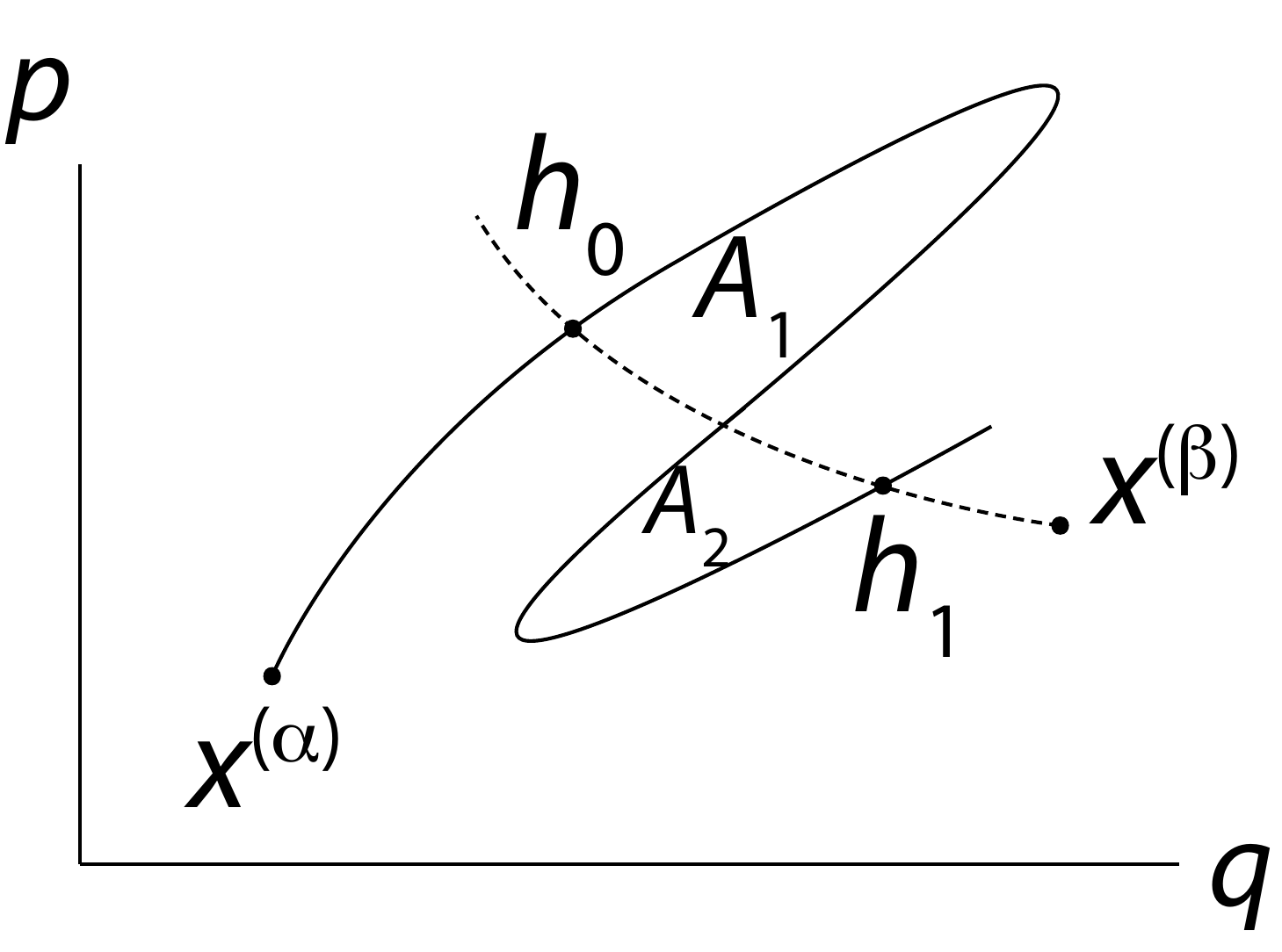}}
 \caption{Schematic partial heteroclinic tangle.  The unstable manifold of $x^{(\alpha)}$ and the stable manifold of $x^{(\beta)}$ intersect at $h_0$, which maps to $h_1$.  Notice that this image requires a second primary heteroclinic orbit, which is not labelled.  The algebraic area $A_{1}-A_{2}$ gives the relative action between $x^{(\beta)}$ and $x^{(\alpha)}$.}
\label{fig:Heteroclinic_3}
\end{figure}

The $k=1$ case is schematically illustrated in Fig.~\ref{fig:Heteroclinic_3}.  
In this case, Eq.~~\eqref{eq:action difference k} reads
\begin{equation}
\label{eq:action difference 1}
\Delta {\cal F}_{x^{(\beta)}x^{(\alpha)}}(1)=  {\cal A}^\circ_{US[h_{0},h_{1}]} = A_1-A_2
\end{equation}
where the last form is using the areas assigned in Fig.~\ref{fig:Heteroclinic_3}.  A homoclinic tangle requires $A_1=A_2$ since $x^{(\alpha)}=x^{(\beta)}$~\cite{MacKay84a}.

\subsection*{Standard map example}
\label{Standard map example 2}

\begin{figure}[ht]
\centering
{\includegraphics[width=8cm]{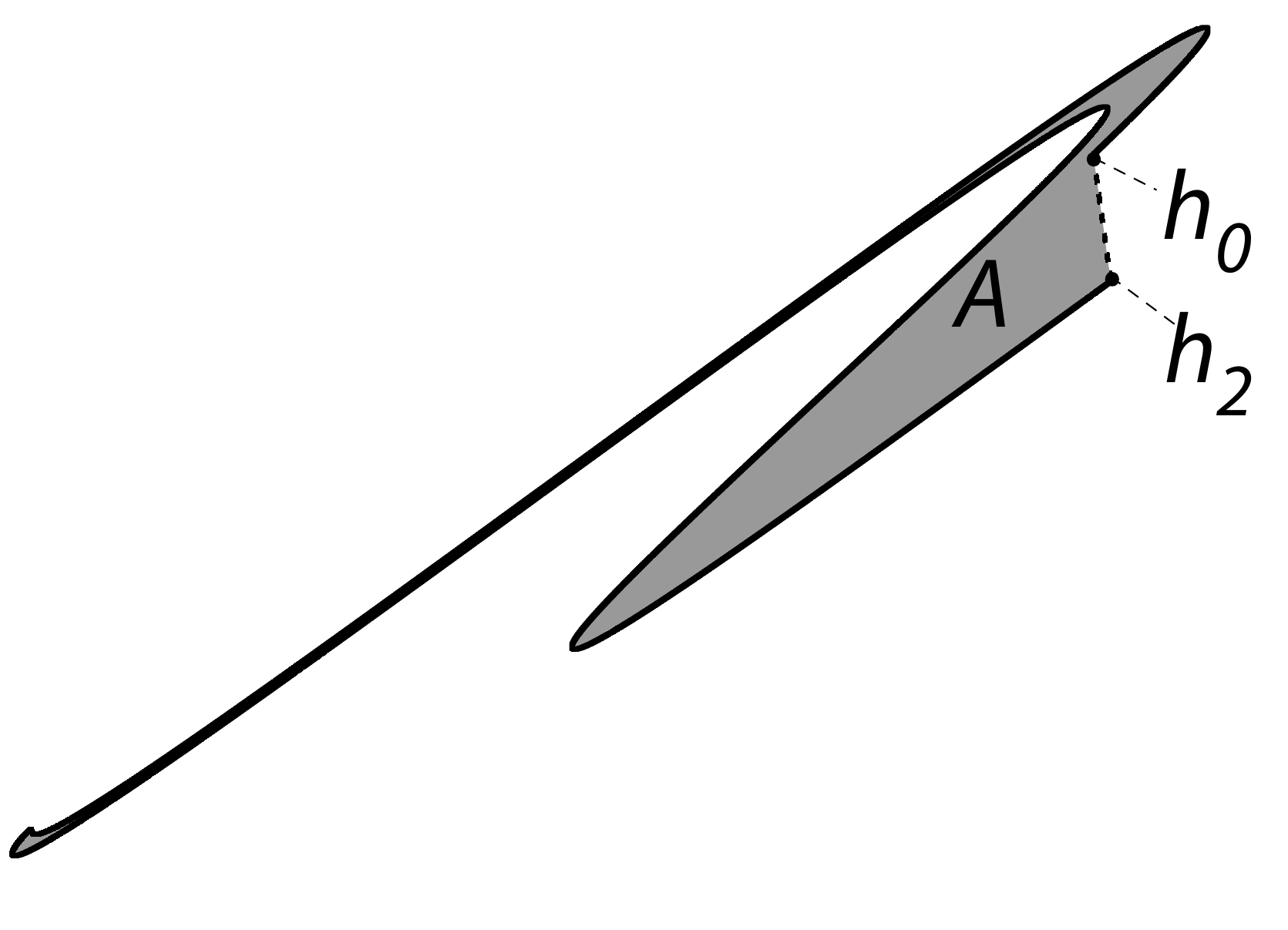}}
 \caption{The fundamental lobe structure of the heteroclinic tangle of $x^{(\alpha)}$ and $x^{(\beta)}$ in the standard map with the torus unfolded. The area of the lobe is enclosed by $U_{x^{(\alpha)}}[h_{0},h_{2}]$ and $S_{x^{(\beta)}}[h_{2},h_{0}]$.  The relative action given by the area is $A=0.835899764985$, which is to be compared with the analytic result using the generating functions of $K/\pi^2$.  The difference is $-6.43 \times 10^{-11}$ showing that the boundaries of $A$ are well determined numerically.  The numerical agreement using double precision is reasonable given the long thin shape of part of the area.}
\label{fig:Heteroclinic_loop}
\end{figure}

Applying Eq.~\eqref{eq:action difference k} to the standard map with the same fixed points as before highlights an intriguing situation due to $x^{(\alpha)}$ being hyperbolic with inversion.  It turns out to be convenient to consider the twice-iterated map $M^2$, under which heteroclinic orbits stay on the same branch of the unstable manifold of $x^{(\alpha)}$.  Therefore, to calculate the action difference between the two fixed points, we consider only the $\Delta k$-even cases.  In addition, there is only one primary heteroclinic orbit, one lobe, and thus one area, not two; see Fig.~\ref{fig:Heteroclinic_loop}.  
The fundamental lobe structure for the heteroclinic tangle does not look like a turnstile as it would for a homoclinic tangle~\cite{Li17}.  Though not visible in the figure, the unstable manifold wraps counterclockwise around the fixed point $x^{(\alpha)}$ in order for this to be possible.

For $k$-even all the heteroclinic points map back onto the same branch of the unstable manifold of $x^{(\alpha)}$.  Therefore, with $k=2$, and $h_{0}$ and $h_{2}$ in Eq.~\eqref{eq:action difference k}:
\begin{equation}
\label{eq:action difference 2}
\Delta {\cal F}_{x^{(\beta)}x^{(\alpha)}}(2) =  {\cal A}^\circ_{US[h_{0}h_{2}]} = A = \frac{K}{\pi^2}
\end{equation} 
where $A$ is defined in Fig.~\ref{fig:Heteroclinic_loop}, and $\frac{K}{\pi^2}$ comes from the generating function of the standard map.  $\Delta {\cal F}_{x^{(\beta)}x^{(\alpha)}}(2)$ is the action difference between the two fixed points under $M^{2}$, the equality is verified to a high accuracy ($\sim 10^{-11}$).     

\section{Satellite periodic orbit actions}
\label{Satellite periodic orbit actions}
In chaotic dynamical systems there is another generic class of unstable periodic orbits that are of great interest.  They are identified as successive points on Moser invariant curves.  Certain sequences of these orbits accumulate on particular homoclinic or heteroclinic orbits~\cite{Ozorio89,Silva87,Moser56,Birkhoff27} and have been referred to as satellite periodic orbits~\cite{Ozorio89}; see Appendix~\ref{Normal form} for more details.  
\subsection{Satellite periodic orbits supported by single Moser curves}
\label{single Moser curve}
\begin{figure}
 \subfigure{
   \label{fig:Ozorio_1}
   \includegraphics[width=7 cm]{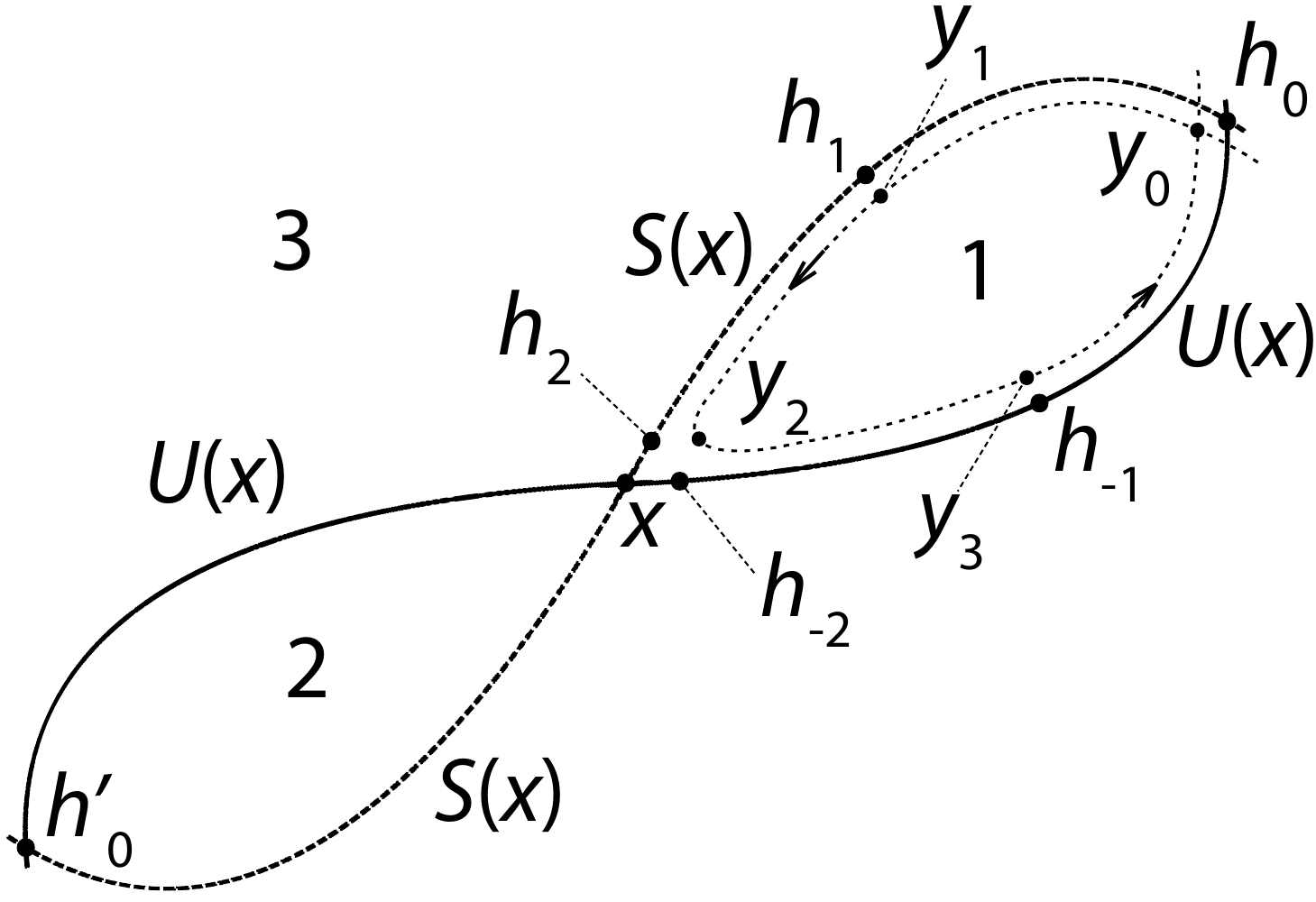}}
 \subfigure{
   \label{fig:Ozorio_2}
   \includegraphics[width=7 cm]{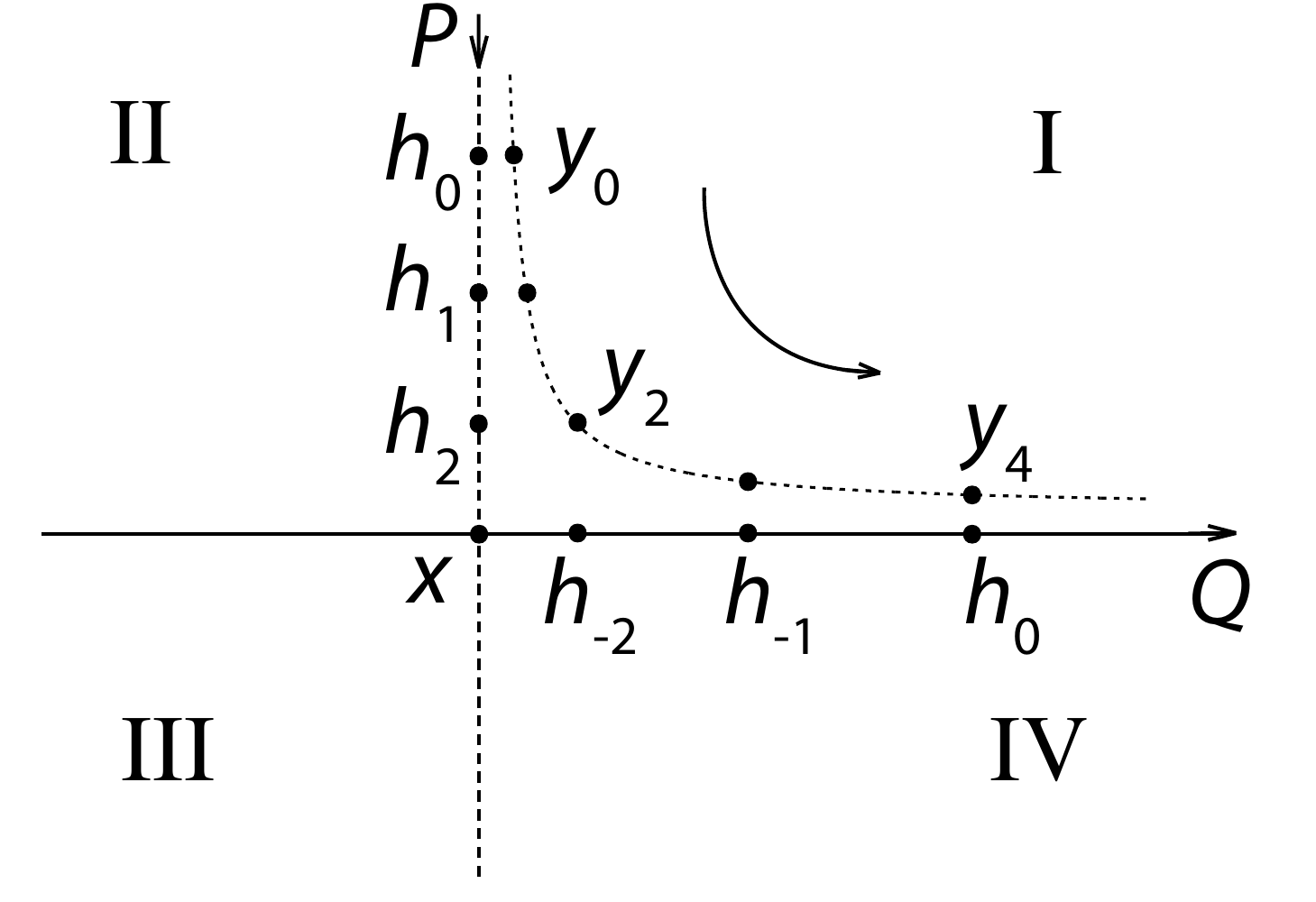}}
\caption{A satellite orbit $\lbrace y_0\rbrace$ of period-$4$ associated with the homoclinic orbit segment $\lbrace h_{-2},\cdots,h_{2}\rbrace$ supported by an invariant Moser curve.  Upper panel: $U(x)$ and $S(x)$ intersect at $h_0$.  The Moser curve which supports a periodic-$4$ satellite orbit $\lbrace y_{0}\rbrace$, thinner dashed line, is plotted inside the complex region.  The orbit segment $\lbrace y_0,y_1,y_2\rbrace$ follows the stable manifold segment $\lbrace h_0,h_1,h_2\rbrace$ for the first $2$ iterations, then switches to the unstable manifold segment, such that $\lbrace y_2,y_3, y_0 \rbrace$ follows $\lbrace h_{-2},h_{-1},h_0 \rbrace$.  $y_2$ is thus a switching point on $\lbrace y_0\rbrace$, where the orbit switches from the future to the past homoclinic segment.  Lower panel:  Normal form coordinates $(Q,P)$.  $h_0$ is on both axes.  $y_4$ and $y_0$ correspond to the same point in phase space.  $y_2$ is the switching point, which is associated either with $h_2$ or $h_{-2}$.    } 
\label{fig:Ozorio}
\end{figure}     
Let us consider a periodic orbit associated with a homoclinic orbit segment and supported by an invariant Moser curve; see Fig.~\ref{fig:Ozorio} for a schematically illustrated example.  The Moser curve extends along $U(x)$ and $S(x)$ out to infinity and converges to them.  Every homoclinic intersection between the manifolds will produce a self-intersection point on the Moser curve.  As argued by Birkhoff~\cite{Birkhoff27} and numerically computed by da Silva Ritter et al.~\cite{Silva87}, special choices can be found for each sufficiently large integer $N$ to make $\lbrace y_{0}\rbrace$ a period-$N$ periodic orbit.  As $N$ increases, the corresponding $y_{0}$ converges to $h_0$, and the homoclinic orbit $\lbrace h_0 \rbrace$ is itself the limiting case of the period-$N$ periodic orbit $\lbrace y_{0}\rbrace$ for $N\to\infty$.  The set  of $y_{0}(N)$ taken from all integer $N$ periodic orbits gives a sequence converging to $h_0$.  In practice, for any homoclinic orbit, $\lbrace h_0 \rbrace$, a truncation into finite segments $\lbrace h_{-l},\cdots,h_0,\cdots,h_k\rbrace$ ($k,l>0$) is possible,  for which a Newton-Raphson search in its neighborhood can be used to construct the satellite orbit of period $(k+l)$ associated with $\lbrace h_{-l},\cdots,h_0,\cdots,h_k\rbrace$.  This provides a convenient way to construct the satellite orbits without the need to calculate the normal form series or the Moser curves.     

\begin{figure}[ht]
\includegraphics[width=8cm]{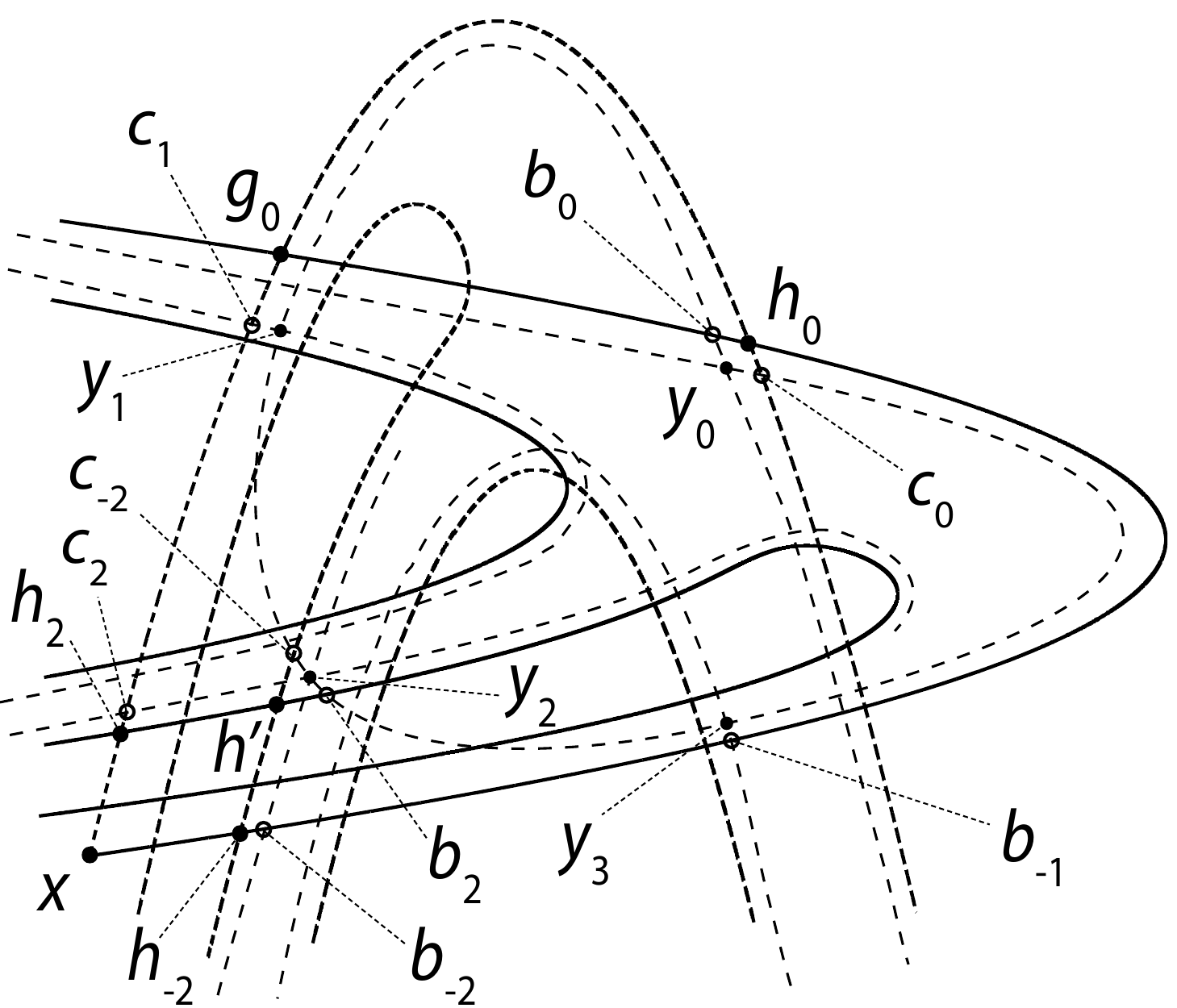}
\caption{(Schematic) Expanded view of the homoclinic tangle in Fig.~\ref{fig:Homoclinic_tangle}.  The manifold segments extending out of the drawing are simply connected, and left out of this figure.  A Moser invariant curve, light dashed line, is shown giving rise to a satellite periodic orbit $\lbrace y_{0} \rbrace$ of period $4$.  The invariant curve has been drawn more distant from  the actual stable/unstable manifolds for illustration purposes.  Every $y_{i}$ is a self-intersection of the Moser curve.  The curve intersects with $U(x)$ [$S(x)$] at $b_{i}$ [$c_{i}$] near its self-intersection at $y_{i} $.}
\label{fig:satellite_X_action}
\end{figure}
The relative action difference between a given $\lbrace h_{0}\rbrace$ and its satellite $\lbrace y_{0}\rbrace$ is determined by a roughly parallelogram shaped region bounded by the manifolds and the Moser curve.  To see how this area arises, consider the homoclinic tangle in Fig.~\ref{fig:satellite_X_action}, which is an expanded view of the tangle in Fig.~\ref{fig:Homoclinic_tangle}.  A Moser invariant curve is drawn which supports a period-$4$ orbit $\lbrace y_{0}\rbrace$, satellite to the homoclinic orbit $\lbrace h_{0}\rbrace$.  $y_2$ is the switching point from the future to the past homoclinic segment.  The orbit segment $\lbrace y_0,y_1,y_2\rbrace$ follows $\lbrace h_0,h_1,h_2\rbrace$, then switches at $y_2$, after which $\lbrace y_2,y_3,y_0 \rbrace$ follows $\lbrace h_{-2},h_{-1},h_0\rbrace$.  
     
The relative-action-area-relation derivation makes direct use of Eq.~\eqref{eq:Meiss92} four times, once for each iteration of the map $M$:
\begin{enumerate}

\item  Starting from the initial point $y_{0}$, and map $M(y_{0})=y_{1}$, follow the path $S[x,c_{0}]+I[c_{0},y_{0}]$; $I[c_{0},y_{0}]$ is the segment of the Moser invariant curve from $c_{0}$ to $y_{0}$.  The path maps to $S[x,c_{1}]+I[c_{1},y_{1}]$.  Substituting the paths into Eq.~\eqref{eq:Meiss92} yields:
\begin{equation}
\label{eq:satellite orbit action y0 to y1}
F_{\lbrace y_0\rbrace}(q_0,q_1)-F_x(q,q)= {\cal A}_{ISI[y_0c_{0}c_1y_1]} \ .
\end{equation}

\item $M(y_{1})=y_{2}$: Let the paths be $S[x,c_{1}]+I[c_{1},y_{1}]$ and $S[x,c_{2}]+I[c_{2},y_{2}]$ giving
\begin{equation}\label{eq:satellite orbit action y1 to y2}
F_{\lbrace y_0\rbrace}(q_1,q_2)-F_x(q,q)= {\cal A}_{ISI[y_1c_{1}c_2y_2]} \ .
\end{equation}

\item  $M(y_{2})=y_{3}$: Let the paths be $U[x,b_{-2}]+I[b_{-2},y_{2}]$ and $U[x,b_{-1}]+I[b_{-1},y_{3}]$ giving
\begin{equation}
\label{eq:satellite orbit action y2 to y3}
F_{\lbrace y_0\rbrace}(q_2,q_3)-F_x(q,q)= {\cal A}_{IUI[y_2b_{-2}b_{-1}y_3]} \ .
\end{equation}

\item $M(y_{3})=y_{0}$:  Let the paths be $U[x,b_{-1}]+I[b_{-1},y_{3}]$ and $U[x,b_{0}]+I[b_{0},y_{0}]$. This gives
\begin{equation}
\label{eq:satellite orbit action y3 to y0}
F_{\lbrace y_0\rbrace}(q_3,q_0)-F_x(q,q)= {\cal A}_{IUI[y_3b_{-1}b_0y_0]} \ .
\end{equation}

\end{enumerate}  

The total relative action is thus
\begin{equation}
\label{eq:satellite orbit total action in compound path}
\Delta {\cal F}_{\lbrace y_0 \rbrace x} = {\cal A}^\circ_{\cal L}
\end{equation}
where the compound closed path $\mathcal{L}$ is

\begin{equation}
\label{eq:compound path L}
\begin{split}
\mathcal{L}= I[y_{0},& c_{0}]+S[c_{0}, c_{2}]+I[c_{2}, y_{2}]+I[y_{2}, b_{-2}]+   \\
& U[b_{-2}, b_{0}]+I[b_{0}, y_{0}]  \ .
\end{split}
\end{equation}

By adding and subtracting certain path segments, it is possible to deform $\cal L$ such that it separates into a path for the relative action of the homoclinic orbit and two parallelogram like correction terms.  This gives the final desired relation between the relative action of the periodic and homoclinic orbits,
\begin{equation}\label{eq:satellite orbit action difference}
\begin{split}
 \Delta {\cal F}_{\lbrace y_{0}\rbrace  x} -& \Delta {\cal F}_{\lbrace h_{0} \rbrace x} \\
&={\cal A}^\circ_{SIIU[xc_2y_2b_{-2}]} - {\cal A}^\circ_{SIIU[h_0c_0y_0b_{0}]}\ .
\end{split}
\end{equation}
The two areas in the above equation resembles two near-parallelograms bounded by the manifolds and the Moser curves.  The satellite orbit action is then:
\begin{equation}\label{eq:satellite orbit action period 4}
\begin{split}
{\cal F}_{\lbrace y_0 \rbrace} =\sum_{n=0}^{3} &F_{\lbrace y_0 \rbrace}(q_{n},q_{n+1})= 4 F_x(q,q) +  {\cal A}^\circ_{US[xh_{0}]}\\ 
&+ {\cal A}^\circ_{SIIU[xc_2y_2b_{-2}]} - {\cal A}^\circ_{SIIU[h_0c_0y_0b_{0}]}
\end{split}
\end{equation}  
where $\Delta {\cal F}_{\lbrace h_{0} \rbrace x} $ is given by area ${\cal A}^\circ_{US[xh_{0}]}$.  Although the $\lbrace h_0\rbrace$ segment used here is a primary homoclinic orbit, with a careful definition of the points $b_i$ and $c_i$ near each orbit point $y_i$, a generalized Eq.~\eqref{eq:satellite orbit action period 4} applies to satellite orbits associated with any homoclinic orbit segment. 
\begin{figure}[ht]
\includegraphics[width=8cm]{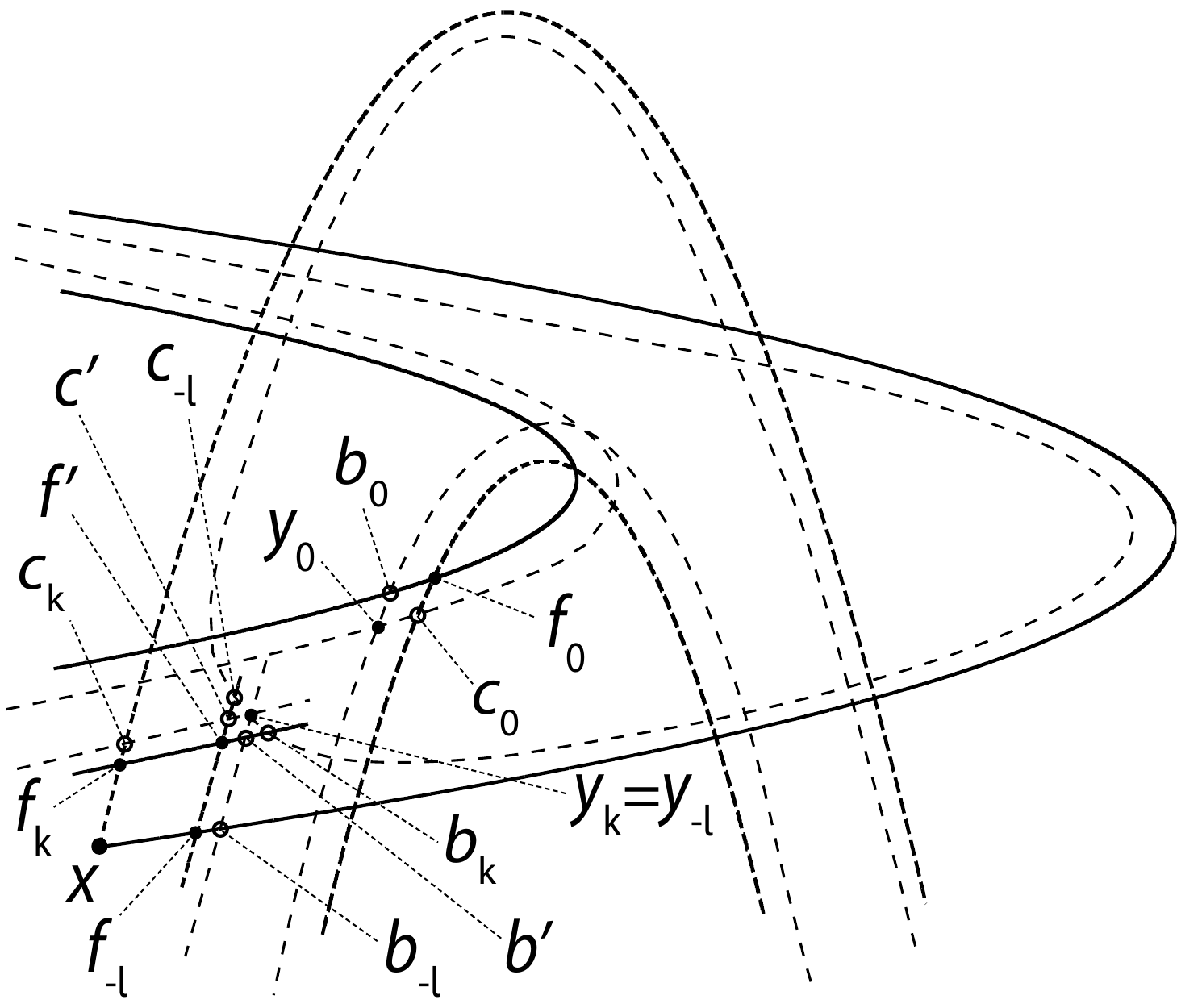}
\caption{(Schematic) Satellite orbit $\lbrace y_0\rbrace$ associated with a non-primary homoclinic orbit segment $\lbrace f_{-l},\cdots,f_0,\cdots,f_{k}\rbrace$.  Since the Moser curve approaches the stable and unstable manifolds, it must intersects with $U(x)$ ($S(x)$) in the same way that $S(x)$ ($U(x)$) does.  It is under this sense that the homoclinic point $f_0$ will force two intersections $b_0$ and $c_0$ between the Moser curve and the unstable/stable manifolds respectively.}
\label{fig:satellite_X_action_general}
\end{figure}
Take the example of Fig.~\ref{fig:satellite_X_action_general}, where a period-$(k+l)$ satellite orbit $\lbrace y_0\rbrace$ is associated with a non-primary homoclinic orbit segment $\lbrace f_{-l},\cdots,f_0,\cdots,f_{k}\rbrace$.  Since the Moser curve approaches the stable and unstable manifolds as it extends along them to infinity, it is forced to make a self-intersection at $y_0$ as the stable and unstable manifolds intersect at $f_0$.  The particular Moser curve is the one for which the $(k+l)^{th}$ mapping of $y_0$ gives back $y_0$.  Therefore, $y_0$ can be thought as being induced by $f_0$.  Following the same logic, define $b_0$ to be the intersection between $U(x)$ and the Moser curve that is induced by $f_0$: as the Moser curve extends along $S(x)$, it intersects with $U(x)$ in the same way that $S(x)$ intersects, so the homoclinic point $f_0$ induces a $b_0$ on the Moser curve.  Similarly, $c_0$ is defined as the intersection between $S(x)$ and the Moser curve that is induced by $f_0$.  All $b_i$ and $c_i$'s can be located in the same way using $f_i$ as the inducing point.  It follows that all previous derivation steps continue to hold with the resulting more general expression of satellite orbit action:
\begin{equation}\label{eq:satellite orbit action general exact}
\begin{split}
{\cal F}_{\lbrace y_0 \rbrace} =\sum_{n=0}^{k+l-1} &F_{\lbrace y_0 \rbrace}(q_{n},q_{n+1})= (k+l) F_x(q,q) +  {\cal A}^\circ_{US[xf_{0}]}\\ 
&+ {\cal A}^\circ_{SIIU[xc_ky_kb_{-l}]} - {\cal A}^\circ_{SIIU[f_0c_0y_0b_{0}]}
\end{split}
\end{equation}  
where $f_0$ can be any homoclinic point.  This formula expresses the satellite action in terms of the fixed point action, the homoclinic relative action, and two four-segmented simple closed curves bounded by stable/unstable manifolds and the Moser curves.  The calculation of the two areas require the construction of the Moser curve, as well as the orbit points $y_0$ and $y_k$, which can be difficult to compute.  However, a simple approximation scheme is possible.  

\subsection{Geometric area approximation}
\label{Geometric area approximation}

Equation~\eqref{eq:satellite orbit action general exact} can be approximated with a wedge product form that only requires the location of the homoclinic points $f_k$ and $f_{-l}$.  In this way, it is possible to calculate the full action of a period-$N$ ($N=k+l$) satellite orbit $\lbrace y_0 \rbrace$ without its reconstruction or its Moser invariant curve.  Assuming the action $F_x(q,q)$ and the area ${\cal A}^\circ_{US[xf_{0}]}=\Delta {\cal F}_{\lbrace f_{0} \rbrace x}$ of some homoclinic orbit point $f_0$ are known, then the first two terms on the right-hand-side of Eq.~\eqref{eq:satellite orbit action general exact} do not depend on knowing $\lbrace y_0 \rbrace$, and only the two areas are needed.  Notice from Fig.~\ref{fig:satellite_X_action_general} that ${\cal A}^\circ_{SIIU[f_0c_0y_0b_{0}]}$ is mapped to ${\cal A}^\circ_{SIIU[f_kc_ky_kb_{k}]}$ under $k$ iterations, so that the areas of the two are identical.  Thus,
\begin{equation}\label{eq:area decomposition 1}
\begin{split}
{\cal A}^\circ_{SIIU[xc_ky_kb_{-l}]} &- {\cal A}^\circ_{SIIU[f_0c_0y_0b_{0}]}\\
&={\cal A}^\circ_{SIIU[xc_ky_kb_{-l}]} -{\cal A}^\circ_{SIIU[f_kc_ky_kb_{k}]}\\
&\approx  {\cal A}^\circ_{SUIU[xf_kb'b_{-l}]}\ .
\end{split}
\end{equation}
The final approximate closed path has only one side which depends on a Moser invariant curve.  Furthermore, as shown in Fig.~\ref{fig:satellite_X_action}, $I[b',b_{-l}]$ is exceedingly close to $S[f^\prime,f_{-l}]$, where $f^\prime$ is a point on a different homoclinic orbit.  Consider that 
\begin{equation}\
\label{eq:area decomposition 2}
 {\cal A}^\circ_{SUIU[xf_kb^{\prime}b_{-l}]} =  {\cal A}^\circ_{SUSU[xf_k f^\prime f_{-l}]} +  {\cal A}^\circ_{UIUS[f^\prime b^{\prime} b_{-l} f_{-l}]}
\end{equation}
and the mean expansion rate of the map is estimated by the positive stability exponent of the fixed point under one iteration of the map,  ${\rm e}^\mu$.  After $k+l$ iteractions, the unstable segment $U[f_{-l},b_{-l}]$ is stretched into $U[f_k,b_k]$ with an expansion factor of roughly ${\rm e}^{(k+l)\mu}$.  This implies that the ratio of areas 
\begin{equation}
\label{eq:ratio}
\frac{ {\cal A}^\circ_{UIUS[f^\prime b^{\prime} b_{-l} f_{-l}]}}{{\cal A}^\circ_{SUSU[xf_k f^\prime f_{-l}]}} \sim O({\rm e}^{-(k+l)\mu})\ .
\end{equation}
For all but the smallest values of $(k+l)$, the small final area term of Eq.~(\ref{eq:area decomposition 2}) can be dropped.  

At this point, one can calculate ${\cal A}^\circ_{SUSU[xf_k f^{\prime} f_{-l}]}$ just by following the manifolds, which is very straightforward.  However, there is a further approximation one can make.  The manifolds are highly constrained in their behaviors in the local neighborhood of $x$.  They must run along nearly parallel, nearly straight lines.  This is approximately a parallelogram with area
\begin{equation}
\label{eq:cross}
{\cal A}^\circ_{SUSU[xf_k f^\prime f_{-l}]}\approx \delta q_{-l}\delta p_k -\delta p_{-l}\delta q_k  =\delta f_{-l}\wedge \delta f_k 
\end{equation}
where $\delta q_k = q_k - q$, $\delta p_k = p_k - p$ and similarly for $(\delta q_{-l},\delta p_{-l})$; i.e.~the $\delta$ coordinates are just those of $f_k$ and $f_{-l}$ relative to $x$.  With this approximation, to a high degree of accuracy the full satellite orbit action ${\cal F}_{\lbrace y_0 \rbrace}$ is determined knowing only $F_x(q,q)$, ${\cal A}^\circ_{US[x,f_{0}]}$, $f_k$ and $f_{-l}$ in general:
\begin{equation}\label{eq:satellite orbit action with approx}
{\cal F}_{\lbrace y_0 \rbrace} \approx (k+l) F_x(q,q) +{\cal A}^\circ_{US[xf_{0}]}+ \delta f_{-l}\wedge \delta f_k,
\end{equation}
where $\lbrace y_0\rbrace$ is the satellite orbit associated with $\lbrace f_{-l},\cdots,f_0,\cdots,f_k\rbrace$.  $y_k$ is the switching point at which the orbit switches from $\lbrace f_0,\cdots,f_k\rbrace$ to $\lbrace f_{-l},\cdots,f_{0}\rbrace$.   

A possible confusion arises from the fact that the same satellite orbit $\lbrace y_0\rbrace$ can also be viewed as associated with any shift in the truncation of the homoclinic orbit: $\lbrace f_{-l+n},\cdots,f_n,\cdots,f_{k+n}\rbrace$, where $n$ is any integer.  Furthermore, for $n\ll l,k$, the Newton iteration using $\lbrace f_{-l+n},\cdots,f_n,\cdots,f_{k+n}\rbrace$ as trail orbit will also converge, and one can verify that it leads to the same satellite orbit as using $\lbrace f_{-l},\cdots,f_0,\cdots,f_k\rbrace$.  Therefore, the choice of the switching point along the satellite orbit seems not unique.  This ambiguity can be resolved by defining the switching point to be the one that minimizes the error from approximation Eq.~\eqref{eq:satellite orbit action with approx} in the original coordiate system, which is the error from replacing ${\cal A}^\circ_{SUSU[xf_k f^\prime f_{-l}]}$ by the wedge product $\delta f_{-l}\wedge \delta f_k$.  In practice, the switching point is easy to identify.  Since the error is the difference between the curvy ``trapezoid'' and its linear interpolation, the minimization is achieved by choosing the orbit point that is ``closest" to the fixed point.  Therefore, in the example of Fig.~\ref{fig:satellite_X_action}, the switching point can be identified graphically to be $y_2$, which is the closest point along $\lbrace y_0\rbrace$ relative to $x$.  By ranging over all possible choices of $f_0$, $l$ and $k$, Eq.~\eqref{eq:satellite orbit action with approx} suffices to calculate the classical actions of all periodic orbits inside the convergence zone. 

\subsection{H\'{e}non map example}
\label{Henon map example}

Consider the action of a satellite orbit in the area-preserving H\'{e}non map \cite{Henon76}:
\begin{equation}
\label{eq:Henon map}
\begin{split}
& p_{n+1} =q_{n}  \\
& q_{n+1} =a-q_n^2-p_n 
\end{split}
\end{equation}  
with parameter value $a=10$.  We have numerically computed a period-$8$ orbit $\lbrace y_0\rbrace$ satellite to one of the primary homoclinic orbit segments $\lbrace h_{-4},\cdots,h_4\rbrace$, where $y_0=(3.1835765543,3.1835765543)$ and $h_0=(3.183580560, 3.183580560)$.  This gives:
\begin{equation}\label{eq:Henon map numerics 1}
{\cal F}_{\lbrace y_0 \rbrace} -8 F_x(q,q) -{\cal A}^\circ_{US[x,h_{0}]}=1.92729\times 10^{-4}
\end{equation}   
whereas the wedge product gives:
\begin{equation}\label{eq:Henon map numerics 2}
\delta h_{-4}\wedge \delta h_{4} = 1.92688\times 10^{-4}\ .
\end{equation}
The difference is in the $4^{th}$ decimal place.  On the other hand, without the knowledge of $\lbrace y_{0}\rbrace$, the full action can be calculated using Eq.~\eqref{eq:satellite orbit action with approx}:
\begin{equation}\label{eq:Henon map numerics 3}
{\cal F}_{\lbrace y_0 \rbrace}=138.940538512
\end{equation}
to be compared with the actual action:
\begin{equation}\label{eq:Henon map numerics 4}
{\cal F}_{\lbrace y_0 \rbrace}=138.940538553\ .
\end{equation}
The relative error equals $3 \times 10^{-10}$, demonstrating the high accuracy of the wedge product approximation.

\subsection{Satellite orbits supported by multiple Moser curves}
\label{multiple invariant curves}

\subsubsection{Homoclinic}
\label{Homoclinic}
Eq. ~\eqref{eq:satellite orbit action with approx} applies to the special cases when the satellite orbit is constructed from one homoclinic orbit segment, and supported by a single Moser curve.  However, for chaotic systems in general, periodic orbits in the convergence zone can be supported by multiple Moser curves, and accumulate alternatively on multiple homoclinic orbit segments.  In this section we generalize the previous results to incorporate these circumstances, giving both the exact and approximate formulas for the classical actions of all periodic orbits within the convergence zone.  

\begin{figure}
  \subfigure{
    \label{fig:J_invariant_curves_homoclinic_1}
    \includegraphics[width=6cm]{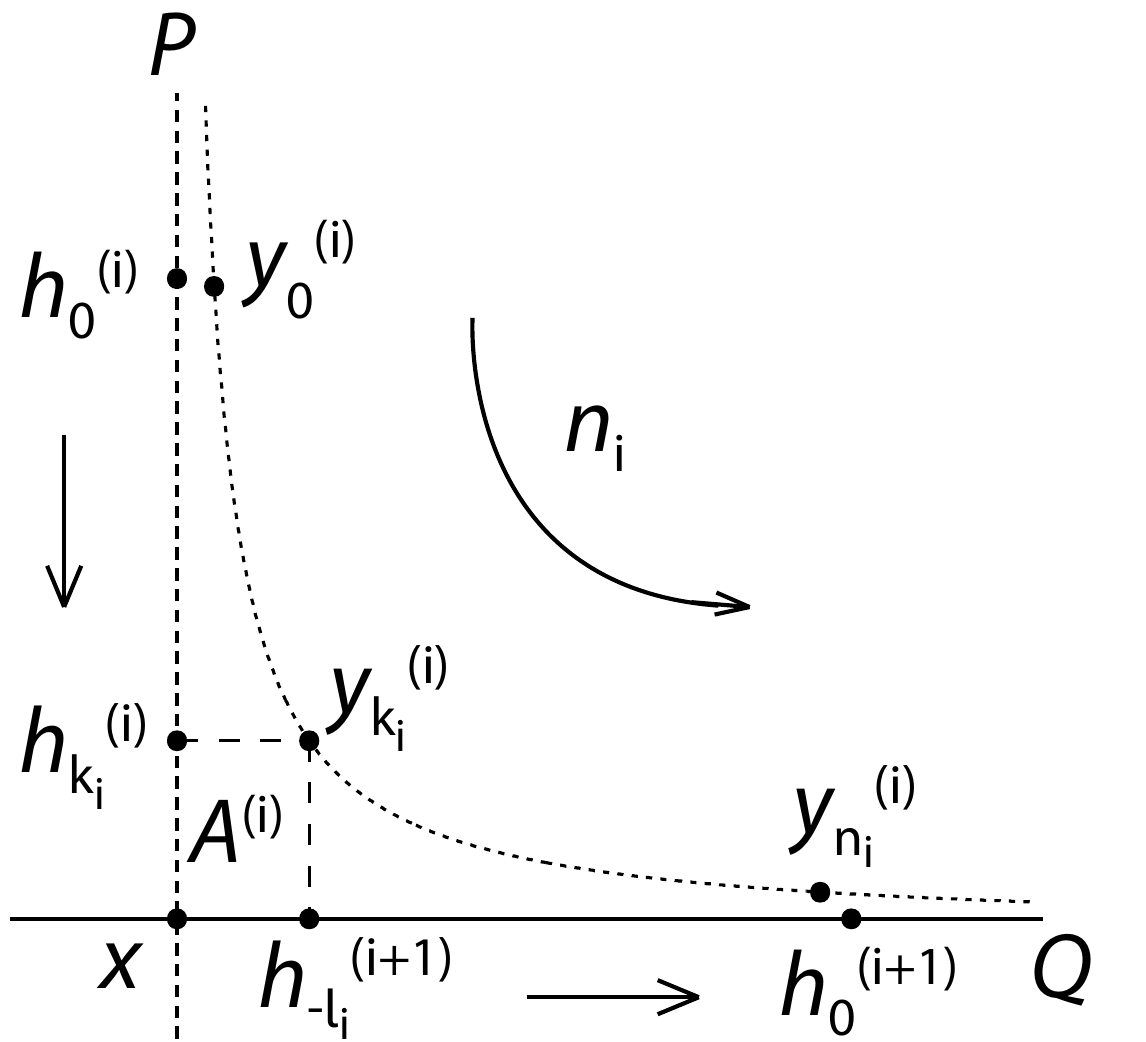}}
    \qquad         
  \subfigure{
    \label{fig:J_invariant_curves_homoclinic_2}
    \includegraphics[width=6.3cm]{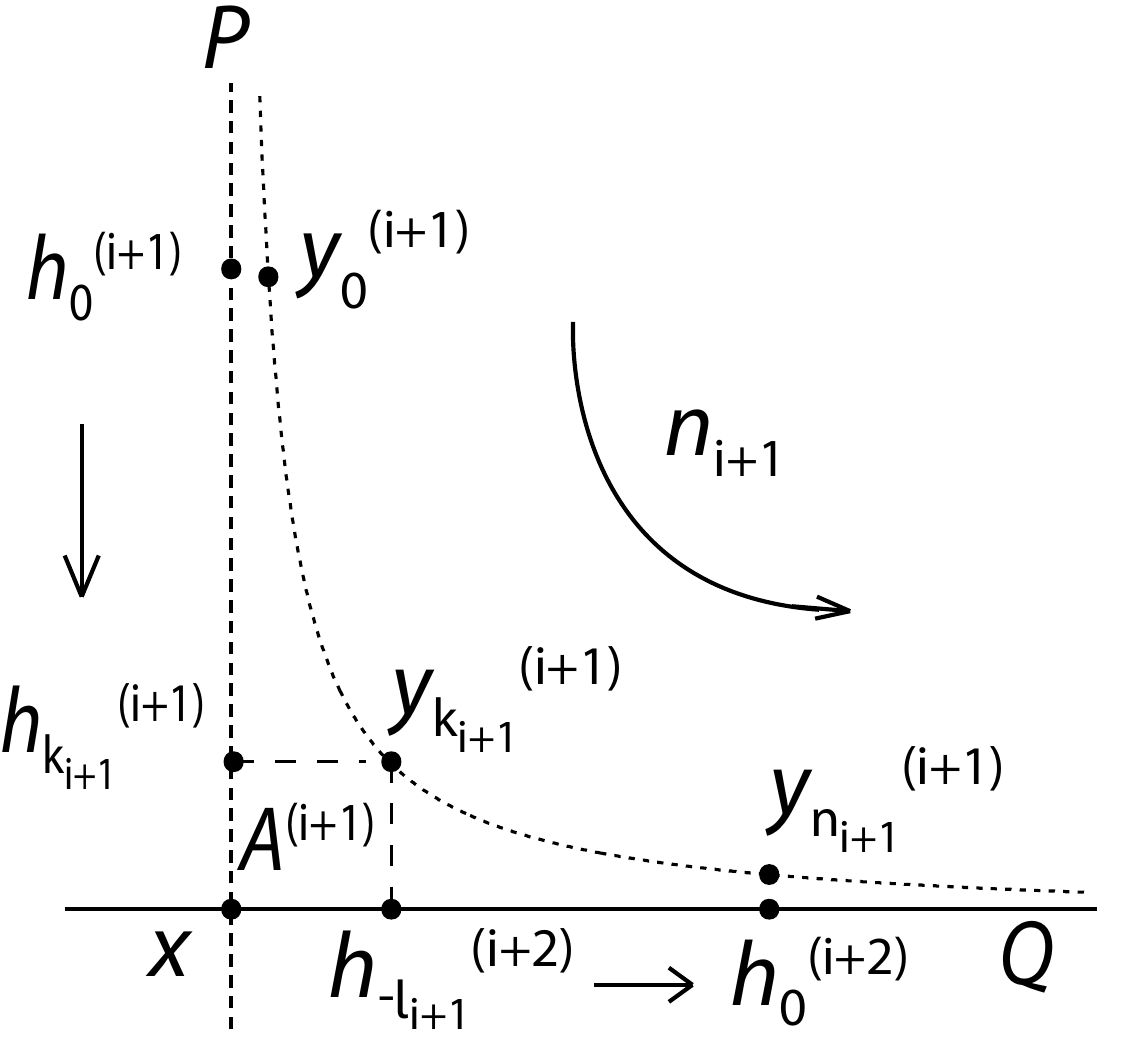}}
\caption{Normal form picture of a satellite orbit accumulating on $J$ homoclinic orbits.  Each $\lbrace h_{0}^{(i)}\rbrace$ is a homoclinic orbit of $x$, and the satellite orbit accumulates on them successively during one period.  Starting from $y_{0}^{(i)}$ in the upper panel, the orbit spends $n_{i}$ iterations on $\mathcal{C}^{(i)}$, which brings it to $y_{n_i}^{(i)}$.  Then it is passed on to $\mathcal{C}^{(i+1)}$ in the lower panel, where $y_{0}^{(i+1)}=y_{n_i}^{(i)}$, and repeat the similar process.  During each $n_{i}$, the orbit spends the first $k_{i}$ iterations accumulating on $\lbrace h_{0}^{(i)}\rbrace$, then switch to accumulate on $\lbrace h_{0}^{(i+1)}\rbrace$ for the rest $l_i$ iterations, where $k_i+l_i=n_i$.} 
\label{fig:J_invariant_curves_homoclinic}
\end{figure}

Suppose we are given $J$ ($J\in \mathbf{Z}^{+}$) homoclinic orbit segments $\lbrace h_{-l_{i-1}}^{(i)},\cdots,h_0^{(i)},\cdots,h_{k_i}^{(i)}\rbrace$ ($i=1,\cdots,J$), and construct a satellite orbit $\lbrace y\rbrace=\bigcup_{i=1}^{J}\lbrace y_0^{(i)},\cdots,y_{n_i}^{(i)}\rbrace$ from them, where $n_i=k_i+l_i$.  The criteria $y_{n_i}^{(i)}=y_0^{(i+1)}$, where $i$ is cyclic in $J$: $1+J=1$, guarantees that $\lbrace y\rbrace$ has primitive period $N=\sum_{1}^{J}n_i$.  As shown by Fig.~\ref{fig:J_invariant_curves_homoclinic}, each segment $\lbrace y_0^{(i)},\cdots,y_{n_i}^{(i)} \rbrace$ is supported by Moser curve $\mathcal{C}^{(i)}$.  The first part $\lbrace y_0^{(i)},\cdots,y_{k_i}^{(i)}\rbrace$ accumulate on the homoclinic segment $\lbrace h_0^{(i)},\cdots,h_{k_i}^{(i)}\rbrace$, and the second part $\lbrace y_{k_i}^{(i)},\cdots,y_{n_i}^{(i)}\rbrace$ accumulate on $\lbrace h_{-l_i}^{(i+1)},\cdots,h_0^{(i+1)}\rbrace$.  

Following similar procedures as Sec.~\ref{single Moser curve}, we can generalize Eq.~\eqref{eq:satellite orbit action with approx} into:
\begin{equation}\label{eq:action full period-n orbit J invariant curves homoclinic}
{\cal F}_{\lbrace y\rbrace}\approx NF_{x}(q,q)+\sum_{i=1}^{J} \Big[ {\cal A}^{\circ}_{US[xh_0^{(i)}]}+(\delta q_{i}\delta p_{i+1}^{\prime}-\delta q_{i+1}^{\prime}\delta p_{i}) \Big]
\end{equation}
where $(\delta q_{i},\delta p_{i})$ is the phase space coordinate of $h_{k_i}^{(i)}$ relative to $x$, and $(\delta q_{i+1}^{\prime},\delta p_{i+1}^{\prime})$ is that of $h_{-l_i}^{(i+1)}$ relative to $x$.  The index $i$ is cyclic in $J$.  Each cross product yields an area $A^{(i)}$ labeled in the figure, which is approximately a parallelogram shaped area in phase space.

\subsubsection{Phase space transport}
\label{Transport}
\begin{figure}
  \subfigure{
    \label{fig:Transport_1}
    \includegraphics[width=6.5cm]{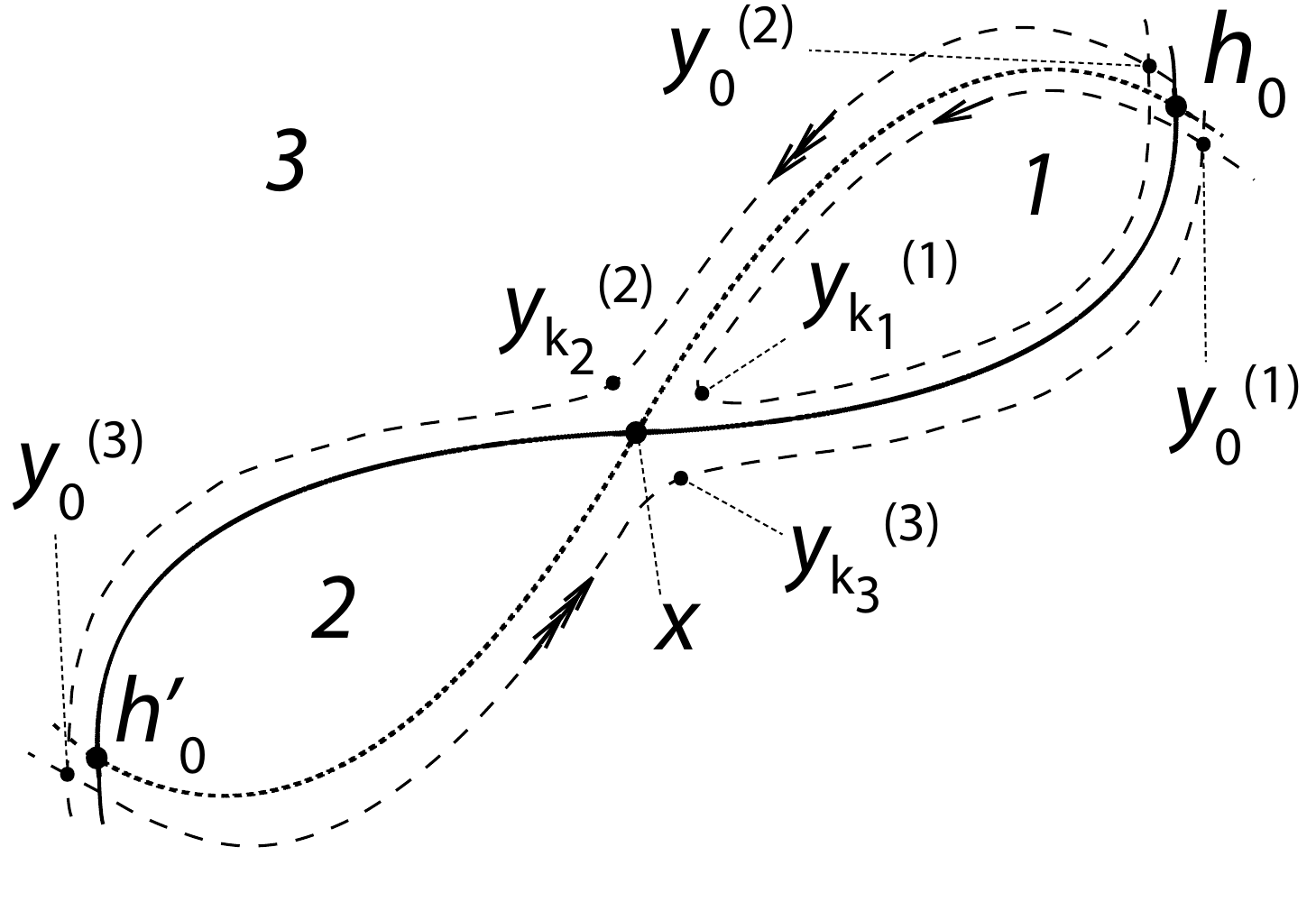}}
    \qquad         
  \subfigure{
    \label{fig:Transport_2}
    \includegraphics[width=6cm]{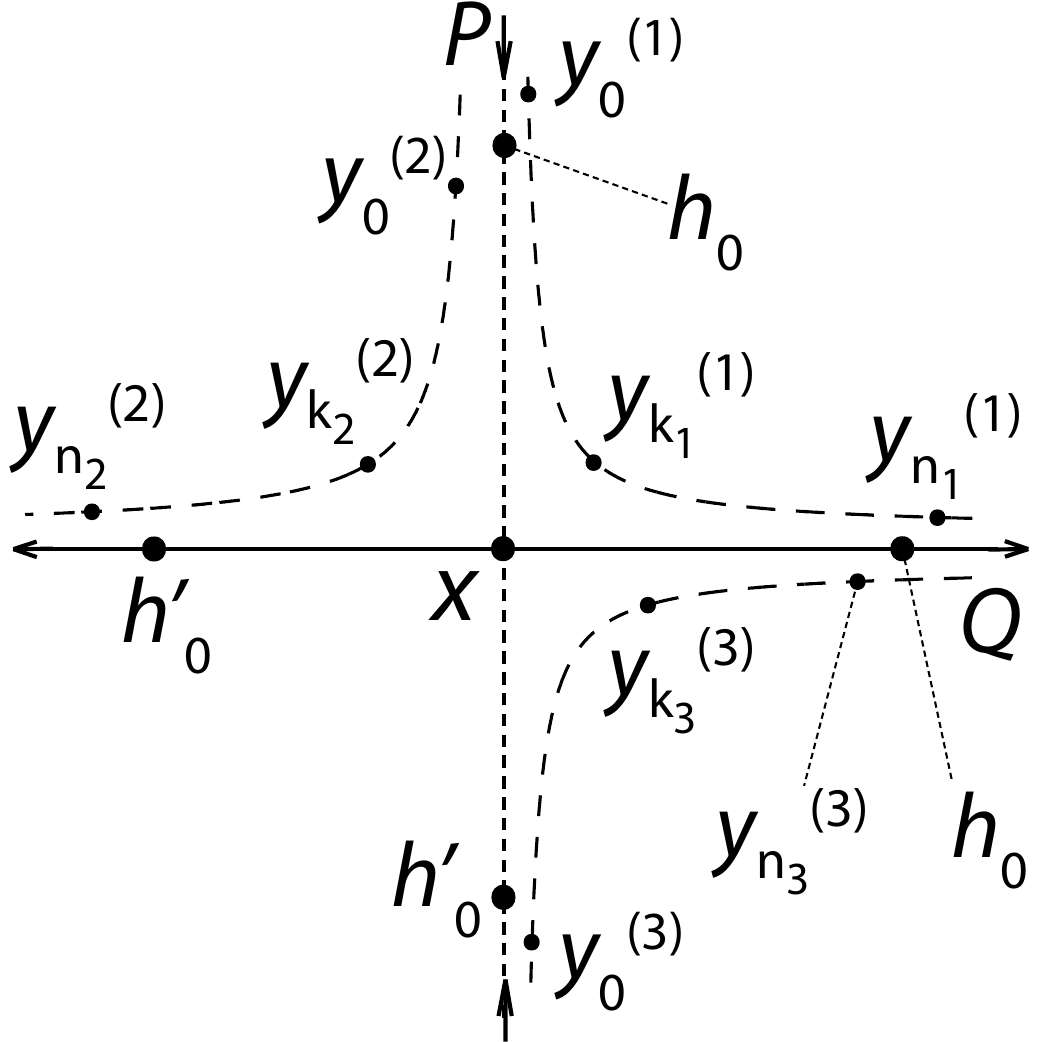}}
\caption{Upper panel: A periodic orbit $\lbrace y\rbrace$ supported by three Moser curves, which transport between region $1$ and $3$.  The single, double and triple arrows indicate the sequence of points along the orbit.  Lower panel: The orbit in normal form coordinates.  Relative positions between orbit points and the homoclinic points determines transition between different quadrants.  For example, in quadrant I, any point that is iterated to the right side of the $h_0$ on $Q$ axis, must transit to quadrant II.  This can be seen from $y_{n_1}^{(1)}$, which coincides with point $y_0^{(2)}$ in quadrant II. } 
\label{fig:Transport}
\end{figure}

Fig.~\ref{fig:J_invariant_curves_homoclinic} may leave the wrong impression that the Moser curves are all in the first quadrant of the normal form coordinates, which, in Fig.~\ref{fig:Ozorio}, is the complex region labeled as ``$1$", bounded by $U[x,h_0]$ and $S[x,h_0]$.  However, the discussions in the previous section apply to Moser curves in all quadrants, and such cases provide the clue of periodic orbit transport between different phase space regions.  Take the example of the homoclinic tangle from Fig.~\ref{fig:Ozorio}, where the the upper and lower branches of $U(x)$ intersect with those of $S(x)$ to form two complex regions, labeled as $1$ and $2$.  The phase space outside the complex regions is labeled as $3$.  A periodic orbit in general may transport between different regions during its motion, as the length of periodic orbits goes to infinity, the measure of region changing ones goes to one.  Thus almost all long orbits will jump between the three regions, and periodic orbit transport becomes prominent.  Take the example from Fig.~\ref{fig:Transport}, where a satellite orbit $\lbrace y\rbrace$ supported by three Moser curves $\mathcal{C}^{(1)}$, $\mathcal{C}^{(2)}$ and $\mathcal{C}^{(3)}$ in quadrant I, II and IV respectively, is schematically plotted.  The orbit spends $n_i$ iterations on $\mathcal{C}^{(i)}$, and is the sequential union of the segments on each curve: $\lbrace y\rbrace=\bigcup_{i=1}^{3}\lbrace y_{0}^{(i)},\cdots,y_{n_i}^{(i)} \rbrace$ with $y_{n_i}^{(i)}=y_0^{(i+1)}$, where $i$ is cyclic in $3$: $3+1=1$.  It starts from region $2$ outside the complex region, then enters region $1$ along $\mathcal{C}^{(1)}$, then to region $2$ along $\mathcal{C}^{(2)}$ and $\mathcal{C}^{(3)}$.  In the normal form coordinates, the orbit transits between quadrant I, II and IV.  The position of points in the normal form coordinates indicate phase space transport:  for any orbit point in quadrant I, if it is above the $h_0$ along the $P$ axis, or to the right side of the $h_0$ on the $Q$ axis, then it indicates this point is outside the complex region, e.g., point $y_0^{(1)}$ or $y_{n_1}^{(1)}$.  Similar results hold for quadrant III with $h'_0$.  There are certain transition rules between the four quadrants, point in quadrant I can only be iterated to quadrant I or II: 
\begin{equation}\label{eq:transition quadrant 1}
\mathrm{I}\to \mathrm{I},\ \mathrm{II}
\end{equation}
and similarly for points in quadrant II, III and IV:
\begin{equation}\label{eq:transition quadrant 2 3 4}
\begin{split}
&\mathrm{II}\to \mathrm{III},\ \mathrm{IV}\\
&\mathrm{III}\to \mathrm{III},\ \mathrm{IV}\\
&\mathrm{IV}\to \mathrm{I},\ \mathrm{II}\ .
\end{split}
\end{equation}
The corresponding transition rules between phase space regions are:
\begin{equation}\label{eq:transition regions}
\begin{split}
&1\to 1,\ 3\\
&2\to 2,\ 3\\
&3\to 1,\ 2,\ 3\ .
\end{split}
\end{equation}
The sequence of regions that a periodic orbit enters, along with the homoclinic points that induced it, may provide a full coding of all periodic orbits within the convergence zone, which is the topic of our future study.

\subsubsection{Heteroclinic}
\label{Heteroclinic}
\begin{figure}[ht]
\centering
{\includegraphics[width=9cm]{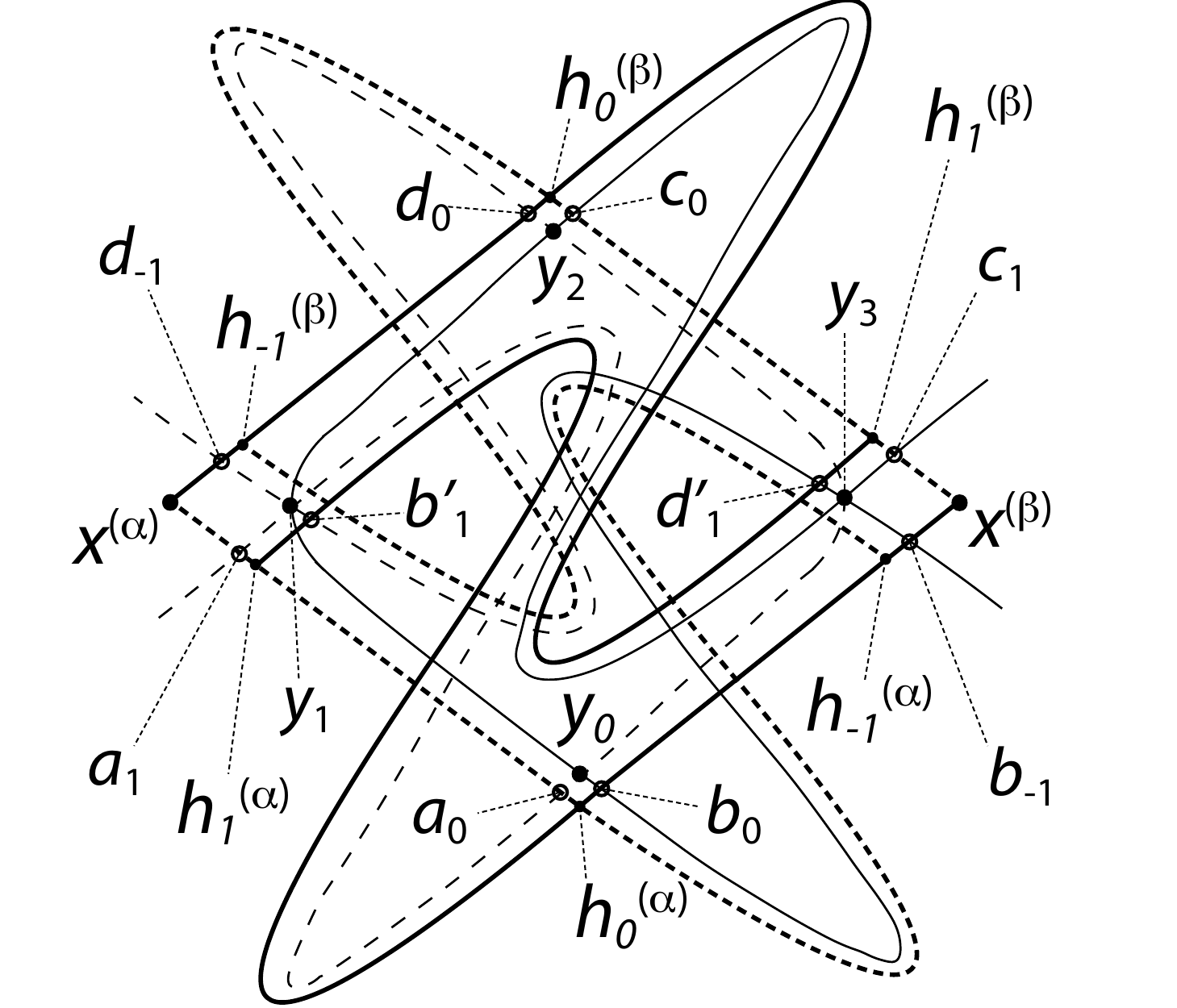}}
 \caption{Period-$4$ orbit $\lbrace y_{0}\rbrace$ supported by $\mathcal{C}^{(\alpha)}$ (thinner solid) and $\mathcal{C}^{(\beta)}$ (thinner dashed).  $h_{0}^{(\alpha)}=S(x^{(\alpha)})\bigcap U(x^{(\beta)})$, and $h_{0}^{(\beta)}=S(x^{(\beta)})\bigcap U(x^{(\alpha)})$.  The point $y_{0}$ is induced by $h_{0}^{(\alpha)}$, and $y_{2}$ is induced by $h_{0}^{(\beta)}$.  $\lbrace y_{0}\rbrace$ accumulates on $\lbrace h_{0}^{(\alpha)}\rbrace$ and $\lbrace h_{0}^{(\beta)}\rbrace$ alternatively in each period. }
\label{fig:Heteroclinic_satellite}
\end{figure}  
Since all periodic orbits within the convergence zone can be treated as satellite orbits of some homoclinic orbit segments, Eq.~\eqref{eq:action full period-n orbit J invariant curves homoclinic} would suffice in principle.  In practice however, it is often convenient to construct periodic orbits from the union of several known heteroclinic orbit segments.  Although such periodic orbits must be satellite to some homoclinic segments, the numerical calculation of those segments could be complicated.  Therefore, a formula relating the classical actions of the periodic to those of the heteroclinics is still desirable.  In the following content, we derive both exact and approximate results of such formula, and apply it to the orbits from the Standard map as an example.        

We begin with an example of two Moser curves from a heteroclinic tangle.  Consider a period-$4$ orbit $\lbrace y_{0} \rbrace$ shown in Fig.~\ref{fig:Heteroclinic_satellite}.  The satellite orbit $\lbrace y_{0}\rbrace$ is supported by Moser curves $\mathcal{C}^{(\alpha)}$ (thinner solid) and $\mathcal{C}^{(\beta)}$ (thinner dashed) in the heteroclinic tangle of fixed points $x^{(\alpha)}=(q^{(\alpha)},p^{(\alpha)})$ and $x^{(\beta)}=(q^{(\beta)},p^{(\beta)})$.  As $S(x^{(\alpha)})$ intersects $U(x^{(\beta)})$ at $h_{0}^{(\alpha)}$, $\mathcal{C}^{(\alpha)}$ and $\mathcal{C}^{(\beta)}$ are also forced to intersect at $y_{0}$, therefore $y_{0}$ can be viewed as induced by $h_{0}^{(\alpha)}$.  For the same reason, $y_{2}$ is induced by $h_{0}^{(\beta)}$.  Starting from $y_{0}$ in the neighborhood of $h_{0}^{(\alpha)}$, the orbit spends two iterations on $\mathcal{C}^{(\alpha)}$, which maps $y_{0}$ into $y_{2}$ in the neighborhood of $h_{0}^{(\beta)}$.  Then the orbit spends another two iterations on $\mathcal{C}^{(\beta)}$, which maps $y_{2}$ back to $y_{0}$.  During one period of its motion, $\lbrace y_{0}\rbrace$ accumulates on $\lbrace h_{0}^{(\alpha)} \rbrace$ and $\lbrace h_{0}^{(\beta)}\rbrace$ alternatively. 

Similar to the homoclinic case, the action function of $\lbrace y_{0}\rbrace$ can be related to the heteroclinic orbit actions via some phase space areas.  The derivation is based on $4$ steps:
\begin{enumerate}

\item Starting from $y_{0}$, and map $M(y_{0})=y_{1}$.  Follow the initial path $S[x^{(\alpha)},a_{0}]+I[a_{0},y_{0}]$, which is mapped to the final path $S[x^{(\alpha)},a_{1}]+I[a_{1},y_{1}]$.  Substituting the paths into Eq.~\eqref{eq:Meiss92} yields:
\begin{equation}\label{eq:satellite orbit action y0 to y1 heteroclinic}
F_{\lbrace y_0\rbrace}(q_0,q_1)-F_{x^{(\alpha)}}(q^{(\alpha)},q^{(\alpha)})= {\cal A}_{ISI[y_0a_{0}a_{1}y_1]} \ .
\end{equation}

\item $M(y_{1})=y_{2}$:  Let  the paths be $U[x^{(\alpha)},d_{-1}]+I[d_{-1},y_{1}]$ and $U[x^{(\alpha)},d_{0}]+I[d_{0},y_{2}]$, then:
 \begin{equation}\label{eq:satellite orbit action y1 to y2 heteroclinic}
F_{\lbrace y_0\rbrace}(q_1,q_2)-F_{x^{(\alpha)}}(q^{(\alpha)},q^{(\alpha)})= {\cal A}_{IUI[y_1d_{-1}d_{0}y_2]} \ .
\end{equation}

\item $M(y_{2})=y_{3}$: Let the paths be $S[x^{(\beta)},c_{0}]+I[c_{0},y_{2}]$ and $S[x^{(\beta)},c_{1}]+I[c_{1},y_{3}]$, then:
\begin{equation}\label{eq:satellite orbit action y2 to y3 heteroclinic}
F_{\lbrace y_0\rbrace}(q_2,q_3)-F_{x^{(\beta)}}(q^{(\beta)},q^{(\beta)})= {\cal A}_{ISI[y_2c_{0}c_{1}y_3]} \ .
\end{equation}

\item $M(y_{3})=y_{0}$: Let the paths be $U[x^{(\beta)},b_{-1}]+I[b_{-1},y_{3}]$ and $U[x^{(\beta)},b_{0}]+I[b_{0},y_{0}]$, then:
\begin{equation}\label{eq:satellite orbit action y3 to y0 heteroclinic}
F_{\lbrace y_0\rbrace}(q_3,q_0)-F_{x^{(\beta)}}(q^{(\beta)},q^{(\beta)})= {\cal A}_{IUI[y_3b_{-1}b_{0}y_0]} \ . 
\end{equation}

\end{enumerate}

The total action is thus:
\begin{equation}\label{eq:satellite orbit total action in compound path heteroclinic}
{\cal F}_{\lbrace y_0 \rbrace}-2F_{x^{(\alpha)}}(q^{(\alpha)},q^{(\alpha)})-2F_{x^{(\beta)}}(q^{(\beta)},q^{(\beta)})= {\cal A}^\circ_{\cal L}
\end{equation}
where the compound path $\cal{L}$ is:
\begin{equation}\label{eq:compound path L heteroclinic}
\begin{split}
{\cal L}=I[&y_{0},a_{0}]+S[a_{0},a_{1}]+I[a_{1},y_{1}]+I[y_{1},d_{-1}]\\
&+U[d_{-1},d_{0}]+I[d_{0},y_{2}]+I[y_{2},c_{0}]+S[c_{0},c_{1}]\\
&\quad +I[c_{1},y_{3}]+I[y_{3},b_{-1}]+U[b_{-1},b_{0}]+I[b_{0},y_{0}] .
\end{split}
\end{equation}
Similar to the previous section, it is possible to deform $\mathcal{L}$ and separate it into paths giving the relative actions of the two heteroclinic orbits and four parallelogram-shaped areas as correction terms: 
\begin{equation}
\label{eq:L deformation heteroclinic}
\begin{split}
 &{\cal A}^\circ_{\cal L}={\cal A}^{\circ}_{USUS[x^{(\alpha)}h_0^{(\beta)}x^{(\beta)}h_0^{(\alpha)}]}-{\cal A}^{\circ}_{UIIS[x^{(\alpha)}d_{-1}y_1a_1]}-\\
 &{\cal A}^{\circ}_{UIIS[x^{(\beta)}b_{-1}y_3c_1]}-{\cal A}^{\circ}_{IUSI[y_0b_{0}h_0^{(\alpha)}a_0]}-{\cal A}^{\circ}_{IUSI[y_2d_{0}h_0^{(\beta)}c_0]}
\end{split}
\end{equation}
where, according to Eq.~\eqref{eq:heteroclinic action past+future} 
\begin{equation}\label{eq:parallelogram into heteroclinic orbit sums}
\begin{split}
{\cal A}^{\circ}_{USUS[x^{(\alpha)}h_0^{(\beta)}x^{(\beta)}h_0^{(\alpha)}]}&=\Delta {\cal F}_{\lbrace h_{0}^{(\beta)}\rbrace^{-} x^{(\alpha)}}+\Delta {\cal F}_{\lbrace h_{0}^{(\beta)}\rbrace^{+} x^{(\beta)}}\\
& +\Delta {\cal F}_{\lbrace h_{0}^{(\alpha)}\rbrace^{-} x^{(\beta)}}+\Delta {\cal F}_{\lbrace h_{0}^{(\alpha)}\rbrace^{+} x^{(\alpha)}}
\end{split}
\end{equation}
and the other four $A^{\circ}$'s are parallelogram-shaped areas giving the corrections from the heteroclinics to the satellite orbit.   

Furthermore, since
\begin{equation}\label{eq:heteroclinic area combination 1}
\begin{split}
{\cal A}^{\circ}_{UIIS[x^{(\alpha)}d_{-1}y_1a_1]}+&{\cal A}^{\circ}_{IUSI[y_0b_{0}h_0^{(\alpha)}a_0]}\\
&\qquad \approx {\cal A}^{\circ}_{UIUS[x^{(\alpha)}d_{-1}b_1^{\prime}h_1^{(\alpha)}]}
\end{split}
\end{equation}
and similarly
\begin{equation}\label{eq:heteroclinic area combination 2}
\begin{split}
{\cal A}^{\circ}_{UIIS[x^{(\beta)}b_{-1}y_3c_1]}+&{\cal A}^{\circ}_{IUSI[y_2d_{0}h_0^{(\beta)}c_0]}\\
&\qquad \approx {\cal A}^{\circ}_{UIUS[x^{(\beta)}b_{-1}d_1^{\prime}h_1^{(\beta)}]}\ ,
\end{split}
\end{equation}
going through similar approximations as Sec.~\ref{Geometric area approximation}, we can express ${\cal A}^{\circ}_{UIUS[x^{(\alpha)}d_{-1}b_1^{\prime}h_1^{(\alpha)}]}$ and $ {\cal A}^{\circ}_{UIUS[x^{(\beta)}b_{-1}d_1^{\prime}h_1^{(\beta)}]}$ as cross products of the corresponding heteroclinic points relative to the fixed points:
\begin{equation}\label{eq:heteroclinic cross product}
\begin{split}
 &{\cal A}^{\circ}_{UIUS[x^{(\alpha)}d_{-1}b_1^{\prime}h_1^{(\alpha)}]}\approx \delta q_{\alpha\alpha}\delta p_{\beta\alpha}-\delta q_{\beta\alpha}\delta p_{\alpha\alpha}\\
 &{\cal A}^{\circ}_{UIUS[x^{(\beta)}b_{-1}d_1^{\prime}h_1^{(\beta)}]}\approx \delta q_{\beta\beta}\delta p_{\alpha\beta}-\delta q_{\alpha\beta}\delta p_{\beta\beta}
 \end{split}
\end{equation}
where the $\delta q$ and $\delta p$'s are the positions of the heteroclinic orbit points $h_{\pm 1}^{(\alpha/\beta)}$ relative to $x^{(\alpha/\beta)}$: 
\begin{equation}\label{eq:heteroclinic cross product relative coordinates}
\begin{split}
&\delta q_{\alpha\alpha}=q_{1}^{(\alpha)}-q^{(\alpha)},\ \delta p_{\alpha\alpha}=p_{1}^{(\alpha)}-p^{(\alpha)}\\
&\delta q_{\beta\alpha}=q_{-1}^{(\beta)}-q^{(\alpha)},\ \delta p_{\beta\alpha}=p_{-1}^{(\beta)}-p^{(\alpha)}\\
&\delta q_{\beta\beta}=q_{1}^{(\beta)}-q^{(\beta)},\ \delta p_{\beta\beta}=p_{1}^{(\beta)}-p^{(\beta)}\\
&\delta q_{\alpha\beta}=q_{-1}^{(\alpha)}-q^{(\beta)},\ \delta p_{\alpha\beta}=p_{-1}^{(\alpha)}-p^{(\beta)}\ .
\end{split}
\end{equation}

Therefore, assuming the knowledge of the fixed point actions $F_{x^{(\alpha)}}(q^{(\alpha)},q^{(\alpha)})$ and $F_{x^{(\beta)}}(q^{(\beta)},q^{(\beta)})$, heteroclinic orbit action sum ${\cal A}^{\circ}_{USUS[x^{(\alpha)}h_0^{(\beta)}x^{(\beta)}h_0^{(\alpha)}]}$ and the coordinates of $h_{\pm 1}^{(\alpha/\beta)}$, the action of the satellite orbit is determined by:
\begin{equation}\label{eq:heteroclinic satellite action 2 moser curves}
\begin{split}
&{\cal F}_{\lbrace y_0 \rbrace}=2F_{x^{(\alpha)}}(q^{(\alpha)},q^{(\alpha)})+2F_{x^{(\beta)}}(q^{(\beta)},q^{(\beta)})+ {\cal A}^\circ_{\cal L}\\
&\approx 2F_{x^{(\alpha)}}(q^{(\alpha)},q^{(\alpha)})+2F_{x^{(\beta)}}(q^{(\beta)},q^{(\beta)})\\
&\quad+{\cal A}^{\circ}_{USUS[x^{(\alpha)}h_0^{(\beta)}x^{(\beta)}h_0^{(\alpha)}]}-(\delta q_{\alpha\alpha}\delta p_{\beta\alpha}-\delta q_{\beta\alpha}\delta p_{\alpha\alpha})\\
&\quad\quad-(\delta q_{\beta\beta}\delta p_{\alpha\beta}-\delta q_{\alpha\beta}\delta p_{\beta\beta})\ .
\end{split}
\end{equation} 

\begin{figure}
  \subfigure{
    \label{fig:J_invariant_curves_1}
    \includegraphics[width=6cm]{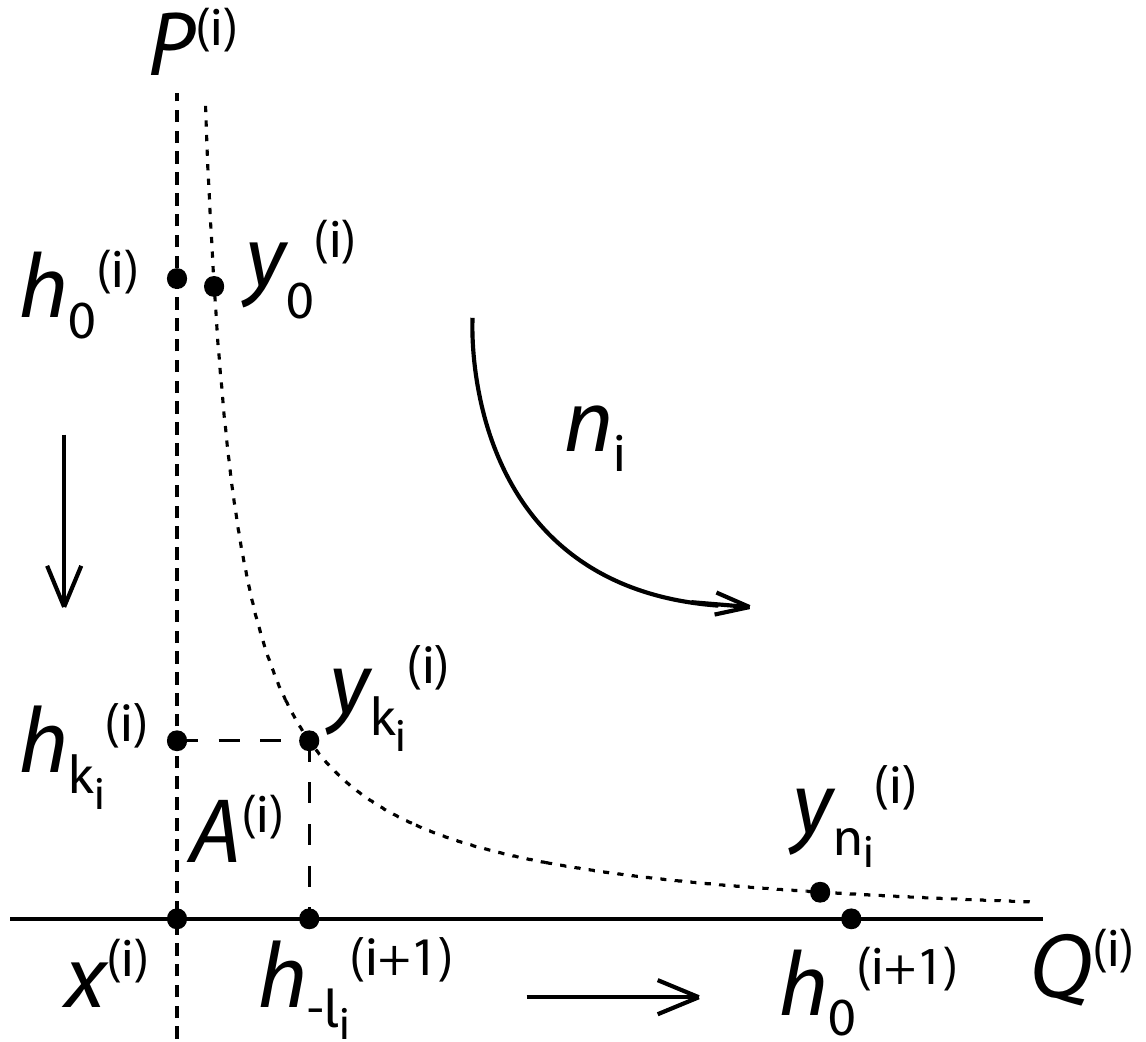}}
    \qquad         
  \subfigure{
    \label{fig:J_invariant_curves_2}
    \includegraphics[width=6.3cm]{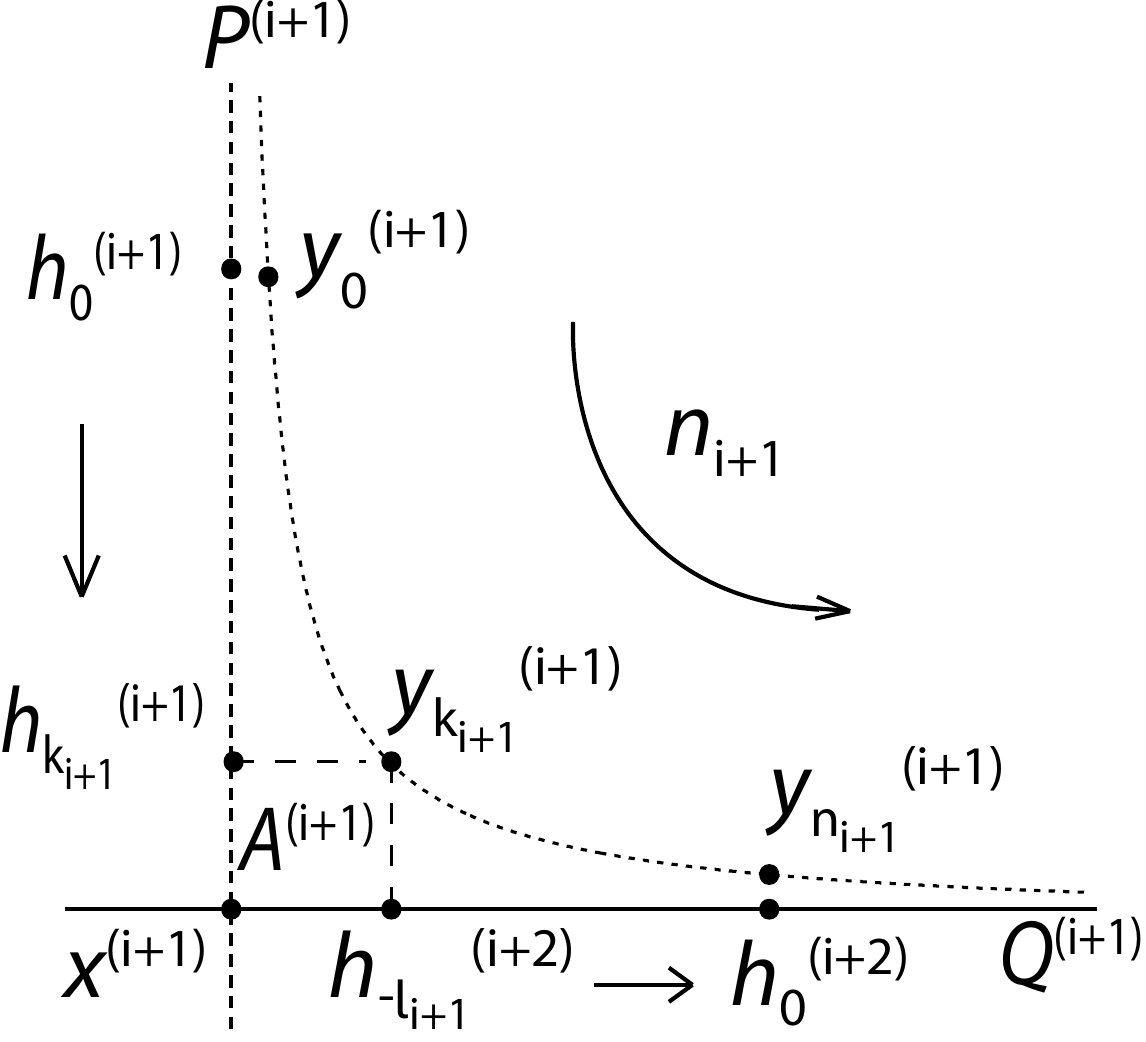}}
\caption{Normal form picture of a satellite orbit accumulating on $J$ heteroclinic orbits.  Each $\lbrace h_{0}^{(i)}\rbrace$ is a heteroclinic orbit connecting $x^{(i-1)}$ to $x^{(i)}$, and the satellite orbit accumulates on them successively during one period.  Starting from $y_{0}^{(i)}$ in the upper panel, the orbit spends $n_{i}$ iterations on $\mathcal{C}^{(i)}$, which brings it to $y_{n_i}^{(i)}$.  Then it is passed on to $\mathcal{C}^{(i+1)}$ in the lower panel, where $y_{0}^{(i+1)}=y_{n_i}^{(i)}$, and repeat the similar process.  During each $n_{i}$, the orbit spends the first $k_{i}$ iterations accumulating on $\lbrace h_{0}^{(i)}\rbrace$, then switch to accumulate on $\lbrace h_{0}^{(i+1)}\rbrace$ for the rest $l_i$ iterations, where $k_i+l_i=n_i$.  } 
\label{fig:J_invariant_curves}
\end{figure}
Eq.~\eqref{eq:heteroclinic satellite action 2 moser curves} can be extended to satellite orbits can be supported by $J$ Moser invariant curves, and induced by $J$ heteroclinic intersections, where $J$ is an arbitrary integer.  This will include all possible constructions of satellite orbits in practice.  As shown in Fig.~\ref{fig:J_invariant_curves}, where we have $J$ hyperbolic fixed points, denoted by $x^{(1)}, x^{(2)}, \cdots, x^{(J)}$.  Each $x^{(i)}$ has its own $S(x^{(i)})$ and $U(x^{(i)})$, which become the axis $P^{(i)}$ and $Q^{(i)}$ respectively in the normal form coordinates.  Consider $J$ heteroclinic intersections between successive fixed points: $h_0^{(i)}=U(x^{(i-1)}) \bigcap S(x^{(i)})$, they generate $J$ heteroclinic orbits $\lbrace h_0^{(i)}\rbrace$ connecting from $x^{(i-1)}$ to $x^{(i)}$.  Study a period-$N$ orbit $\lbrace y \rbrace$ induced by the $J$ heteroclinic intersections, and accumulates on the $J$ heteroclinic orbits alternatively during each period.  Then $\lbrace y \rbrace$ is supported by $J$ invariant curves $\mathcal{C}^{(i)}$, which carries it from the neighborhood of $h_0^{(i)}$ to the neighborhood of $h_0^{(i+1)}$ under $n_i$ iterations of the map.    The period $N$ is then:
\begin{equation}\label{eq:n into ni homoclinic}
N=\sum_{i=1}^{J}n_{i}.
\end{equation}      
The action of $\lbrace y \rbrace$ is a direct generalization of Eq.~\eqref{eq:heteroclinic satellite action 2 moser curves}:
\begin{equation}\label{eq:action full period-n orbit J invariant curves}
\begin{split}
{\cal F}_{\lbrace y\rbrace}&\approx \sum_{i=1}^{J} \Big[ n_{i}F_{x^{(i)}}(q^{(i)},q^{(i)})+{\cal A}_{US[x^{(i)}h_0^{(i+1)}x^{(i+1)}]}\\
&\qquad\qquad-(\delta q_{ii}\delta p_{(i+1)i}-\delta q_{(i+1)i}\delta p_{ii}) \Big]
\end{split}
\end{equation}
where $(\delta q_{ii},\delta p_{ii})$ is the phase space coordinate of $h_{k_i}^{(i)}$ relative to $x^{(i)}$, and $(\delta q_{(i+1)i},\delta p_{(i+1)i})$ is that of $h_{-l_i}^{(i+1)}$ relative to $x^{(i)}$.  Refer to Fig.~\ref{fig:J_invariant_curves} for the definition of these points.  Each cross product yields the area $A^{(i)}$ labeled in the figure, which maps to a parallelogram-shaped area in phase space.  The index $i$ is cyclic in $J$: $1+J=1$.   Notice that the $J$ fixed points do not have to be distinct, in cases of coinciding fixed points some heteroclinic orbits reduce to homoclinic orbits, but the same formula is still valid.  Eq.~\eqref{eq:action full period-n orbit J invariant curves} applies to any satellite periodic orbit constructed by arbitrary combinations of homoclinic/heteroclinic orbits.  It only requires the prior knowledge of the actions associated with each fixed point, and the homoclinic/heteroclinic orbit segments that constructed the periodic orbit.      

\subsubsection{Standard map example}
\label{Standard map satellite example}
Using the standard map define in Eq.~\eqref{eq:kicked rotor}, we have numerically calculated two heteroclinic intersections $h_0^{(\alpha)}=S(x^{(\alpha)})\bigcap U(x^{(\beta)})$ and $h_0^{(\beta)}=S(x^{(\beta)})\bigcap U(x^{(\alpha)})$, where
\begin{equation}
\begin{split}
h_0^{(\alpha)}&=( -7.662754300\times 10^{-2},-0.518797853)\\
 h_0^{(\beta)}&=(0.442170310,0.518797853)
\end{split}
\end{equation}
and their heteroclinic orbit segments $\lbrace h_{-2}^{(\alpha/\beta)},\cdots,h_{2}^{(\alpha/\beta)}\rbrace$.  A period-$8$ orbit is constructed from the two heteroclinic segments, with initial point $y_0=(0.442153628,0.518385726)$.  By setting $\alpha=1$ and $\beta=2$ for the indexes of Eq.~\eqref{eq:action full period-n orbit J invariant curves}, the parallelogram areas can be calculated using the orbit actions:
\begin{equation}
\begin{split}
\sum_{i=1}^{2}&(\delta q_{ii}\delta p_{(i+1)i}-\delta q_{(i+1)i}\delta p_{ii})\approx \\
&4F_{x^{(1)}}(q^{(1)},q^{(1)})+4F_{x^{(2)}}(q^{(2)},q^{(2)})\\
&\qquad+{\cal A}^{\circ}_{USUS[x^{(1)}h_0^{(2)}x^{(2)}h_0^{(1)}]}-{\cal F}_{\lbrace y\rbrace}\\
&\qquad\qquad\qquad =1.8662\times 10^{-4}
\end{split}
\end{equation}
while direct calculation of the cross product gives:
\begin{equation}
\sum_{i=1}^{2}(\delta q_{ii}\delta p_{(i+1)i}-\delta q_{(i+1)i}\delta p_{ii})=1.8685\times 10^{-4}\ .
\end{equation}
They match to the $3^{rd}$ decimal place.  On the other hand, without the knowledge of its orbit points, the satellite action can be calculated using Eq.~\eqref{eq:action full period-n orbit J invariant curves}:
\begin{equation}
{\cal F}_{\lbrace y\rbrace}\approx 0.2585909
\end{equation}   
while the same quantity can be calculated with the knowledge of its orbit:
\begin{equation}
{\cal F}_{\lbrace y\rbrace}= 0.2585911\ .
\end{equation}   
The results match to the $7^{th}$ decimal place, demonstrating the high accuracy of Eq.~\eqref{eq:action full period-n orbit J invariant curves}.

\section{Implications on Sieber-Richter orbit pairs}
\label{Implications Sieber-Richter}

\begin{figure}[ht]
\centering
{\includegraphics[width=8cm]{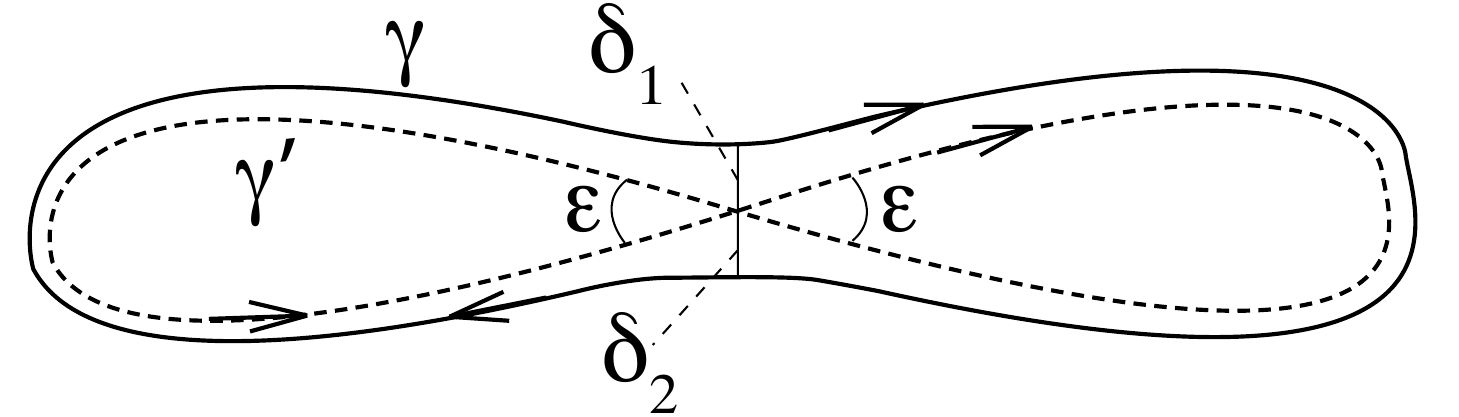}}
 \caption{Sieber-Richter orbit pair.  $\gamma$ is the non-crossing orbit, $\gamma'$ is the $8$-shaped orbit with a crossing in the middle.  $\epsilon$ is the small opening angle of the crossing.  $\delta_{1}$ and $\delta_{2}$ are the distances from the parallel segments on $\gamma$ to the crossing point.  The trajectories of $\gamma$ and $\gamma'$ get exponentially close on the left and right loops, only deviate from each other in center encounter region.}
\label{fig:Sieber-Richter}
\end{figure}  

The concept of Sieber-Richter orbit pair was introduced in detail in \cite{Sieber01}, and later generalized in \cite{Muller04,Turek05,Muller05}.  An in-depth discussion can be found in Chap.~9 of \cite{Haake10}.  Such orbit pairs, as labeled by $\gamma$ and $\gamma '$ in Fig.~\ref{fig:Sieber-Richter}, provide a general mechanism for action correlations between periodic orbit pairs in time-reversible systems that nearly trace each other in the configurations space.  
\begin{figure}
 \subfigure{
   \label{fig:Sieber-Richter_phase_space_a}
   \includegraphics[width=8cm]{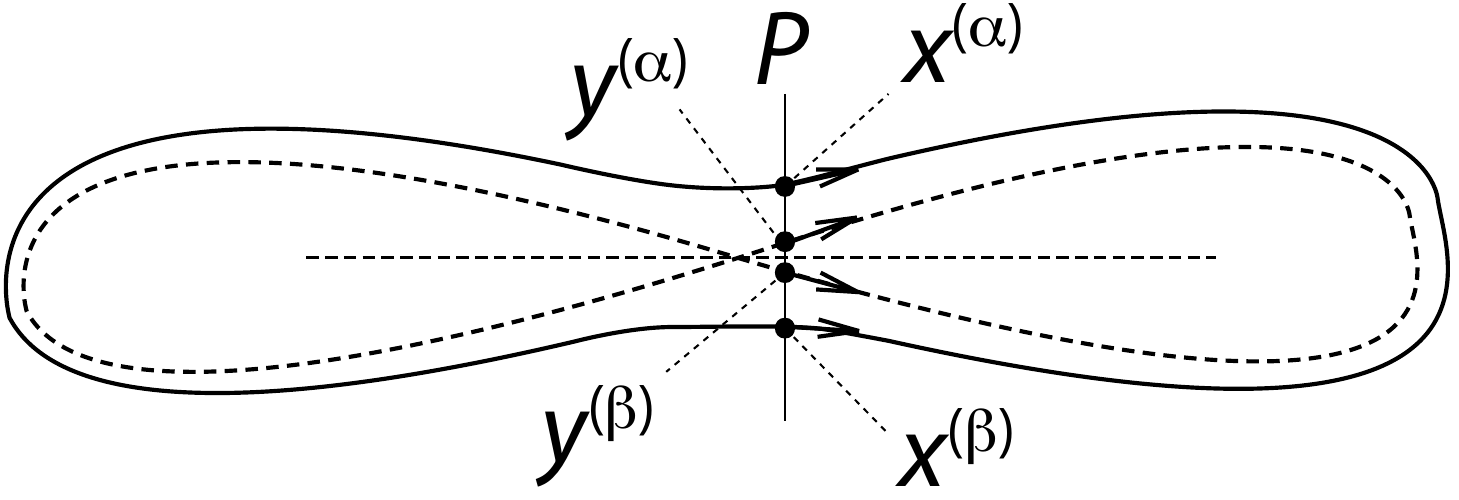}}
   \par\bigskip 
 \subfigure{
   \label{fig:Sieber-Richter_phase_space_b}
   \includegraphics[width=8cm]{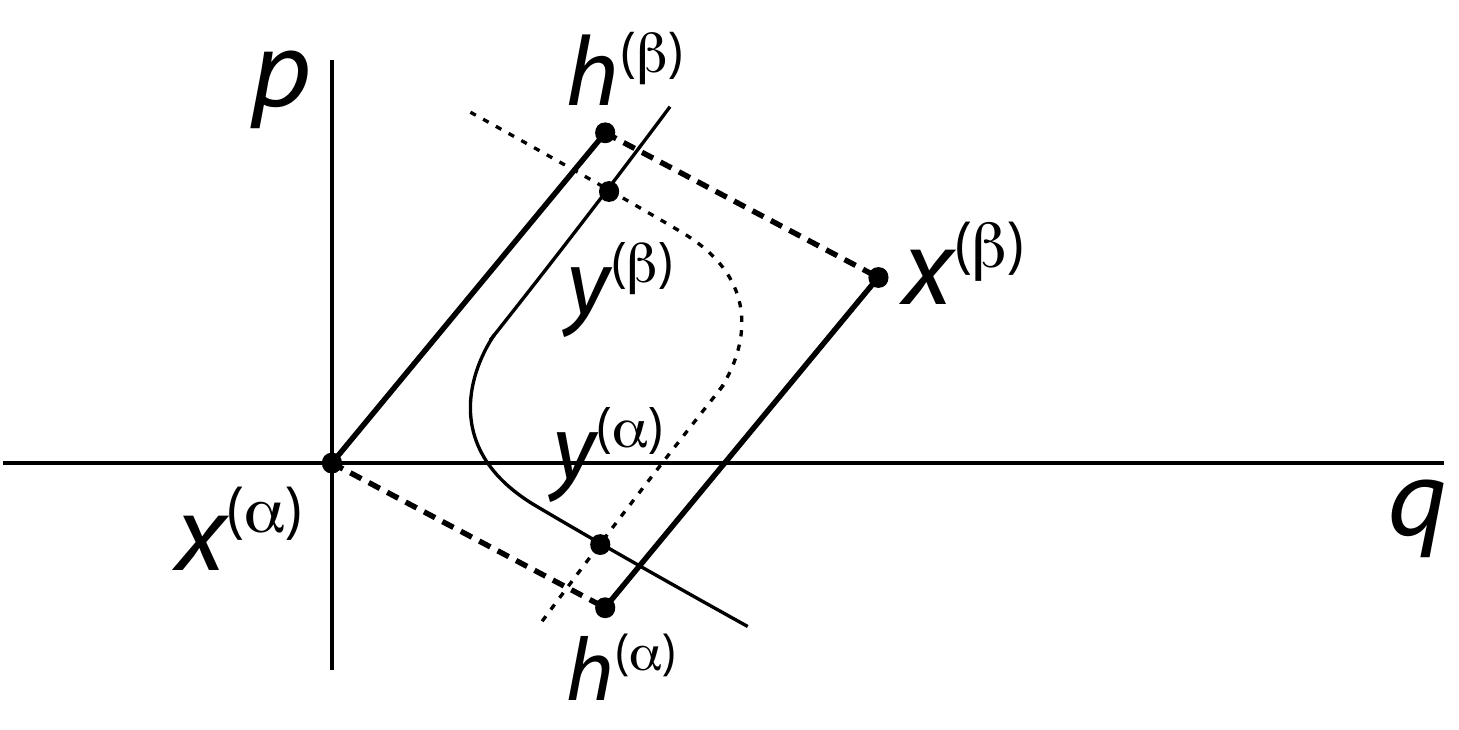}}
\caption{Upper panel: Poincar\'{e} surface of section $P$ goes through the crossing point and perpendicular to the horizontal direction.  $\gamma$, $\tau\gamma$, $\gamma'$ and $\tau\gamma'$ becomes the fixed piercing points $x^{(\alpha)}$, $x^{(\beta)}$, $y^{(\alpha)}$ and $y^{(\beta)}$ on $P$.  Lower panel: Canonical coordinates on $P$.  The unstable (resp. stable) manifolds of $x^{(\alpha)}$ and $x^{(\beta)}$ are plotted in solid (dashed) lines, forming a complex region of the heteroclinic tangle.  $y^{(\alpha)}$ and $y^{(\beta)}$ are period-$1$ satellite orbits induced by $h^{(\alpha)}$ and $h^{(\beta)}$, supported by two invariant curves $\mathcal{C}^{(\alpha)}$ (lighter solid) and $\mathcal{C}^{(\beta)}$ (lighter dashed). } 
\label{fig:Sieber-Richter_phase_space}
\end{figure}
The inner $8$-shaped orbit $\gamma'$ makes a small angle ($\epsilon$) self-crossing.  Linearization with respect to $\gamma'$ allows the existence of a partner orbit $\gamma$, which stays exponentially close to $\gamma'$ in the left and right loops, only to differ from it in the middle of the $8$-shaped $\gamma'$.  As seen from the figure, the $\gamma$ orbit avoids the narrow crossing and forms a pair of parallel segments in the middle region, which is called $2$-$\mathit{encounters}$, meaning that two segments from $\gamma$ come nearly parallel to each other without self-crossing in the middle of the $8$-shaped $\gamma'$.  As shown in Fig.~\ref{fig:Sieber-Richter_phase_space}, chose the direction of the angular bisector of $\epsilon$ to be the ``horizontal'' direction (horizontal dashed line).  The Poincar\'{e} surface of section $P$ is chosen to be the plane that goes through the crossing point and perpendicular to the horizontal direction.  The returning map is defined as successive incidents that an orbit pierces through the surface of section from left to right.  Under this map, the orbit $\gamma$ reduces to a single piercing point $x^{(\alpha)}$, and $\gamma'$ reduces to a piercing point $y^{(\alpha)}$.  The time-reversal of $\gamma$, denoted by $\tau\gamma$, reduces to $x^{(\beta)}$, and the time reversal $\tau\gamma'$ reduces to $y^{(\beta)}$.  Note that $y^{(\alpha)}$ and $y^{(\beta)}$ actually coincide with the crossing point of $\gamma'$, we are only plotting them as separated for clarity.   

The phase space diagram for the returning map on $P$ is demonstrated by the lower panel of Fig.~\ref{fig:Sieber-Richter_phase_space}.  Placing $x^{(\alpha)}$ at the origin, and let $p$ be the relative momentum between $\gamma'$ and $\gamma$ at $y^{(\alpha)}$.  Let $\theta_{1}$ and $\theta_{2}$ be the angles between the horizontal line and the tangents to the outer orbit at $x^{(\alpha)}$ and $x^{(\beta)}$, then:
\begin{equation}
\label{eq:Sieber-Richter coordinates}
\begin{split}
&x^{(\beta)}=\big(\delta_{1}+\delta_{2},p(\theta_{1}-\theta_{2})\big)\\
&y^{(\alpha)}=\big(\delta_{1},p(\theta_{1}-\epsilon/2)\big)\\
&y^{(\beta)}=\big(\delta_{1},p(\theta_{1}+\epsilon/2)\big).
\end{split}
\end{equation} 
Two heteroclinic intersections are identified in the figure, which are $h^{(\alpha)}=S(x^{(\alpha)})\bigcap U(x^{(\beta)})$ and $h^{(\beta)}=S(x^{(\beta)})\bigcap U(x^{(\alpha)})$.  Since $h^{(\alpha)}$ (resp. $h^{(\beta)}$) is exponentially close to $y^{(\alpha)}$ (resp. $y^{(\beta)}$), it is legitimate to approximate:
\begin{equation}\label{eq:Sieber-Richter coordinates heteroclinic points}
h^{(\alpha)}\approx \big(\delta_{1},p(\theta_{1}-\epsilon/2)\big),\qquad h^{(\beta)}\approx\big(\delta_{1},p(\theta_{1}+\epsilon/2)\big).
\end{equation} 
Previous researches from \cite{Sieber01,Muller04,Turek05,Muller05} has treated $h^{(\alpha)}$ and $y^{(\alpha)}$ (resp. $h^{(\beta)}$ and $y^{(\beta)}$) as same points, implicitly replacing $\gamma'$ with a sum of two heteroclinic orbits $\lbrace h^{(\alpha)}\rbrace$ and $\lbrace h^{(\beta)}\rbrace$, to obtain an expression for the action difference between $\gamma$ and $\gamma'$ to the leading order:
\begin{equation}\label{eq:Sieber-Richter action difference parallelogram}
S_{\gamma'}-S_{\gamma}\approx {\cal A}^{\circ}_{USUS[x^{(\alpha)}h^{(\beta)}x^{(\beta)}h^{(\alpha)}]}=\frac{p\epsilon}{2}(\delta_{1}+\delta_{2}).
\end{equation}  
where ${\cal A}^{\circ}_{USUS[x^{(\alpha)}h^{(\beta)}x^{(\beta)}h^{(\alpha)}]}$ is the sum of relative actions of $\lbrace h^{(\alpha)}\rbrace$ and $\lbrace h^{(\beta)}\rbrace$:
\begin{equation}\label{eq:Sieber-Richter area strict interpretation}
\begin{split}
&{\cal A}^{\circ}_{USUS[x^{(\alpha)}h^{(\beta)}x^{(\beta)}h^{(\alpha)}]}=\Delta {\cal F}_{\lbrace h^{(\beta)}\rbrace^{-} x^{(\alpha)}}+\Delta {\cal F}_{\lbrace h^{(\beta)}\rbrace^{+} x^{(\beta)}}\\
&\qquad\qquad\qquad+\Delta {\cal F}_{\lbrace h^{(\alpha)}\rbrace^{-} x^{(\beta)}}+\Delta {\cal F}_{\lbrace h^{(\alpha)}\rbrace^{+} x^{(\alpha)}}\ .
\end{split}
\end{equation}

\begin{figure}[ht]
\centering
{\includegraphics[width=7cm]{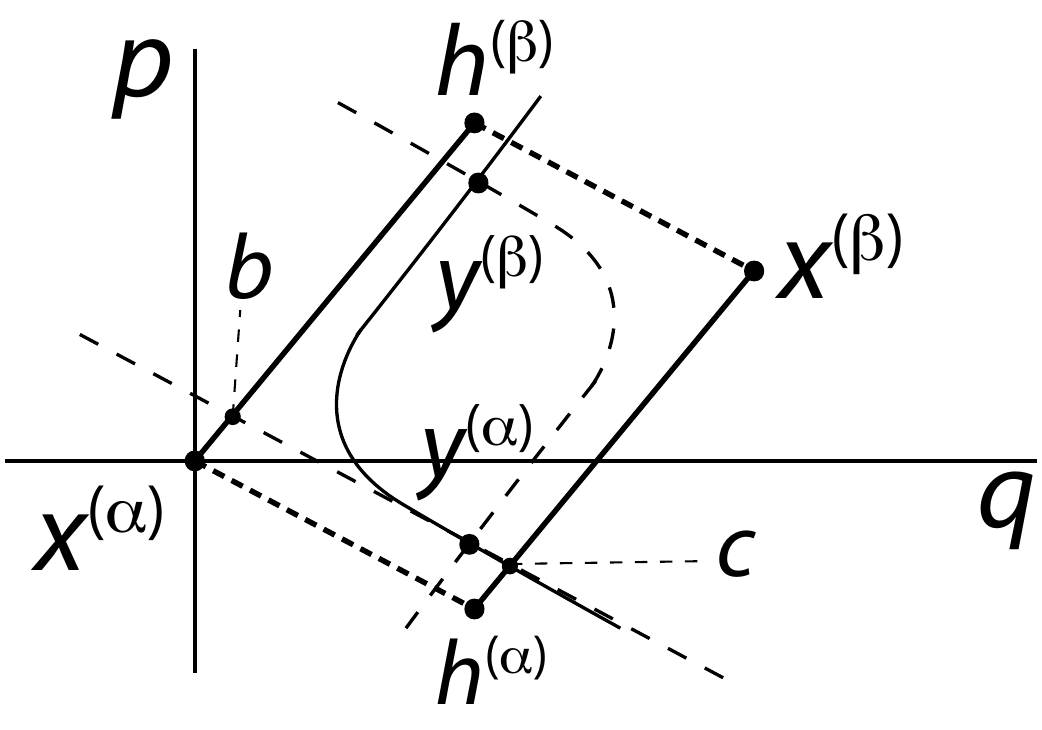}}
 \caption{Parallelograms for the Sieber-Richter pair.  This figure is the special case of Fig.~\ref{fig:Heteroclinic_satellite} in which $n_{\alpha}=1$, $n_{\beta}=0$.  ${\cal A}_{UIUS[x^{(\alpha)}bch^{(\alpha)}]}\approx {\cal A}^{\circ}_{USUS[x^{(\alpha)}h^{(\beta)}x^{(\beta)}h^{(\alpha)}]}\cdot e^{-\kappa}$.}
\label{fig:Sieber-Richter_parallelogram}
\end{figure}

Choosing $\gamma$ as the center orbit, then $\gamma'$ is just a satellite orbit that ``circulates'' in the heteroclinic tangle of $\gamma$ and $\tau\gamma$.  This is shown by the lower panel of Fig.~\ref{fig:Sieber-Richter_phase_space}, where $\gamma$ (resp. $\tau\gamma$) is represented by $x^{(\alpha)}$ (resp. $x^{(\beta)}$), and $\gamma'$ (resp. $\tau\gamma'$) represented by $y^{(\alpha)}$ (resp. $y^{(\beta)}$).  $y^{(\alpha)}$ and $y^{(\beta)}$ are intersections between two Moser curves (lighter dashed) from the heteroclinic tangle, induced by $h^{(\alpha)}$ and $h^{(\beta)}$ respectively.  Therefore, $\lbrace y^{(\alpha)}\rbrace$ (resp.  $\lbrace y^{(\beta)}\rbrace$) are just period-$1$ satellite orbits supported by two invariant curves under the Poincar\'{e} return map, giving $\gamma'$ (resp. $\tau\gamma'$) in the continuous time flow.  This is a special case of a heteroclinic orbit satellite with period $1$.  Refer to Fig.~\ref{fig:Sieber-Richter_parallelogram} for the geometry.  A slight modification to Eq.~\eqref{eq:satellite orbit total action in compound path heteroclinic} by setting $n_{\alpha}=1$, $n_{\beta}=0$ yields
\begin{equation}
\label{eq:Sieber-Richter action difference exact}
\begin{split}
{\cal F}_{\lbrace y^{(\alpha)}\rbrace}-F_{x^{(\alpha)}}(q^{(\alpha)},q^{(\alpha)})={\cal A}^{\circ}_{USUI[bh^{(\beta)}x^{(\beta)}c]}
\end{split}
\end{equation}
thus:
\begin{equation}\label{eq:Sieber-Richter action difference S gamma' - S gamma}
\begin{split}
S_{\gamma'}-S_{\gamma}&={\cal A}^{\circ}_{USUI[bh^{(\beta)}x^{(\beta)}c]}\\
&\approx {\cal A}^{\circ}_{USUS[x^{(\alpha)}h^{(\beta)}x^{(\beta)}h^{(\alpha)}]}\cdot (1-e^{-\kappa})\\
&=\frac{p\epsilon}{2}(\delta_{1}+\delta_{2})\cdot (1-e^{-\kappa}).
\end{split}
\end{equation}
Previous researches implicitly replaced the satellite orbit by the sum of two heteroclinic orbits, thus the $e^{-\kappa}$ correction was absent.

\section{Conclusion}
\label{Conclusions}

The information about classical actions associated with homoclinic, heteroclinic and periodic orbits that come into various semiclassical sum rules play an important role in the study of quantum chaotic dynamical systems.  Although the orbit actions can be calculated from the generating functions, the relations and correlations amongst their values cannot be discovered without an analysis of the type given in this paper.  Furthermore, in the asymptotic limit of semiclassical mechanics, the actions must be known to high precision to understand the interferences that arise in quantum dynamics, and that otherwise requires the accurate determination of long orbit segments.  Since any initial deviation due to the machine precision will diverge exponentially, it is a priori difficult to compute periodic orbits with long periods.  The analysis given here gives an explicit mechanism from which correlations could emerge and avoids the numerical difficulties by making the detailed long orbit calculations unnecessary.   

One interesting example is given by the heteroclinic tangle of the standard map, which arises from the two unstable fixed points of the map on the $p=0$ line.  One of the fixed points is hyperbolic with reflection, which generates a single lobe fundamental structure in the tangle under a double iteration of the map.  This lobe's area must equal twice the action difference of the fixed points, a nontrivial relation to imagine without generating Eqs.~(\ref{eq:heteroclinic action past+future},\ref{eq:heteroclinic action past+future-k}).

For fully chaotic systems, the convergence zone can cover most of the accessible phase space~\cite{Harsoula15}, and in that case nearly all of the periodic orbits fall into the category of satellite orbits, to which our analysis applies.  Action differences between any pair of the satellite periodic orbits or between them and particular homoclinic (heteroclinic) orbits follow naturally.  The simple, rather accurate geometric approximation involving a wedge product generates expressions that do not require the construction of the orbits or Moser invariant curves, only short sections of the stable and unstable manifolds (very simple and stable to calculate) and the endpoints of the homoclinic segments concerned.  The error of this approximation scheme decreases exponentially as the length of the orbit increases and the instability exponent of the system increases.  

All Moser invariant curves intersect in the untransformed phase space and as shown in~\cite{Birkhoff27,Silva87,Ozorio89}, some satellite orbits lie on more than one Moser curve.  In those cases, the actions of the satellite orbits are also related to homoclinic (heteroclinic) orbit actions, and possibly multiple fixed point actions~\cite{Ozorio89}.  It is of significant interest to understand the connections of the resulting multiple possible action relations, and this subject is left for future publication.

\acknowledgments

The authors gratefully acknowledge that Eq.~(19) (which relates the action difference of two periodic orbits with the loop structure of their heteroclinic tangle) was derived jointly with Akira Shudo, Hiromitsu Harada, Kensuke Yoshida during a productive visit to Tokyo Metropolitan University and also gratefully acknowledge support for the travel.  

\appendix
 
\section{MacKay-Meiss-Percival action principle}
\label{MacKay-Meiss-Percival}

The MacKay-Meiss-Percival action principle discussed in this section was first developed in \cite{MacKay84a} for transport theory.  A comprehensive review can be found in \cite{Meiss92}.  Generalization of the original principle beyond the ``twist" and area-preserving conditions is discussed in \cite{Easton91}.  A higher-dimensional generalization using generating $1$-forms and phase space volume forms is discussed in \cite{Lomeli09}.    

Consider an arbitrary point $a_{0}=(q_{0},p_{0})$ and its orbit $\lbrace a_{0}\rbrace$ in phase space .  The twist condition indicates the existence of a generating (action) function $F(q_{n},q_{n+1})$ which brings $a_{n}$ into $a_{n+1}$ under the mapping $M$, such that:
\begin{equation}
\begin{split}
&p_{n}=-\partial F/\partial q_{n}\\ 
&p_{n+1}=\partial F/\partial q_{n+1}.
\end{split}
\end{equation}

The total action ${\cal F}$ is the sum:
\begin{equation}
\label{eq:full orbit action in general appendix}
{\cal F}=\sum_{n=-\infty}^{\infty}F(q_{n},q_{n+1}).
\end{equation}

\begin{figure}[ht]
\centering
{\includegraphics[width=8cm]{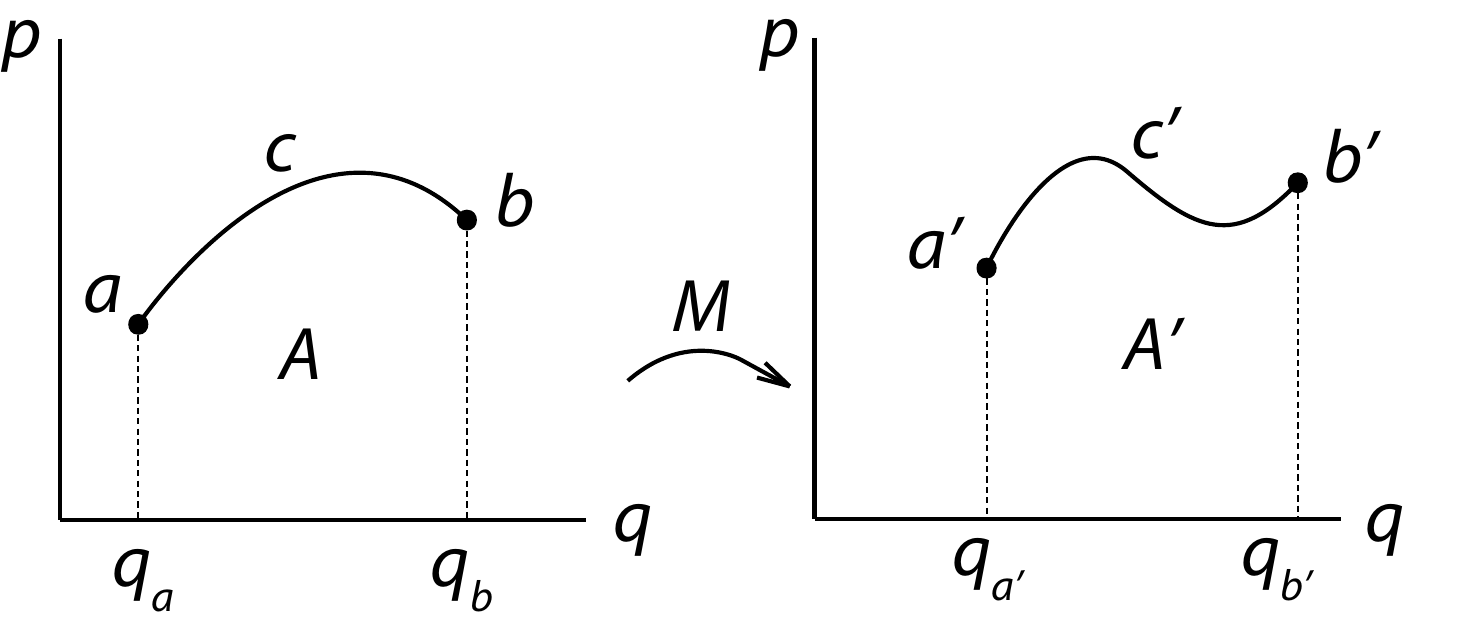}}
 \caption{$a$ and $b$ are arbitrary points and $c$ is a curve connecting them.  $a'=M(a)$, $b'=M(b)$ and $c'=M(c)$.  Then: $A'-A=F(q_{b},q_{b'})-F(q_{a},q_{a'})$. }
\label{fig:Area_under_curve}
\end{figure}
\begin{figure}[ht]
\centering
{\includegraphics[width=4cm]{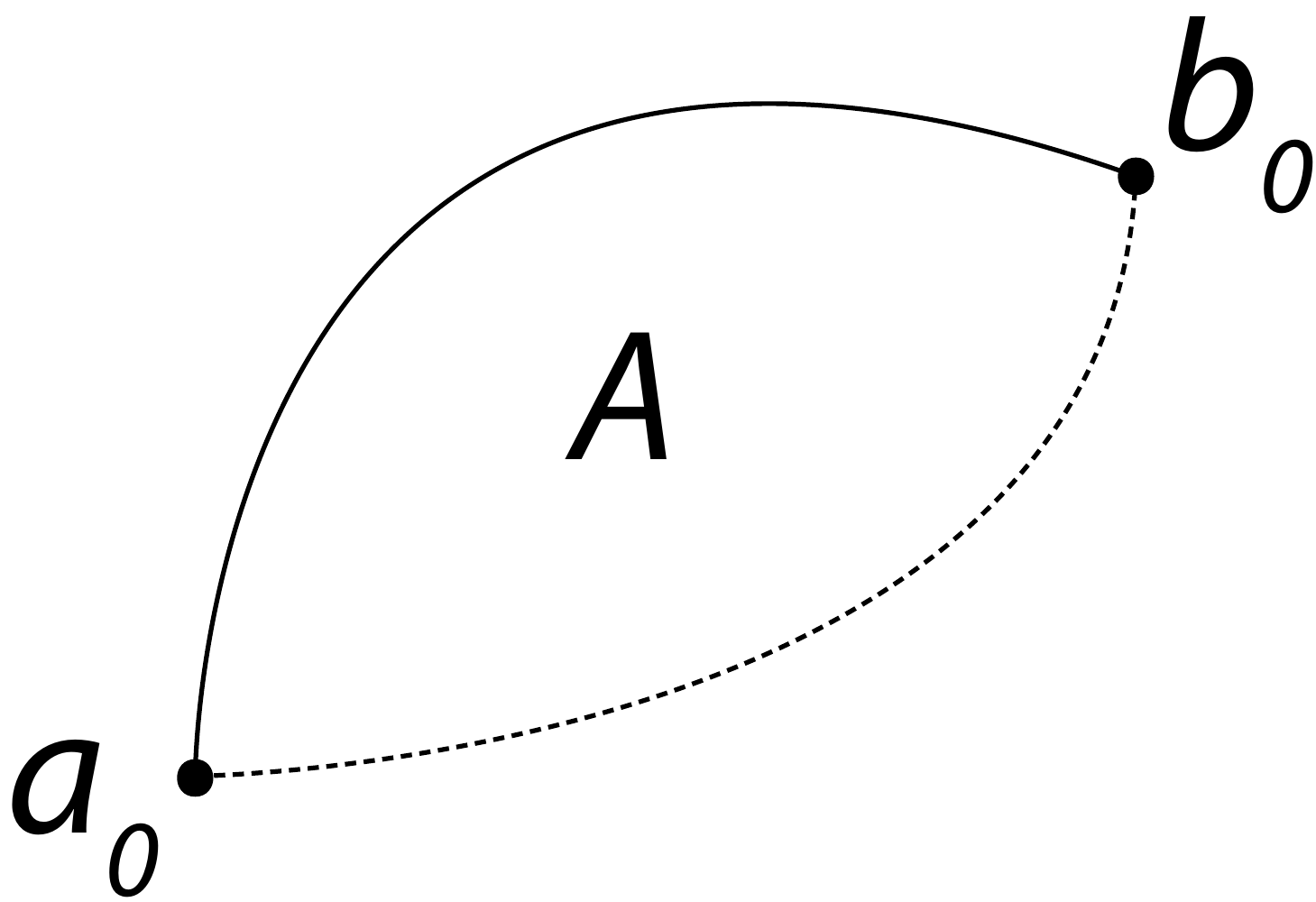}}
 \caption{$a_{0}$ and $b_{0}$ is a homoclinic pair.  They are connected by an unstable segment $U[a_{0},b_{0}]$ (solid) and a stable segment $S[b_{0},a_{0}]$ (dashed).  Then the action difference between the homoclinic orbit pair is $\Delta {\cal F}_{\lbrace b_0\rbrace \lbrace a_0\rbrace}=A$.    }
\label{fig:Homoclinic_pair_action}
\end{figure}

  
The central step to obtain the MacKay-Meiss-Percival action principle is demonstrated by Fig.~39 along with Eq.~(5.6) in~\cite{Meiss92}.  Shown here in Fig.~\ref{fig:Area_under_curve} are two arbitrary points $a=(q_{a},p_{a})$, $b=(q_{b},p_{b})$ and their images $a'=M(a)$, $b'=M(b)$.  Let $c$ be an arbitrary curve connecting $a$ and $b$, which is mapped to a curve $c'=M(c)$ connecting $a'$ and $b'$.  Let $A$ and $A'$ denote the algebraic area under $c$ and $c'$ respectively.  Then the difference between these areas is
\begin{equation}\label{eq:Meiss92}
\begin{split}
A'-A&=\int_{c'}p\mathrm{d}q-\int_{c}p\mathrm{d}q\\
&=F(q_{b},q_{b'})-F(q_{a},q_{a'})
\end{split}
\end{equation}
i.e., the difference between the two algebraic areas gives the difference between the action functions for one iteration of the map.  Starting from this, MacKay $\mathit{et}$ $\mathit{al.}$ \cite{MacKay84a} derived a formula relating the action difference between a pair of homoclinic orbits to the phase space area of a region bounded by stable and unstable manifolds, as demonstrated by Fig.~\ref{fig:Homoclinic_pair_action}.
In this Figure, $a_{0}$ and $b_{0}$ is a pair of homoclinic points:
\begin{equation}\label{eq:asymptotic pair}
a_{\pm\infty} \to b_{\pm\infty}\ .
\end{equation}
There exist unstable and stable manifolds connecting the two points shown by the solid and dashed curves.  Those manifolds could be the manifolds of other fixed points, or manifolds associated with $a_{0}$ and $b_{0}$ themselves.  Let $U[a_{0},b_{0}]$ and $S[b_{0},a_{0}]$ be the corresponding segments, then the action difference between $\lbrace a_{0}\rbrace$ and $\lbrace b_{0}\rbrace$ is given by:
\begin{equation}\label{eq:homoclinic action difference appendix}
\begin{split}
\Delta {\cal F}_{\lbrace b_0\rbrace \lbrace a_0\rbrace}&=\sum_{n=-\infty}^{\infty}\big[ F_{\lbrace b_0\rbrace}(q_n,q_{n+1})-F_{\lbrace a_0\rbrace}(q_{n},q_{n+1}) \big]\\
&=\int\limits_{U[a_{0},b_{0}]}p\mathrm{d}q+\int\limits_{S[b_{0},a_{0}]}p\mathrm{d}q=A
\end{split}
\end{equation}
where $A$ denotes the area shown in Fig.~\ref{fig:Homoclinic_pair_action}.

\section{Normal form coordinates, Moser invariant curves and satellite periodic orbits}
\label{Normal form}

There are infinite families of unstable periodic obits accumulating on every homoclinic orbit \cite{Silva87,Ozorio89}.  These orbits are supported by Moser invariant curves, with the orbit points being successive self- or mutual-intersections between the invariant curves.  The existence of such curves and orbits is a consequence of the Birkhoff-Moser theorem \cite{Birkhoff27,Moser56,Silva87}.  If the Poincar\'{e} map is invertible and analytic, there exists an analytic transformation  (normal form transformation) from the normal form coordinates $(Q,P)$ to the neighborhood of stable and unstable manifolds of the hyperbolic fixed point, for which the map takes the simple form:  
\begin{equation}\label{eq:normal form}
\begin{split}
&Q_{n+1}=\Lambda(Q_{n}P_{n})\cdot Q_{n}\\
&P_{n+1}=[\Lambda(Q_{n}P_{n})]^{-1}\cdot P_{n}
\end{split}
\end{equation}
where $\Lambda(Q_{n}P_{n})$ is a polynomial function of the product $Q_{n}P_{n}$ \cite{Harsoula15}: 
\begin{equation}\label{eq:Lambda}
\Lambda(QP)=\lambda+w_{2}\cdot (QP)+w_{3}\cdot (QP)^{2}+\cdots
\end{equation}
with $\lambda=e^{\mu}$, where $\mu$ is the Lyapunov exponent of the fixed point.  The normal form convergence zone was first proved by Moser~\cite{Moser56} to be a small disk-shaped region centered at the fixed point, and later proved by da Silva Ritter $\mathit{et.}$ $\mathit{al.}$~\cite{Silva87} to extend along the stable and unstable manifolds into infinity.  The extended convergence zone follows hyperbolae to the manifolds  (``gets exponentially close'' the further out along the manifolds).  The stable and unstable manifolds are just images of the $P$ and $Q$ axes respectively under the normal form transformation.  Every homoclinic intersection point in phase space is mapped to two points $H_{P}=(0,P_{H})$ and $H_{Q}=(Q_{H},0)$.  

All points inside the extended convergence zone near the $Q$ or $P$ axis move along invariant hyperbolas, which are mapped to Moser invariant curves in phase space.  Being confined in the extended convergence zone, the Moser invariant curves also get exponentially close to the stable and unstable manifolds while extending along them outward to infinity.  In fact, as shown by~\cite{Harsoula15}, the convergence zone can be quantified using the outermost Moser curve with the largest $QP$ product.  

Self- and mutual-intersections between certain Moser invariant curves give rise to infinite families of periodic orbits.  A simple example is shown in Fig.~\ref{fig:Ozorio}.  Since the Moser invariant curve (dotted line) extends along $S(x)$ and $U(x)$, intersections between $S(x)$ and $U(x)$ will ``force" it to make self-intersections.  Its topological behavior is thus determined by the topology of the homoclinic tangle.  For example, when $S(x)$ and $U(x)$ make an intersection $h_0$, it is forced to it to make a self-intersection at $y_{0}$.  Thus, one can say that $y_{0}$ is induced by $h_0$.  Special choices of Moser curve can be found for each large enough integer $N$ to make $\lbrace y_{0}\rbrace$ a period-$N$ orbit.  The detailed numerical technique is demonstrated in~\cite{Silva87}, where the position of $y_{0}$ is explicitly calculated using a linearization in the neighborhood of the homoclinic point.  The upper panel of Fig.~\ref{fig:Ozorio} shows a period-$4$ orbit $\lbrace y_{0}\rbrace$.  The lower panel is the picture in $(Q,P)$.  Under $4$ iterations, $y_{0}$ is mapped along the hyperbola into $y_{4}$.  Under the normal form transformation, the $P$ and $Q$ axis become $S(x)$ and $U(x)$ respectively, folding back to intersect each other at $h_0$.  The invariant hyperbolae fold in the same way, with the image of $y_{4}$ being identical to one of the images of $y_{0}$ due to their being at the self-intersection point.  The solution for $y_{0}$ is unique for every period $N$.  As $N$ becomes larger, $y_{0}$ gets closer to $h_0$.  The homoclinic orbit $\lbrace h_0\rbrace$ is the limiting case of the period-$N$ orbit $\lbrace y_{0}\rbrace$ when $N\to\infty$:
\begin{equation}
\lim_{N\to\infty}\lbrace y_{0}\rbrace=\lbrace h_0\rbrace.
\end{equation}
The terminology of~\cite{Silva87,Ozorio89}, refers to these $\lbrace y_{0}\rbrace$ as satellite orbits induced by $h_0$.  

\bibliography{classicalchaos,quantumchaos}

\begin{thebibliography}{48}%
\makeatletter
\providecommand \@ifxundefined [1]{%
 \@ifx{#1\undefined}
}%
\providecommand \@ifnum [1]{%
 \ifnum #1\expandafter \@firstoftwo
 \else \expandafter \@secondoftwo
 \fi
}%
\providecommand \@ifx [1]{%
 \ifx #1\expandafter \@firstoftwo
 \else \expandafter \@secondoftwo
 \fi
}%
\providecommand \natexlab [1]{#1}%
\providecommand \enquote  [1]{``#1''}%
\providecommand \bibnamefont  [1]{#1}%
\providecommand \bibfnamefont [1]{#1}%
\providecommand \citenamefont [1]{#1}%
\providecommand \href@noop [0]{\@secondoftwo}%
\providecommand \href [0]{\begingroup \@sanitize@url \@href}%
\providecommand \@href[1]{\@@startlink{#1}\@@href}%
\providecommand \@@href[1]{\endgroup#1\@@endlink}%
\providecommand \@sanitize@url [0]{\catcode `\\12\catcode `\$12\catcode
  `\&12\catcode `\#12\catcode `\^12\catcode `\_12\catcode `\%12\relax}%
\providecommand \@@startlink[1]{}%
\providecommand \@@endlink[0]{}%
\providecommand \url  [0]{\begingroup\@sanitize@url \@url }%
\providecommand \@url [1]{\endgroup\@href {#1}{\urlprefix }}%
\providecommand \urlprefix  [0]{URL }%
\providecommand \Eprint [0]{\href }%
\providecommand \doibase [0]{http://dx.doi.org/}%
\providecommand \selectlanguage [0]{\@gobble}%
\providecommand \bibinfo  [0]{\@secondoftwo}%
\providecommand \bibfield  [0]{\@secondoftwo}%
\providecommand \translation [1]{[#1]}%
\providecommand \BibitemOpen [0]{}%
\providecommand \bibitemStop [0]{}%
\providecommand \bibitemNoStop [0]{.\EOS\space}%
\providecommand \EOS [0]{\spacefactor3000\relax}%
\providecommand \BibitemShut  [1]{\csname bibitem#1\endcsname}%
\let\auto@bib@innerbib\@empty
\bibitem [{\citenamefont {Poincar\'e}(1899)}]{Poincare99}%
  \BibitemOpen
  \bibfield  {author} {\bibinfo {author} {\bibfnamefont {H.}~\bibnamefont
  {Poincar\'e}},\ }\href@noop {} {\emph {\bibinfo {title} {Les m\'ethodes
  nouvelles de la m\'ecanique c\'eleste}}},\ Vol.~\bibinfo {volume} {3}\
  (\bibinfo  {publisher} {Gauthier-Villars et fils},\ \bibinfo {address}
  {Paris},\ \bibinfo {year} {1899})\BibitemShut {NoStop}%
\bibitem [{\citenamefont {So}(2007)}]{So07}%
  \BibitemOpen
  \bibfield  {author} {\bibinfo {author} {\bibfnamefont {P.}~\bibnamefont
  {So}},\ }\href@noop {} {\bibfield  {journal} {\bibinfo  {journal}
  {Scholarpedia}\ }\textbf {\bibinfo {volume} {2}},\ \bibinfo {pages} {1353}
  (\bibinfo {year} {2007})}\BibitemShut {NoStop}%
\bibitem [{\citenamefont {Gutzwiller}(1971)}]{Gutzwiller71}%
  \BibitemOpen
  \bibfield  {author} {\bibinfo {author} {\bibfnamefont {M.~C.}\ \bibnamefont
  {Gutzwiller}},\ }\href@noop {} {\bibfield  {journal} {\bibinfo  {journal}
  {J.~Math.~Phys.}\ }\textbf {\bibinfo {volume} {12}},\ \bibinfo {pages} {343}
  (\bibinfo {year} {1971})},\ \bibinfo {note} {and references
  therein}\BibitemShut {NoStop}%
\bibitem [{\citenamefont {Balian}\ and\ \citenamefont
  {Bloch}(1971)}]{Balian71}%
  \BibitemOpen
  \bibfield  {author} {\bibinfo {author} {\bibfnamefont {R.}~\bibnamefont
  {Balian}}\ and\ \bibinfo {author} {\bibfnamefont {C.}~\bibnamefont {Bloch}},\
  }\href@noop {} {\bibfield  {journal} {\bibinfo  {journal} {Ann.~Phys.
  (N.Y.)}\ }\textbf {\bibinfo {volume} {63}},\ \bibinfo {pages} {592} (\bibinfo
  {year} {1971})}\BibitemShut {NoStop}%
\bibitem [{\citenamefont {Berry}\ and\ \citenamefont {Tabor}(1976)}]{Berry76}%
  \BibitemOpen
  \bibfield  {author} {\bibinfo {author} {\bibfnamefont {M.~V.}\ \bibnamefont
  {Berry}}\ and\ \bibinfo {author} {\bibfnamefont {M.}~\bibnamefont {Tabor}},\
  }\href@noop {} {\bibfield  {journal} {\bibinfo  {journal}
  {Proc.~R.~Soc.~Lond.~A}\ }\textbf {\bibinfo {volume} {349}},\ \bibinfo
  {pages} {101} (\bibinfo {year} {1976})}\BibitemShut {NoStop}%
\bibitem [{\citenamefont {Du}\ and\ \citenamefont
  {Delos}(1988{\natexlab{a}})}]{Du88a}%
  \BibitemOpen
  \bibfield  {author} {\bibinfo {author} {\bibfnamefont {M.~L.}\ \bibnamefont
  {Du}}\ and\ \bibinfo {author} {\bibfnamefont {J.~B.}\ \bibnamefont {Delos}},\
  }\href@noop {} {\bibfield  {journal} {\bibinfo  {journal} {Phys.~Rev.~A}\
  }\textbf {\bibinfo {volume} {38}},\ \bibinfo {pages} {1896} (\bibinfo {year}
  {1988}{\natexlab{a}})}\BibitemShut {NoStop}%
\bibitem [{\citenamefont {Du}\ and\ \citenamefont
  {Delos}(1988{\natexlab{b}})}]{Du88b}%
  \BibitemOpen
  \bibfield  {author} {\bibinfo {author} {\bibfnamefont {M.~L.}\ \bibnamefont
  {Du}}\ and\ \bibinfo {author} {\bibfnamefont {J.~B.}\ \bibnamefont {Delos}},\
  }\href@noop {} {\bibfield  {journal} {\bibinfo  {journal} {Phys.~Rev.~A}\
  }\textbf {\bibinfo {volume} {38}},\ \bibinfo {pages} {1913} (\bibinfo {year}
  {1988}{\natexlab{b}})}\BibitemShut {NoStop}%
\bibitem [{\citenamefont {Friedrich}\ and\ \citenamefont
  {Wintgen}(1989)}]{Friedrich89}%
  \BibitemOpen
  \bibfield  {author} {\bibinfo {author} {\bibfnamefont {H.}~\bibnamefont
  {Friedrich}}\ and\ \bibinfo {author} {\bibfnamefont {D.}~\bibnamefont
  {Wintgen}},\ }\href@noop {} {\bibfield  {journal} {\bibinfo  {journal}
  {Phys.~Rep.}\ }\textbf {\bibinfo {volume} {183}},\ \bibinfo {pages} {37}
  (\bibinfo {year} {1989})}\BibitemShut {NoStop}%
\bibitem [{\citenamefont {Tomsovic}\ and\ \citenamefont
  {Heller}(1991)}]{Tomsovic91b}%
  \BibitemOpen
  \bibfield  {author} {\bibinfo {author} {\bibfnamefont {S.}~\bibnamefont
  {Tomsovic}}\ and\ \bibinfo {author} {\bibfnamefont {E.~J.}\ \bibnamefont
  {Heller}},\ }\href@noop {} {\bibfield  {journal} {\bibinfo  {journal}
  {Phys.~Rev.~Lett.}\ }\textbf {\bibinfo {volume} {67}},\ \bibinfo {pages}
  {664} (\bibinfo {year} {1991})}\BibitemShut {NoStop}%
\bibitem [{\citenamefont {Tomsovic}\ and\ \citenamefont
  {Heller}(1993)}]{Tomsovic93}%
  \BibitemOpen
  \bibfield  {author} {\bibinfo {author} {\bibfnamefont {S.}~\bibnamefont
  {Tomsovic}}\ and\ \bibinfo {author} {\bibfnamefont {E.~J.}\ \bibnamefont
  {Heller}},\ }\href@noop {} {\bibfield  {journal} {\bibinfo  {journal}
  {Phys.~Rev.~E}\ }\textbf {\bibinfo {volume} {47}},\ \bibinfo {pages} {282}
  (\bibinfo {year} {1993})}\BibitemShut {NoStop}%
\bibitem [{\citenamefont {Tanner}\ \emph {et~al.}(1991)\citenamefont {Tanner},
  \citenamefont {Scherer}, \citenamefont {Bogomonly}, \citenamefont
  {Eckhardt},\ and\ \citenamefont {Wintgen}}]{Tanner91}%
  \BibitemOpen
  \bibfield  {author} {\bibinfo {author} {\bibfnamefont {G.}~\bibnamefont
  {Tanner}}, \bibinfo {author} {\bibfnamefont {P.}~\bibnamefont {Scherer}},
  \bibinfo {author} {\bibfnamefont {E.~B.}\ \bibnamefont {Bogomonly}}, \bibinfo
  {author} {\bibfnamefont {B.}~\bibnamefont {Eckhardt}}, \ and\ \bibinfo
  {author} {\bibfnamefont {D.}~\bibnamefont {Wintgen}},\ }\href@noop {}
  {\bibfield  {journal} {\bibinfo  {journal} {Phys.~Rev.~Lett.}\ }\textbf
  {\bibinfo {volume} {67}},\ \bibinfo {pages} {2410} (\bibinfo {year}
  {1991})}\BibitemShut {NoStop}%
\bibitem [{\citenamefont {Cvitanovi\'{c}}\ and\ \citenamefont
  {Eckhardt}(1989)}]{Cvitanovic89}%
  \BibitemOpen
  \bibfield  {author} {\bibinfo {author} {\bibfnamefont {P.}~\bibnamefont
  {Cvitanovi\'{c}}}\ and\ \bibinfo {author} {\bibfnamefont {B.}~\bibnamefont
  {Eckhardt}},\ }\href@noop {} {\bibfield  {journal} {\bibinfo  {journal}
  {Phys.~Rev.~Lett.}\ }\textbf {\bibinfo {volume} {63}},\ \bibinfo {pages}
  {823} (\bibinfo {year} {1989})}\BibitemShut {NoStop}%
\bibitem [{\citenamefont {Berry}\ and\ \citenamefont
  {Keating}(1990)}]{Berry90}%
  \BibitemOpen
  \bibfield  {author} {\bibinfo {author} {\bibfnamefont {M.~V.}\ \bibnamefont
  {Berry}}\ and\ \bibinfo {author} {\bibfnamefont {J.~P.}\ \bibnamefont
  {Keating}},\ }\href@noop {} {\bibfield  {journal} {\bibinfo  {journal}
  {J.~Phys.~A}\ }\textbf {\bibinfo {volume} {23}},\ \bibinfo {pages} {4839}
  (\bibinfo {year} {1990})}\BibitemShut {NoStop}%
\bibitem [{\citenamefont {Argaman}\ \emph {et~al.}(1993)\citenamefont
  {Argaman}, \citenamefont {Dittes}, \citenamefont {Doron}, \citenamefont
  {Keating}, \citenamefont {Kitaev}, \citenamefont {Sieber},\ and\
  \citenamefont {Smilansky}}]{Argaman93}%
  \BibitemOpen
  \bibfield  {author} {\bibinfo {author} {\bibfnamefont {N.}~\bibnamefont
  {Argaman}}, \bibinfo {author} {\bibfnamefont {F.-M.}\ \bibnamefont {Dittes}},
  \bibinfo {author} {\bibfnamefont {E.}~\bibnamefont {Doron}}, \bibinfo
  {author} {\bibfnamefont {J.~P.}\ \bibnamefont {Keating}}, \bibinfo {author}
  {\bibfnamefont {A.~Y.}\ \bibnamefont {Kitaev}}, \bibinfo {author}
  {\bibfnamefont {M.}~\bibnamefont {Sieber}}, \ and\ \bibinfo {author}
  {\bibfnamefont {U.}~\bibnamefont {Smilansky}},\ }\href@noop {} {\bibfield
  {journal} {\bibinfo  {journal} {Phys.~Rev.~Lett.}\ }\textbf {\bibinfo
  {volume} {71}},\ \bibinfo {pages} {4326} (\bibinfo {year}
  {1993})}\BibitemShut {NoStop}%
\bibitem [{\citenamefont {Ozorio~de Almeida}(1989)}]{Ozorio89}%
  \BibitemOpen
  \bibfield  {author} {\bibinfo {author} {\bibfnamefont {A.~M.}\ \bibnamefont
  {Ozorio~de Almeida}},\ }\href@noop {} {\bibfield  {journal} {\bibinfo
  {journal} {Nonlinearity}\ }\textbf {\bibinfo {volume} {2}},\ \bibinfo {pages}
  {519} (\bibinfo {year} {1989})}\BibitemShut {NoStop}%
\bibitem [{\citenamefont {Bogomolny}(1992)}]{Bogomolny92}%
  \BibitemOpen
  \bibfield  {author} {\bibinfo {author} {\bibfnamefont {E.~B.}\ \bibnamefont
  {Bogomolny}},\ }\href@noop {} {\bibfield  {journal} {\bibinfo  {journal}
  {Chaos}\ }\textbf {\bibinfo {volume} {2}},\ \bibinfo {pages} {5} (\bibinfo
  {year} {1992})}\BibitemShut {NoStop}%
\bibitem [{\citenamefont {Sieber}\ and\ \citenamefont
  {Richter}(2001)}]{Sieber01}%
  \BibitemOpen
  \bibfield  {author} {\bibinfo {author} {\bibfnamefont {M.}~\bibnamefont
  {Sieber}}\ and\ \bibinfo {author} {\bibfnamefont {K.}~\bibnamefont
  {Richter}},\ }\href@noop {} {\bibfield  {journal} {\bibinfo  {journal}
  {Physica Scripta}\ }\textbf {\bibinfo {volume} {T90}},\ \bibinfo {pages}
  {128} (\bibinfo {year} {2001})}\BibitemShut {NoStop}%
\bibitem [{\citenamefont {M\"{u}ller}\ \emph {et~al.}(2004)\citenamefont
  {M\"{u}ller}, \citenamefont {Heusler}, \citenamefont {Braun}, \citenamefont
  {Haake},\ and\ \citenamefont {Altland}}]{Muller04}%
  \BibitemOpen
  \bibfield  {author} {\bibinfo {author} {\bibfnamefont {S.}~\bibnamefont
  {M\"{u}ller}}, \bibinfo {author} {\bibfnamefont {S.}~\bibnamefont {Heusler}},
  \bibinfo {author} {\bibfnamefont {P.}~\bibnamefont {Braun}}, \bibinfo
  {author} {\bibfnamefont {F.}~\bibnamefont {Haake}}, \ and\ \bibinfo {author}
  {\bibfnamefont {A.}~\bibnamefont {Altland}},\ }\href@noop {} {\bibfield
  {journal} {\bibinfo  {journal} {Phys.~Rev.~Lett.}\ }\textbf {\bibinfo
  {volume} {93}},\ \bibinfo {pages} {014103} (\bibinfo {year}
  {2004})}\BibitemShut {NoStop}%
\bibitem [{\citenamefont {Turek}\ \emph {et~al.}(2005)\citenamefont {Turek},
  \citenamefont {Spehner}, \citenamefont {M\"{u}ller},\ and\ \citenamefont
  {Richter}}]{Turek05}%
  \BibitemOpen
  \bibfield  {author} {\bibinfo {author} {\bibfnamefont {M.}~\bibnamefont
  {Turek}}, \bibinfo {author} {\bibfnamefont {D.}~\bibnamefont {Spehner}},
  \bibinfo {author} {\bibfnamefont {S.}~\bibnamefont {M\"{u}ller}}, \ and\
  \bibinfo {author} {\bibfnamefont {K.}~\bibnamefont {Richter}},\ }\href@noop
  {} {\bibfield  {journal} {\bibinfo  {journal} {Phys.~Rev.~E}\ }\textbf
  {\bibinfo {volume} {71}},\ \bibinfo {pages} {016210} (\bibinfo {year}
  {2005})}\BibitemShut {NoStop}%
\bibitem [{\citenamefont {M\"{u}ller}\ \emph {et~al.}(2005)\citenamefont
  {M\"{u}ller}, \citenamefont {Heusler}, \citenamefont {Braun}, \citenamefont
  {Haake},\ and\ \citenamefont {Altland}}]{Muller05}%
  \BibitemOpen
  \bibfield  {author} {\bibinfo {author} {\bibfnamefont {S.}~\bibnamefont
  {M\"{u}ller}}, \bibinfo {author} {\bibfnamefont {S.}~\bibnamefont {Heusler}},
  \bibinfo {author} {\bibfnamefont {P.}~\bibnamefont {Braun}}, \bibinfo
  {author} {\bibfnamefont {F.}~\bibnamefont {Haake}}, \ and\ \bibinfo {author}
  {\bibfnamefont {A.}~\bibnamefont {Altland}},\ }\href@noop {} {\bibfield
  {journal} {\bibinfo  {journal} {Phys.~Rev.~E}\ }\textbf {\bibinfo {volume}
  {72}},\ \bibinfo {pages} {046207} (\bibinfo {year} {2005})}\BibitemShut
  {NoStop}%
\bibitem [{\citenamefont {Moser}(1956)}]{Moser56}%
  \BibitemOpen
  \bibfield  {author} {\bibinfo {author} {\bibfnamefont {J.}~\bibnamefont
  {Moser}},\ }\href@noop {} {\bibfield  {journal} {\bibinfo  {journal}
  {Commun.~Pure Appl.~Math.}\ }\textbf {\bibinfo {volume} {9}},\ \bibinfo
  {pages} {673} (\bibinfo {year} {1956})}\BibitemShut {NoStop}%
\bibitem [{\citenamefont {da~Silva~Ritter}\ \emph {et~al.}(1987)\citenamefont
  {da~Silva~Ritter}, \citenamefont {Ozorio~de Almeida},\ and\ \citenamefont
  {Douady}}]{Silva87}%
  \BibitemOpen
  \bibfield  {author} {\bibinfo {author} {\bibfnamefont {G.~L.}\ \bibnamefont
  {da~Silva~Ritter}}, \bibinfo {author} {\bibfnamefont {A.~M.}\ \bibnamefont
  {Ozorio~de Almeida}}, \ and\ \bibinfo {author} {\bibfnamefont
  {R.}~\bibnamefont {Douady}},\ }\href@noop {} {\bibfield  {journal} {\bibinfo
  {journal} {Physica~D}\ }\textbf {\bibinfo {volume} {29}},\ \bibinfo {pages}
  {181} (\bibinfo {year} {1987})}\BibitemShut {NoStop}%
\bibitem [{\citenamefont {Birkhoff}(1927)}]{Birkhoff27}%
  \BibitemOpen
  \bibfield  {author} {\bibinfo {author} {\bibfnamefont {G.~D.}\ \bibnamefont
  {Birkhoff}},\ }\href@noop {} {\bibfield  {journal} {\bibinfo  {journal} {Acta
  Math.}\ }\textbf {\bibinfo {volume} {50}},\ \bibinfo {pages} {359} (\bibinfo
  {year} {1927})}\BibitemShut {NoStop}%
\bibitem [{\citenamefont {Harsoula}\ \emph {et~al.}(2015)\citenamefont
  {Harsoula}, \citenamefont {Contopoulos},\ and\ \citenamefont
  {Efthymiopoulos}}]{Harsoula15}%
  \BibitemOpen
  \bibfield  {author} {\bibinfo {author} {\bibfnamefont {M.}~\bibnamefont
  {Harsoula}}, \bibinfo {author} {\bibfnamefont {G.}~\bibnamefont
  {Contopoulos}}, \ and\ \bibinfo {author} {\bibfnamefont {C.}~\bibnamefont
  {Efthymiopoulos}},\ }\href@noop {} {\bibfield  {journal} {\bibinfo  {journal}
  {J.~Phys.~A: Math.~Theor.}\ }\textbf {\bibinfo {volume} {48}},\ \bibinfo
  {pages} {135102} (\bibinfo {year} {2015})},\ \bibinfo {note}
  {arXiv:1502.00664 [nlin.CD]}\BibitemShut {NoStop}%
\bibitem [{\citenamefont {Contopoulos}\ and\ \citenamefont
  {Harsoula}(2015)}]{Contopoulos15}%
  \BibitemOpen
  \bibfield  {author} {\bibinfo {author} {\bibfnamefont {G.}~\bibnamefont
  {Contopoulos}}\ and\ \bibinfo {author} {\bibfnamefont {M.}~\bibnamefont
  {Harsoula}},\ }\href@noop {} {\bibfield  {journal} {\bibinfo  {journal}
  {J.~Phys.~A: Math.~Theor.}\ }\textbf {\bibinfo {volume} {48}},\ \bibinfo
  {pages} {335101} (\bibinfo {year} {2015})}\BibitemShut {NoStop}%
\bibitem [{\citenamefont {H\'enon}(1969)}]{Henon69}%
  \BibitemOpen
  \bibfield  {author} {\bibinfo {author} {\bibfnamefont {M.}~\bibnamefont
  {H\'enon}},\ }\href@noop {} {\bibfield  {journal} {\bibinfo  {journal}
  {Quart.~Appl.~Math.}\ }\textbf {\bibinfo {volume} {27}},\ \bibinfo {pages}
  {291} (\bibinfo {year} {1969})}\BibitemShut {NoStop}%
\bibitem [{\citenamefont {O'Connor}\ \emph {et~al.}(1992)\citenamefont
  {O'Connor}, \citenamefont {Tomsovic},\ and\ \citenamefont
  {Heller}}]{Oconnor92}%
  \BibitemOpen
  \bibfield  {author} {\bibinfo {author} {\bibfnamefont {P.~W.}\ \bibnamefont
  {O'Connor}}, \bibinfo {author} {\bibfnamefont {S.}~\bibnamefont {Tomsovic}},
  \ and\ \bibinfo {author} {\bibfnamefont {E.~J.}\ \bibnamefont {Heller}},\
  }\href@noop {} {\bibfield  {journal} {\bibinfo  {journal} {Physica D}\
  }\textbf {\bibinfo {volume} {55}},\ \bibinfo {pages} {340} (\bibinfo {year}
  {1992})}\BibitemShut {NoStop}%
\bibitem [{\citenamefont {Creagh}\ \emph {et~al.}(1990)\citenamefont {Creagh},
  \citenamefont {Robbins},\ and\ \citenamefont {Littlejohn}}]{Creagh90}%
  \BibitemOpen
  \bibfield  {author} {\bibinfo {author} {\bibfnamefont {S.~C.}\ \bibnamefont
  {Creagh}}, \bibinfo {author} {\bibfnamefont {J.~M.}\ \bibnamefont {Robbins}},
  \ and\ \bibinfo {author} {\bibfnamefont {R.~G.}\ \bibnamefont {Littlejohn}},\
  }\href@noop {} {\bibfield  {journal} {\bibinfo  {journal} {Phys.~Rev.~A}\
  }\textbf {\bibinfo {volume} {42}},\ \bibinfo {pages} {1907} (\bibinfo {year}
  {1990})}\BibitemShut {NoStop}%
\bibitem [{\citenamefont {Esterlis}\ \emph {et~al.}(2014)\citenamefont
  {Esterlis}, \citenamefont {Haggard}, \citenamefont {Hedeman},\ and\
  \citenamefont {Littlejohn}}]{Esterlis14}%
  \BibitemOpen
  \bibfield  {author} {\bibinfo {author} {\bibfnamefont {I.}~\bibnamefont
  {Esterlis}}, \bibinfo {author} {\bibfnamefont {H.~M.}\ \bibnamefont
  {Haggard}}, \bibinfo {author} {\bibfnamefont {A.}~\bibnamefont {Hedeman}}, \
  and\ \bibinfo {author} {\bibfnamefont {R.~G.}\ \bibnamefont {Littlejohn}},\
  }\href@noop {} {\bibfield  {journal} {\bibinfo  {journal} {Europhys. ~Lett.}\
  }\textbf {\bibinfo {volume} {106}},\ \bibinfo {pages} {50002} (\bibinfo
  {year} {2014})}\BibitemShut {NoStop}%
\bibitem [{\citenamefont {Mao}\ \emph {et~al.}(1992)\citenamefont {Mao},
  \citenamefont {Shaw},\ and\ \citenamefont {Delos}}]{Mao92}%
  \BibitemOpen
  \bibfield  {author} {\bibinfo {author} {\bibfnamefont {J.-M.}\ \bibnamefont
  {Mao}}, \bibinfo {author} {\bibfnamefont {J.}~\bibnamefont {Shaw}}, \ and\
  \bibinfo {author} {\bibfnamefont {J.~B.}\ \bibnamefont {Delos}},\ }\href@noop
  {} {\bibfield  {journal} {\bibinfo  {journal} {J.~Stat.~Phys.}\ }\textbf
  {\bibinfo {volume} {68}},\ \bibinfo {pages} {51} (\bibinfo {year}
  {1992})}\BibitemShut {NoStop}%
\bibitem [{\citenamefont {MacKay}\ \emph {et~al.}(1984)\citenamefont {MacKay},
  \citenamefont {Meiss},\ and\ \citenamefont {Percival}}]{MacKay84a}%
  \BibitemOpen
  \bibfield  {author} {\bibinfo {author} {\bibfnamefont {R.~S.}\ \bibnamefont
  {MacKay}}, \bibinfo {author} {\bibfnamefont {J.~D.}\ \bibnamefont {Meiss}}, \
  and\ \bibinfo {author} {\bibfnamefont {I.~C.}\ \bibnamefont {Percival}},\
  }\href@noop {} {\bibfield  {journal} {\bibinfo  {journal} {Physica~D}\
  }\textbf {\bibinfo {volume} {13}},\ \bibinfo {pages} {55} (\bibinfo {year}
  {1984})}\BibitemShut {NoStop}%
\bibitem [{\citenamefont {Wiggins}(1992)}]{Wiggins92}%
  \BibitemOpen
  \bibfield  {author} {\bibinfo {author} {\bibfnamefont {S.}~\bibnamefont
  {Wiggins}},\ }\href@noop {} {\emph {\bibinfo {title} {Chaotic Transport in
  Dynamical Systems}}}\ (\bibinfo  {publisher} {Springer},\ \bibinfo {address}
  {New York},\ \bibinfo {year} {1992})\BibitemShut {NoStop}%
\bibitem [{\citenamefont {Easton}(1986)}]{Easton86}%
  \BibitemOpen
  \bibfield  {author} {\bibinfo {author} {\bibfnamefont {R.}~\bibnamefont
  {Easton}},\ }\href@noop {} {\bibfield  {journal} {\bibinfo  {journal}
  {Trans.~Am.~Math.~Soc.}\ }\textbf {\bibinfo {volume} {294}},\ \bibinfo
  {pages} {2} (\bibinfo {year} {1986})}\BibitemShut {NoStop}%
\bibitem [{\citenamefont {Rom-Kedar}(1990)}]{Rom-Kedar90}%
  \BibitemOpen
  \bibfield  {author} {\bibinfo {author} {\bibfnamefont {V.}~\bibnamefont
  {Rom-Kedar}},\ }\href@noop {} {\bibfield  {journal} {\bibinfo  {journal}
  {Physica~D}\ }\textbf {\bibinfo {volume} {43}},\ \bibinfo {pages} {229}
  (\bibinfo {year} {1990})}\BibitemShut {NoStop}%
\bibitem [{\citenamefont {Meiss}(1992)}]{Meiss92}%
  \BibitemOpen
  \bibfield  {author} {\bibinfo {author} {\bibfnamefont {J.~D.}\ \bibnamefont
  {Meiss}},\ }\href@noop {} {\bibfield  {journal} {\bibinfo  {journal}
  {Rev.~Mod.~Phys.}\ }\textbf {\bibinfo {volume} {64}},\ \bibinfo {pages} {795}
  (\bibinfo {year} {1992})}\BibitemShut {NoStop}%
\bibitem [{\citenamefont {Bensimon}\ and\ \citenamefont
  {Kadanoff}(1984)}]{Bensimon84}%
  \BibitemOpen
  \bibfield  {author} {\bibinfo {author} {\bibfnamefont {D.}~\bibnamefont
  {Bensimon}}\ and\ \bibinfo {author} {\bibfnamefont {L.~P.}\ \bibnamefont
  {Kadanoff}},\ }\href@noop {} {\bibfield  {journal} {\bibinfo  {journal}
  {Physica~D}\ }\textbf {\bibinfo {volume} {13}},\ \bibinfo {pages} {82}
  (\bibinfo {year} {1984})}\BibitemShut {NoStop}%
\bibitem [{\citenamefont {Mitchell}\ \emph
  {et~al.}(2003{\natexlab{a}})\citenamefont {Mitchell}, \citenamefont
  {Handley}, \citenamefont {Tighe}, \citenamefont {Delos},\ and\ \citenamefont
  {Knudson}}]{Mitchell03a}%
  \BibitemOpen
  \bibfield  {author} {\bibinfo {author} {\bibfnamefont {K.~A.}\ \bibnamefont
  {Mitchell}}, \bibinfo {author} {\bibfnamefont {J.~P.}\ \bibnamefont
  {Handley}}, \bibinfo {author} {\bibfnamefont {B.}~\bibnamefont {Tighe}},
  \bibinfo {author} {\bibfnamefont {J.~B.}\ \bibnamefont {Delos}}, \ and\
  \bibinfo {author} {\bibfnamefont {S.~K.}\ \bibnamefont {Knudson}},\
  }\href@noop {} {\bibfield  {journal} {\bibinfo  {journal} {Chaos}\ }\textbf
  {\bibinfo {volume} {13}},\ \bibinfo {pages} {880} (\bibinfo {year}
  {2003}{\natexlab{a}})}\BibitemShut {NoStop}%
\bibitem [{\citenamefont {Mitchell}\ \emph
  {et~al.}(2003{\natexlab{b}})\citenamefont {Mitchell}, \citenamefont
  {Handley}, \citenamefont {Delos},\ and\ \citenamefont
  {Knudson}}]{Mitchell03b}%
  \BibitemOpen
  \bibfield  {author} {\bibinfo {author} {\bibfnamefont {K.~A.}\ \bibnamefont
  {Mitchell}}, \bibinfo {author} {\bibfnamefont {J.~P.}\ \bibnamefont
  {Handley}}, \bibinfo {author} {\bibfnamefont {J.~B.}\ \bibnamefont {Delos}},
  \ and\ \bibinfo {author} {\bibfnamefont {S.~K.}\ \bibnamefont {Knudson}},\
  }\href@noop {} {\bibfield  {journal} {\bibinfo  {journal} {Chaos}\ }\textbf
  {\bibinfo {volume} {13}},\ \bibinfo {pages} {892} (\bibinfo {year}
  {2003}{\natexlab{b}})}\BibitemShut {NoStop}%
\bibitem [{\citenamefont {Mitchell}\ and\ \citenamefont
  {Delos}(2006)}]{Mitchell06}%
  \BibitemOpen
  \bibfield  {author} {\bibinfo {author} {\bibfnamefont {K.~A.}\ \bibnamefont
  {Mitchell}}\ and\ \bibinfo {author} {\bibfnamefont {J.~B.}\ \bibnamefont
  {Delos}},\ }\href@noop {} {\bibfield  {journal} {\bibinfo  {journal}
  {Physica~D}\ }\textbf {\bibinfo {volume} {221}},\ \bibinfo {pages} {170}
  (\bibinfo {year} {2006})}\BibitemShut {NoStop}%
\bibitem [{\citenamefont {Mitchell}\ \emph {et~al.}(2004)\citenamefont
  {Mitchell}, \citenamefont {Handley}, \citenamefont {Tighe}, \citenamefont
  {Flower},\ and\ \citenamefont {Delos}}]{Mitchell04}%
  \BibitemOpen
  \bibfield  {author} {\bibinfo {author} {\bibfnamefont {K.~A.}\ \bibnamefont
  {Mitchell}}, \bibinfo {author} {\bibfnamefont {J.~P.}\ \bibnamefont
  {Handley}}, \bibinfo {author} {\bibfnamefont {B.}~\bibnamefont {Tighe}},
  \bibinfo {author} {\bibfnamefont {A.}~\bibnamefont {Flower}}, \ and\ \bibinfo
  {author} {\bibfnamefont {J.~B.}\ \bibnamefont {Delos}},\ }\href@noop {}
  {\bibfield  {journal} {\bibinfo  {journal} {Phys.~Rev.~Lett.}\ }\textbf
  {\bibinfo {volume} {92}},\ \bibinfo {pages} {073001} (\bibinfo {year}
  {2004})}\BibitemShut {NoStop}%
\bibitem [{\citenamefont {Novick}\ \emph {et~al.}(2012)\citenamefont {Novick},
  \citenamefont {Keeler}, \citenamefont {Giefer},\ and\ \citenamefont
  {Delos}}]{Novick12a}%
  \BibitemOpen
  \bibfield  {author} {\bibinfo {author} {\bibfnamefont {J.}~\bibnamefont
  {Novick}}, \bibinfo {author} {\bibfnamefont {M.~L.}\ \bibnamefont {Keeler}},
  \bibinfo {author} {\bibfnamefont {J.}~\bibnamefont {Giefer}}, \ and\ \bibinfo
  {author} {\bibfnamefont {J.~B.}\ \bibnamefont {Delos}},\ }\href@noop {}
  {\bibfield  {journal} {\bibinfo  {journal} {Phys.~Rev.~E}\ }\textbf {\bibinfo
  {volume} {85}},\ \bibinfo {pages} {016205} (\bibinfo {year}
  {2012})}\BibitemShut {NoStop}%
\bibitem [{\citenamefont {Novick}\ and\ \citenamefont
  {Delos}(2012)}]{Novick12b}%
  \BibitemOpen
  \bibfield  {author} {\bibinfo {author} {\bibfnamefont {J.}~\bibnamefont
  {Novick}}\ and\ \bibinfo {author} {\bibfnamefont {J.~B.}\ \bibnamefont
  {Delos}},\ }\href@noop {} {\bibfield  {journal} {\bibinfo  {journal}
  {Phys.~Rev.~E}\ }\textbf {\bibinfo {volume} {85}},\ \bibinfo {pages} {016206}
  (\bibinfo {year} {2012})}\BibitemShut {NoStop}%
\bibitem [{\citenamefont {Chirikov}(1979)}]{Chirikov79}%
  \BibitemOpen
  \bibfield  {author} {\bibinfo {author} {\bibfnamefont {B.~V.}\ \bibnamefont
  {Chirikov}},\ }\href@noop {} {\bibfield  {journal} {\bibinfo  {journal}
  {Phys.~Rep.}\ }\textbf {\bibinfo {volume} {52}},\ \bibinfo {pages} {263}
  (\bibinfo {year} {1979})}\BibitemShut {NoStop}%
\bibitem [{\citenamefont {Li}\ and\ \citenamefont {Tomsovic}(2017)}]{Li17}%
  \BibitemOpen
  \bibfield  {author} {\bibinfo {author} {\bibfnamefont {J.}~\bibnamefont
  {Li}}\ and\ \bibinfo {author} {\bibfnamefont {S.}~\bibnamefont {Tomsovic}},\
  }\href@noop {} {\bibfield  {journal} {\bibinfo  {journal} {J.~Phys.~A:
  Math.~Theor.}\ }\textbf {\bibinfo {volume} {50}},\ \bibinfo {pages} {135101}
  (\bibinfo {year} {2017})},\ \bibinfo {note} {arXiv:1507.06455
  [nlin.CD]}\BibitemShut {NoStop}%
\bibitem [{\citenamefont {H\'enon}(1976)}]{Henon76}%
  \BibitemOpen
  \bibfield  {author} {\bibinfo {author} {\bibfnamefont {M.}~\bibnamefont
  {H\'enon}},\ }\href@noop {} {\bibfield  {journal} {\bibinfo  {journal}
  {Comm.~Math.~Phys.}\ }\textbf {\bibinfo {volume} {50}},\ \bibinfo {pages}
  {69} (\bibinfo {year} {1976})}\BibitemShut {NoStop}%
\bibitem [{\citenamefont {Haake}(2010)}]{Haake10}%
  \BibitemOpen
  \bibfield  {author} {\bibinfo {author} {\bibfnamefont {F.}~\bibnamefont
  {Haake}},\ }\href@noop {} {\emph {\bibinfo {title} {Quantum signatures of
  chaos, third edition}}}\ (\bibinfo  {publisher} {Springer},\ \bibinfo
  {address} {Heidelberg},\ \bibinfo {year} {2010})\BibitemShut {NoStop}%
\bibitem [{\citenamefont {Easton}(1991)}]{Easton91}%
  \BibitemOpen
  \bibfield  {author} {\bibinfo {author} {\bibfnamefont {R.}~\bibnamefont
  {Easton}},\ }\href@noop {} {\bibfield  {journal} {\bibinfo  {journal}
  {Nonlinearity}\ }\textbf {\bibinfo {volume} {4}},\ \bibinfo {pages} {583}
  (\bibinfo {year} {1991})}\BibitemShut {NoStop}%
\bibitem [{\citenamefont {Lomel\'{i}}\ and\ \citenamefont
  {Meiss}(2009)}]{Lomeli09}%
  \BibitemOpen
  \bibfield  {author} {\bibinfo {author} {\bibfnamefont {H.~E.}\ \bibnamefont
  {Lomel\'{i}}}\ and\ \bibinfo {author} {\bibfnamefont {J.~D.}\ \bibnamefont
  {Meiss}},\ }\href@noop {} {\bibfield  {journal} {\bibinfo  {journal}
  {Nonlinearity}\ }\textbf {\bibinfo {volume} {22}},\ \bibinfo {pages} {1761}
  (\bibinfo {year} {2009})}\BibitemShut {NoStop}%
\end{thebibliography}%

\end{document}